\documentclass[longauth]{aa}  
\usepackage{natbib} % nota bene: to get the .bbl file for the ArXiv submission, download the project, recompile it on the local machine and add the file to the overleaf project. 
\usepackage{txfonts}
\usepackage{graphicx}
\newcommand{\chisq} {$\chi$$^2$}
\newcommand{\Msun}{\>{\rm M_{\odot}}}

\newcommand{\num} {$\nu_{\rm max}$\xspace}

\newcommand{\dnu} {$\Delta\nu$\xspace}
\newcommand{\TESS} {TESS\xspace}
\newcommand{\Tess} {TESS\xspace}

\newcommand{\Gaia} {\textit{Gaia}\xspace}
\newcommand{\gaia} {\textit{Gaia}\xspace}
\newcommand{\GaiaDR} {\textit{Gaia}\,DR3\xspace}

\newcommand{\Kepler} {\textit{Kepler}\xspace}
\newcommand{\kepler} {\textit{Kepler}\xspace}

\def\dnu{$\Delta\nu$\xspace}
\def\dn1{$\delta\nu_{01}$\xspace}
\def\dn2{$\delta\nu_{02}$\xspace}

\newcommand{\Fig}[1]{Fig.\,\ref{#1}\xspace}

\usepackage[colorlinks=true,
                        linkcolor=cyan,
                        citecolor=cyan, 
                        filecolor=cyan, 
                        urlcolor=cyan]{hyperref}

\def\gmag{\ensuremath{G}\xspace}
\def\gbp{\ensuremath{G_\mathrm{BP}}\xspace}
\def\grp{\ensuremath{G_\mathrm{RP}}\xspace}

\begin{document} 

\title{Constraining stellar and orbital co-evolution through ensemble seismology of solar-like oscillators in binary systems\thanks{Tables \ref{tab:A1}, \ref{tab:B1}, and \ref{tab:ApogeeBinaryDetection} are only 
available in electronic form
at the CDS via anonymous ftp to {cdsarc.cds.unistra.fr} (130.79.128.5)
or via https://cdsarc.cds.unistra.fr/cgi-bin/qcat?J/A+A/}}
\subtitle{A census of oscillating red giants and dwarf stars in \GaiaDR binaries
}

\titlerunning{Constraining stellar and orbital co-evolution of \GaiaDR binary systems hosting solar-like oscillators}
\authorrunning{Beck\,et\,al.}

\author{P.\,G.\,Beck\inst{\ref{inst:IAC},\ref{inst:ULL},\ref{inst:Graz}} %\orcid{0000-0003-4745-2242}
\and D.\,H.\,Grossmann\inst{\ref{inst:Graz}}%\orcid{0000-0001-6529-9769}
\and L.\,Steinwender\inst{\ref{inst:Graz},\ref{inst:Philly}} %0000-0002-5110-9669
\and L.\,S.\,Schimak\inst{\ref{inst:Graz}} %0009-0006-8575-2106
\and N.\,Muntean\inst{\ref{inst:Graz}} 
\and M.\,Vrard\inst{\ref{inst:OSU},\ref{inst:OSUCosmology}}, \\
R.\,A.\,Patton\inst{\ref{inst:OSU},\ref{inst:OSUCosmology}}
\and J.\,Merc\inst{\ref{inst:Prague}}
\and S.\,Mathur\inst{\ref{inst:IAC},\ref{inst:ULL}} 
\and R.\,A.\,Garcia\inst{\ref{inst:CEA}}
\and M.\,H.\,Pinsonneault\inst{\ref{inst:OSU},\ref{inst:OSUCosmology}} 
\and D.\,M.\,Rowan\inst{\ref{inst:OSU},\ref{inst:OSUCosmology}} P.\,Gaulme\inst{\ref{inst:TLS}}, \\
C.\,Allende\,Prieto\inst{\ref{inst:IAC},\ref{inst:ULL}}
\and K.\,Z.\,Arellano-C\'ordova\inst{\ref{inst:UT},\ref{inst:edi}} %\orcid{0000-0002-2644-3518}
\and L.\,Cao\inst{\ref{inst:OSU},\ref{inst:OSUCosmology}}
\and E.\,Corsaro\inst{\ref{inst:Catania}}
\and O.\,Creevey\inst{\ref{inst:Nice}}
\and K.\,M.\,Hambleton\inst{\ref{inst:Philly}}, \\
A.\,Hanslmeier\inst{\ref{inst:Graz}}
\and B.\,Holl\inst{\ref{inst:Geneva}}
\and J.\,Johnson\inst{\ref{inst:OSU},\ref{inst:OSUCosmology}} 
\and S.\,Mathis\inst{\ref{inst:CEA}} 
\and D.\,Godoy-Rivera\inst{\ref{inst:IAC},\ref{inst:ULL}} %0000-0003-4556-1277
\and S.\,Símon-Díaz\inst{\ref{inst:IAC},\ref{inst:ULL}}
\and J.\,C.\,Zinn\inst{\ref{inst:Joel}}
} 

%P. G. Beck, D. H. Grossmann, L. Steinwender, L. S. Schimak, N. Muntean, M. Vrard, R. A. Patton, J. Merc, S. Mathur, R. A. Garcia, M. H. Pinsonneault, D. M. Rowan, P. Gaulme, C. Allende Prieto, K. Z. Arellano-C\'ordova, L. Cao, E. Corsaro, O. Creevey, K. M. Hambleton, A. Hanslmeier, B. Holl, J. Johnson, S. Mathis, D. Godoy-Rivera, S. Símon-Díaz, J. Zinn

\institute{Instituto de Astrof\'{\i}sica de Canarias, E-38200 La Laguna, Tenerife, Spain \label{inst:IAC}
\email{paul.beck@iac.es}
\and Departamento de Astrof\'{\i}sica, Universidad de La Laguna, E-38206 La Laguna, Tenerife, Spain \label{inst:ULL}
\and Institut für Physik, Karl-Franzens Universität Graz, Universitätsplatz 5/II, NAWI Graz, 8010 Graz, Austria \label{inst:Graz}
\and Department of Astrophysics and Planetary Science, Villanova University, 800 East Lancaster Avenue, Villanova, PA 19085, USA \label{inst:Philly}
\and Department of Astronomy, The Ohio State University, Columbus, OH 43210, USA \label{inst:OSU}
\and Center for Cosmology and AstroParticle Physics, The Ohio State University,
191 West Woodruff Avenue, Columbus, OH 43210 \label{inst:OSUCosmology}
\and Astronomical Institute, Faculty of Mathematics and Physics, Charles University, V Holešovičkách 2, 180 00 Prague, Czechia\label{inst:Prague}
\and Universit\'e Paris-Saclay, Universit\'e Paris Cit\'e, CEA, CNRS, AIM, 91191, Gif-sur-Yvette, France \label{inst:CEA}
\and Th\"{u}ringer Landessternwarte, Sternwarte 5, 07778 Tautenburg, Germany \label{inst:TLS}
\and Department of Astronomy, The University of Texas at Austin, 2515 Speedway, Stop C1400, Austin, TX 78712, USA \label{inst:UT}
\and Institute for Astronomy, University of Edinburgh, Royal Observatory, Edinburgh EH9 3HJ, UK \label{inst:edi}
\and INAF - Osservatorio Astrofisico di Catania, Via S. Sofia 78, I-95123, Catania, Italy \label{inst:Catania}
\and Université Côte d’Azur, Observatoire de la Côte d’Azur, CNRS,
Laboratoire Lagrange, Bd de l’Observatoire, CS 34229, 06304 Nice
Cedex 4, France \label{inst:Nice}
\and  Department of Astronomy, University of Geneva, Chemin Pegasi 51, CH-1290 Versoix, Switzerland \label{inst:Geneva}
\and Department of Physics and Astronomy, California State University Long Beach, Long Beach, CA 90840, USA \label{inst:Joel}
}

\date{submitted: May 4%(be with you)
, 2023; accepted: October 25, 2023}

\abstract
%Context.
{Binary systems constitute a valuable astrophysics tool for testing our understanding of stellar structure and evolution. Systems containing at least one oscillating component are interesting in this regard because asteroseismology offers independent parameters for the oscillating component that aid in the analysis. Systems of particular interest include those with known inclinations. With $\sim$0.8 million binary candidates, the two-body orbit catalog (TBO) of \Gaia Data\,Release\,3 (DR3) substantially increases the number of known binaries and the quality of the astrometric data available for them. 
}
%Aims.
{To enlarge the sample of these astrophysically valuable benchmark objects, we searched for new binary system candidates identified in the \GaiaDR TBO, for which one component has a detection of solar-like oscillations reported in the literature.}
%Methods. 
{We cross-matched the TBO, the full non-single star (NSS) and eclipsing binary catalogs from \GaiaDR with catalogs of confirmed solar-like oscillators in the main-sequence and red-giant phase from the NASA\,\Kepler mission and stars in the Southern Continuous Viewing Zone of NASA\,TESS. The wealth of seismic information is used to characterize the oscillating primary. To test the completeness and robustness of the values reported in the TBO catalog, we performed a similar analysis on stars of the Ninth Catalog of Spectroscopic Binary Orbits (SB9).
}
%Results. 
{The analysis of the SB9 reveals an overall completeness factor for the \Gaia TBO catalog of up to $\sim$30\% providing reliable orbital parameters for $\geq$\,90\,\% of the systems below P$_\mathrm{orb,SB9}$\,$\lesssim$250\,d.
We obtained new 954 unique binary system candidates from \GaiaDR, which host solar-like oscillators, of which we found 45 stars in binary candidates to be on the main sequence and 909 in the red giant phase. Additionally, we report 918 oscillators in potentially long-periodic systems. We present the seismic properties of the full sample and test whether the reported orbital periods are physically possible. 
For 146 giants, the evolutionary state has been determined from their mixed-mode period spacing, showing
a clear trend to long periodic and less eccentric systems in the advanced phases of stellar evolution.
Two new eclipsing binary systems, hosting a red-giant primary were found. For another 146 systems hosting oscillating stars, the values for the orbital inclination were found in the TBO.
Of 181 TBO candidate systems observed multiple times with APOGEE, 149 (82\%) are confirmed as binaries from radial-velocity (RV) measurement. }
%Conclusions.
{
We conclude that the grand majority of the orbital elements reported in the TBO catalog are physically reasonable and realistic. This finding increases the number included in the sample of known solar-like oscillators in binary systems by an order of magnitude. The large fraction of confirmed binaries from APOGEE RV measurements indicates that the TBO catalog is robust.
We suggest that due to instrumental noise, the seismically inferred masses and radii of stars observed with the TESS satellite and with an excess of oscillation power of  $\nu_\mathrm{max}$\,$\lesssim$\,30\,$\mu$Hz could be significantly overestimated.
The differences in the distributions of the orbital period and eccentricity are due to the accumulative effect of the equilibrium tide acting in these evolved binary systems. }
\keywords{Asteroseismology 
$-$ (Stars:) binaries: spectroscopic
$-$ Stars: late-type $-$ Stars: oscillations (including pulsations).}

\let\linenumbers\nolinenumbers\nolinenumbers

\maketitle

%\linenumbers\modulolinenumbers[5]
%\nolinenumbers

% I leave this comment in, for historic reaseons, to remind us upon the time when we thought that those 60 pages would fint in a 4 page letter:
%\textbf{Limitations for a letter: 3000 words max, 4-5 pages long, unlimited supplementary material.}

\section{Introduction}

\begin{figure}[t!]
    \centering
    %\vspace{-8mm}
    \includegraphics[width=\columnwidth%,height=80mm
    ]{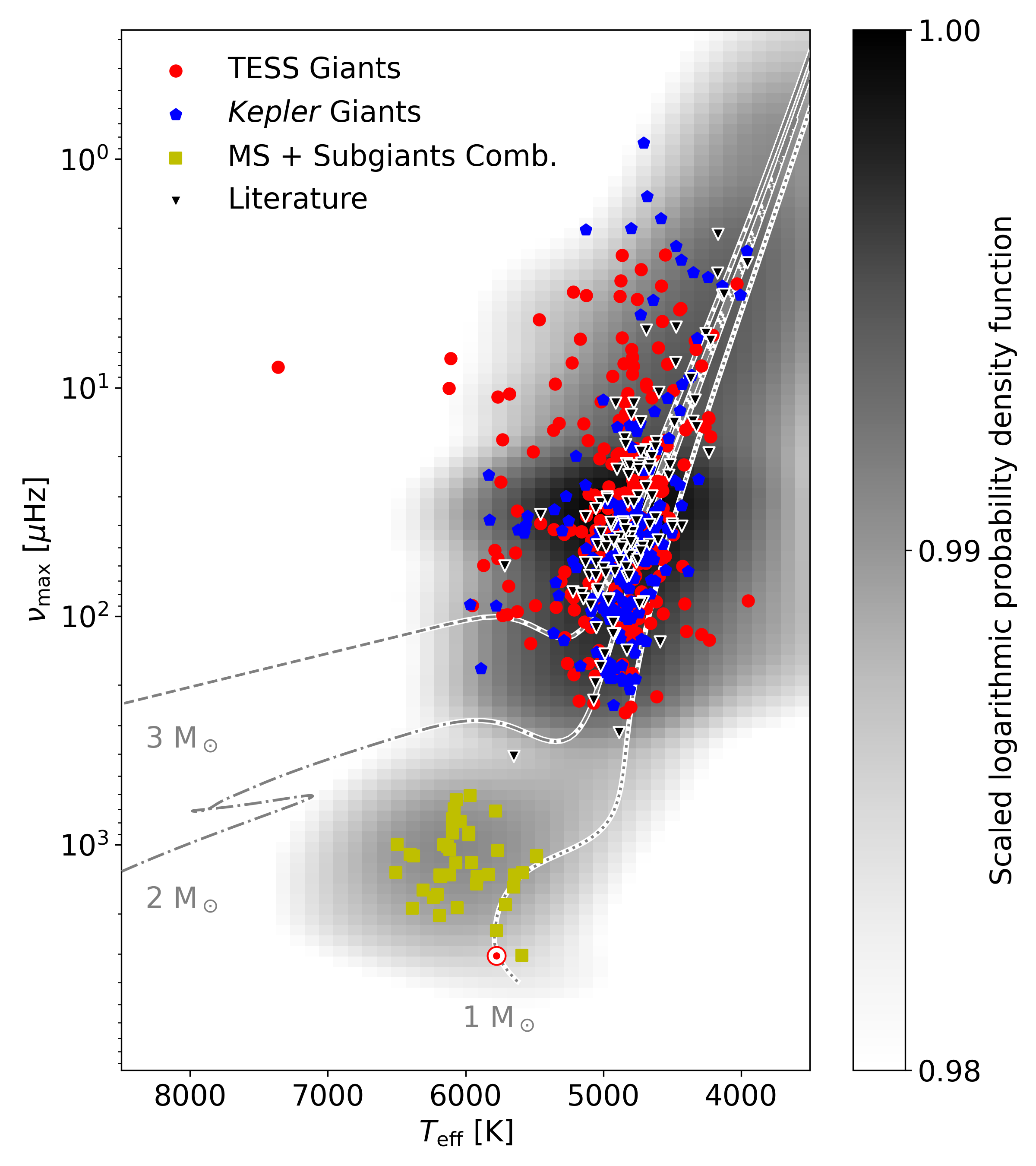}
    \caption{Distribution of the sample in the seismic HRD with the frequency of the oscillation-power excess on the left vertical axis. The colored symbols show binary candidates reported in \GaiaDR hosting giant or main-sequence solar-like oscillators.
        Red marks red giants with asteroseismic solutions obtained from TESS data. 
        Blue denotes red giants with asteroseismic solutions from the \kepler mission.
        Yellow shows oscillating solar-like main sequence stars and subgiants, observed with \Kepler.
        In black triangles, the literature sample is depicted.
        Only systems with APOGEE temperatures are shown.
        For the literature sample, all targets with asteroseismic solutions are shown, independently of whether a solution in the \GaiaDR TBO catalog exists.
        With grey lines, the evolutionary paths for stars of $1\,\Msun$, $2\,\Msun$, and $3\,\Msun$ and with solar metallicity are shown.
        The grey probability distribution, underlying the scatter and curves, is the distribution of all targets with asteroseismic solutions of the above-mentioned catalogs.
    \label{fig:seismicHRD} }
\end{figure}

Among the 5000 stars visible to the naked eye, about 2000 are known to be multiple-star systems. Naked-eye stars are a small fraction of all stars in the Milky Way, but they are reasonably representative of the incidence of binarity, which is estimated to be between 50 and almost 100\% \citep[e.g.,][]{Eggleton_2006}. Some systems are close enough to be in contact and most are far apart enough to evolve almost independently. Binary systems are known with orbital periods as short as 0.2 days or as long as thousands of years.
Stars in multiple systems are also precious test benches for testing our understanding of stellar structure and evolution \citep{MoeStefano2017, Offner2022}. 
While their components may vary significantly in temperature, luminosity, radius, and lithium abundance, both components are identical in their initial conditions, age, and distance \citep[e.g.,][]{Prsa2018}. 

The advent of space telescopes, such as Convection, Rotation
et Transits planétaires  \citep[CoRoT][]{Baglin2006}, \Kepler \citep{Borucki2010} and its refurbished K2 mission \citep{Howell2014}, and Transiting
Exoplanet Survey Satellite \citep[\textit{}TESS][]{Ricker2014} have allowed for the detection of solar-like oscillations in late-type stars. These convection-driven oscillations provide a frequency pattern that allows for direct identification of the spherical degree of the oscillation mode \citep*[see monograph by][and references therein]{Aerts2010}, providing optimal input information for stellar modeling. 
For stars where an individual mode fitting is not feasible, the information content from the power spectrum can be summarized by two characteristic frequencies. These frequencies can then be combined through scaling relations to yield stellar mass and radius \citep{Brown1991,Kjeldsen1995}.
The detection and exploitation of non-radial modes in giant stars by  \cite{deRidder2009}, followed by those  of dipole-mixed modes \citep{Beck2011,Beck2012, bedding2011, Mosser2011a}, have unlocked the full potential of the asteroseismic analysis of red-giant stars.
The constraint set by stellar binarity, amended with the independent information on stellar structure and its properties, offers a unique opportunity for testing the complex microscopic and macroscopic physics involved in building stellar models \citep[e.g.,][]{Beck2018Asterix,LiT2018,Johnston2021}.

The number of known binaries with solar-like oscillating components is still limited. From photometry, binaries are only detectable if their orbital geometry leads to eclipses or if the hydrostatic readjustment due to the tidal interaction of the stars at periastron gives rise to significant flux modulations \citep{kumar1995}. The latter systems are colloquially referred to as ``heartbeat stars.'' In a recent effort, \citet{Beck2022SB9} made an inventory of the red-giant oscillators that belong to binary systems (eclipsing and otherwise) in the literature, while also searching for more in the Ninth Catalog of Spectroscopic Binary Orbits \citep[SB9,][]{Pourbaix2004}, leading to a total of 182 oscillating red giants in binary systems with resolved orbital parameters  (see Fig.\,\ref{fig:seismicHRD}). 
Following a different approach, \cite{Hon2022} provided a catalog of seismic parameters of members of 99 binary systems, whereby the red giant primary is spatially resolved from the secondary. Because of their very long orbits, the orbital periods are hardly known for these systems. 

An area of particular interest are the double-lined spectroscopic binaries (SB2), where spectral lines from both components are detectable, with a known inclination angle of 90\,degrees, for which stellar masses and radii can be accurately determined. \cite{Gaulme2016} used ten of these systems to test the accuracy and precision of the asteroseismic scaling relations, suggesting an offset of up to 15\% in mass and 5\% in radius. \cite{Themessl2018} and \cite{Benbakoura2021} added another four eclipsing systems, increasing the count of these high-value targets to 14 \citep[see][for a complete list]{Beck2022SB9}. 
For a robust analysis of the comparison of dynamical and seismic masses, a substantially larger number of systems with known inclination angles is necessary. 

The third data release (DR3) of the ESA \Gaia mission \citep[][]{GaiaMissionMainReference2016,GaiaDR3Vallenari2022} is the first \Gaia\,DR to include a specific analysis for non-single stars \citep[see][Siopis et\,al.\ subm., Damerdji et\,al.\ subm., and Gosset et\,al.\ subm., for catalogs and details of the analysis]{Arenou2022, Holl2022,
Halbwachs2022, Mowlavi2022}. 
In the \Gaia data, a multiple-star system can be identified with one or several of the following: 
$(i)$ astrometric measurements, 
$(ii)$ detections of periodic photometric dimming caused by eclipses or phase effects,
$(iii)$ radial velocity measurements obtained by the high-resolution spectrometric channel, and $(iv)$  by the low-resolution spectro-photometry from SED fitting. From 34\,months or 1034\,days of observations of the whole sky, \cite{Arenou2022} reports about 814\,000 binaries in \GaiaDR, resulting from \Gaia astrometric \citep{Halbwachs2022, Holl2022}, and radial velocity orbital fitting (Gosset, et al., 2023; Damerdji, et al. 2023). This catalog contains only a small subset of the 2.2 million eclipsing binaries\footnote{see the more complete \texttt{gaiadr3.vari\_eclipsing\_binary} catalog} detected by photometry \cite{Mowlavi2022}, for which an orbital solution has been computed (Siopis et al. subm.). In the optimal case, astrometric solutions offer the possibility to obtain the inclination of the orbital plane in the sky, which allows us to measure the masses of their components and utilize non-eclipsing systems for calibrating asteroseismology. This new catalog of non-single stars released by \Gaia constitutes a major change in the known inventory of such types of systems. 

It will take a long time to go through the data from the past CoRoT, \textit{Kepler}, and K2 missions, the current TESS, and the forthcoming Chinese \textit{Earth\,2.0} \citep{Earch2pointO2022}, ESA PLATO \citep{Rauer2014}, and the NASA \textit{Roman} \citep{Johnson2020, huber2023roman} missions, which are expected to be operational by the end of the decade. Before exploring the archived photometric data, a natural first step consists of checking which of the known stellar pulsators are listed as non-single stars in the \GaiaDR catalog.
This work presents the successful search for solar-like oscillating stars in binary systems, revealed through photometric, spectroscopic, and astrometric solutions in the \GaiaDR. 

The paper is structured as follows. 
In Sect.\,\ref{sec:Search}, we describe the selection of the sample red-giant stars and our search for orbital solutions reported in the \GaiaDR. 
The general asteroseismic properties of the found sample of binary candidates with oscillating red-giant primaries are described in Sect.\,\ref{sec:GeneralSampleProperties}. 
The history of tidal interaction and stellar activity and rotation is analyzed in Sect.\,\ref{sec:tides}. 
In Sect.\,\ref{sec:EclipsingInclinations}, we present an eclipsing binary system, listed in the \GaiaDR eclipsing binary catalog and report oscillations and updated orbital period from TESS photometry. Furthermore, we present 146 solar-like oscillators in systems with \Gaia inclinations. In this chapter, we discuss why eclipsing binaries are difficult to find, resulting in their small number compared to the known binary population. 
Independent confirmation for 181 reported binary candidates through RV variations, mainly from \textit{Apache Point Observatory Galactic Evolution Experiment} \citep[APOGEE, ][]{Majewski2017} is provided in Sect.\,\ref{sec:RadialVelocities}. 
Section\,\ref{sec:extendedScienceCase} discusses symbiotic binaries and giants with anomalous peaks in the power-spectral density (PSD), two related science cases (including red-giant binaries) onto which the data from  \GaiaDR shed new light.

%=======================================
\section{Sample selection and  binary statistics \label{sec:Search}}

Using the \textit{Simbad} module in the \textit{AstroPy} package \citep[and references therein]{astropy2018}, we created a table of cross-identifiers. This list was then used to query the inventory of the two-body orbit (TBO) catalog of \GaiaDR \citep{Arenou2022} on the \Gaia data archive\footnote{https://gea.esac.esa.int/archive/}. 

\subsection{Completeness and robustness of \GaiaDR binary candidate solutions}

The documentation of the \GaiaDR TBO stresses that the reported systems are 'only' binary system candidates. To better understand the completeness, robustness, and quality metrics of the orbital solutions provided in the catalog, particularly for the outliers, we explore the TBO inventory for a well-known sample of binaries. A commonly chosen gold-star sample is SB9, created and curated by the team of \cite{Pourbaix2004}, which we used to assess the quality and completeness of the \GaiaDR solutions. In this subsection, we only review the takeaway results of our comparison. For  details of the full analysis, we refer to Appendix\,\ref{sec:SB9}.
\begin{figure}[t!]
    \centering
    \includegraphics[width=\columnwidth,height=100mm]{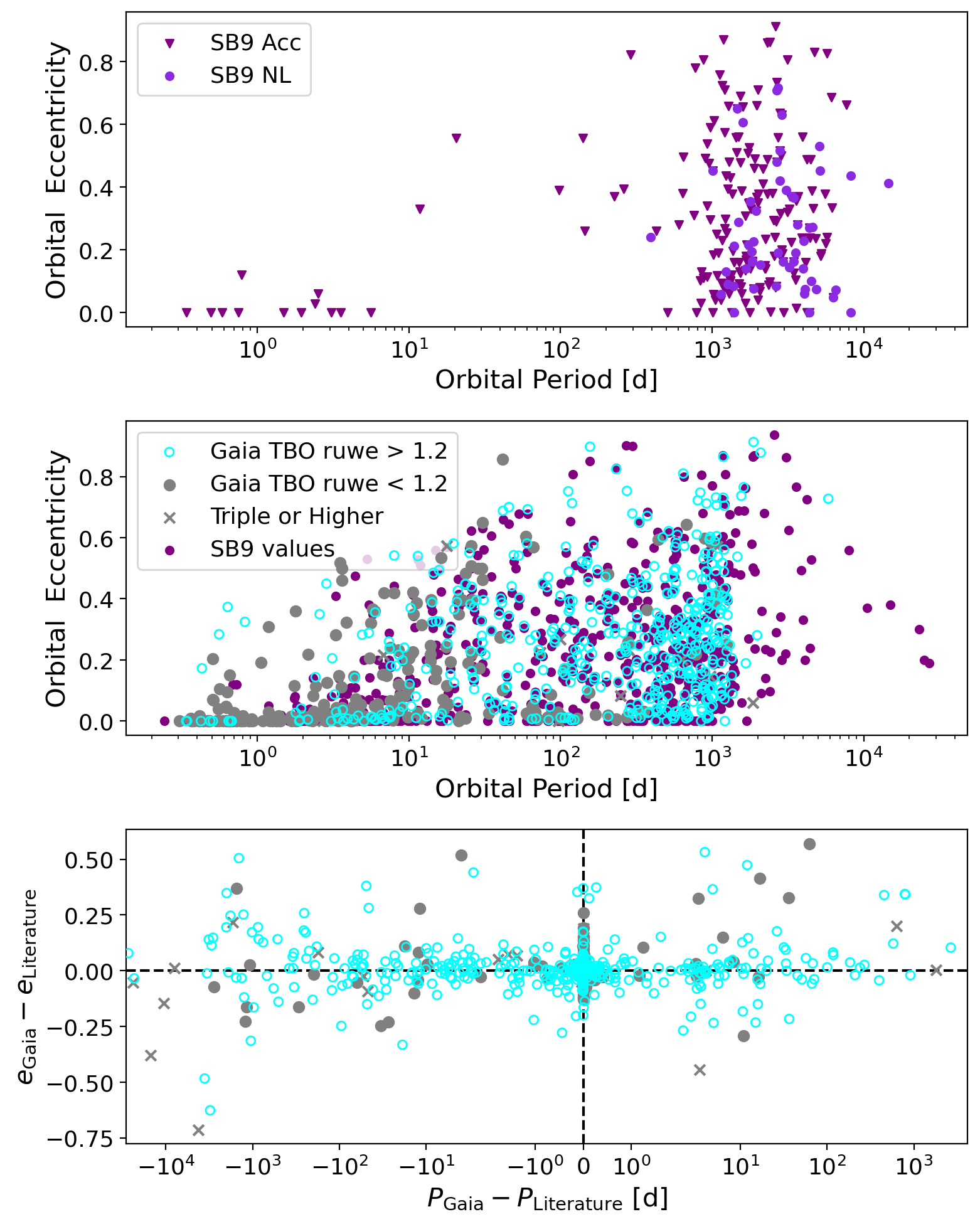}
     \caption{Comparison of period and eccentricity from \GaiaDR TBO and the SB9 catalog. 
     The top panel depicts the SB9 values of the orbital parameters of systems for which only non-linear (violet dots) or acceleration solutions (purple triangles) are reported in the \Gaia TBO. The middle panel shows 729 systems that have a full solution for the orbital parameters in both catalogs. The solution reported in the SB9 catalog is shown in purple. The DR3 solutions for the same set of systems are separated based on their ruwe value (ruwe< 1.2: grey dots; ruwe> 1.2: cyan circles). Triple or higher order systems are marked with a cross. The bottom panel depicts the absolute residuals for the period and eccentricity between the two sets of solutions.
     \label{fig:SB9residuals}}
\end{figure}

In the SB9, we identified the 3413 unique objects. Because the efficiency of the methods for detecting binaries is dependent on the (integrated) object's magnitude \citep[see Fig.\,10 in][]{Arenou2022}, we constructed a magnitude-limited sample between the \Gaia magnitudes 4\,$\lesssim$\,G\,[mag]\,$\lesssim$\,13 to asses a completeness factor, bias corrected for objects outside the detection range of \Gaia. Because many SB9-systems have orbits longer than the baseline of \GaiaDR, we further limited the sample to $P_{\rm orb, SB9}$\,$\leq$\,1\,100\,d. From the 2343 systems in the period-magnitude-limited SB9 sample 668 were identified as binary systems (Fig.\,\ref{fig:SB9residuals}, middle panel). This corresponds to a completeness factor of 28.5\%. 
Additional 241 SB9 systems are present in the catalog of non-linear solutions \citep{Arenou2022}, whose orbital elements could not be resolved from the current data in \GaiaDR and most likely indicate systems with orbital periods longer than the baseline of \GaiaDR. Indeed the comparison in Fig.\,\ref{fig:SB9residuals} (top panel) confirms this expectation. Taking these detections into account increases the overall completeness factor to 38.8\%.

The middle panel of Fig.\,\ref{fig:SB9residuals} shows the $e$-$P$ plane of all 715 systems for which both parameters (the eccentricity and period) are provided in the SB9 and the \GaiaDR catalog. We consider a solution reliable, if the residual between the values listed in SB9 and \GaiaDR (bottom panel of Fig.\,\ref{fig:SB9residuals}) differ by less than 10\%. For such solutions, we find a good agreement for the eccentricities with a mean residual \hbox{of 0.011 and a standard deviation of 0.104}.

A more complex picture is found for the orbital periods. In the range of P$_\mathrm{orb,SB9}$\,$\lesssim$250\,d, and 250\,$\leq$\,P$_\mathrm{orb,SB9}$ [d]\,$\leq$500, we find that $\sim$90\% and $\sim$99\% of all systems have periods with residuals better than 10\%, respectively. Interestingly, with a completeness factor of 41.0\% the longer period range yields a higher detection rate than the short periodic range with 24.9\%.  The borders of the intervals were chosen to resemble one-fourth and half of the timebase covered by \GaiaDR. The lower percentage of reliable solutions in the short periodic range can be explained with a larger number of short-periodic solutions for long periodic systems (Fig.\,\ref{fig:SB9comparison}). 
For $P_{\rm orb,DR3}$\,$\geq$\,500, the number of reliable solutions drops to $\sim$74\%, which contains a significant number of significantly underestimated orbital periods of systems with $P_{\rm orb,SB9}$\,$\geq$\,1\,100\,d. The comparison of the orbital elements in the SB9 and \GaiaDR catalog are depicted in the bottom panel of Fig.\,\ref{fig:SB9residuals}, Fig.\,\ref{fig:ePcomparison}, and discussed in more detail in App.\,\ref{sec:SB9}.

\begin{figure}[t!]
    \centering
    \includegraphics[width=\columnwidth,height=60mm]{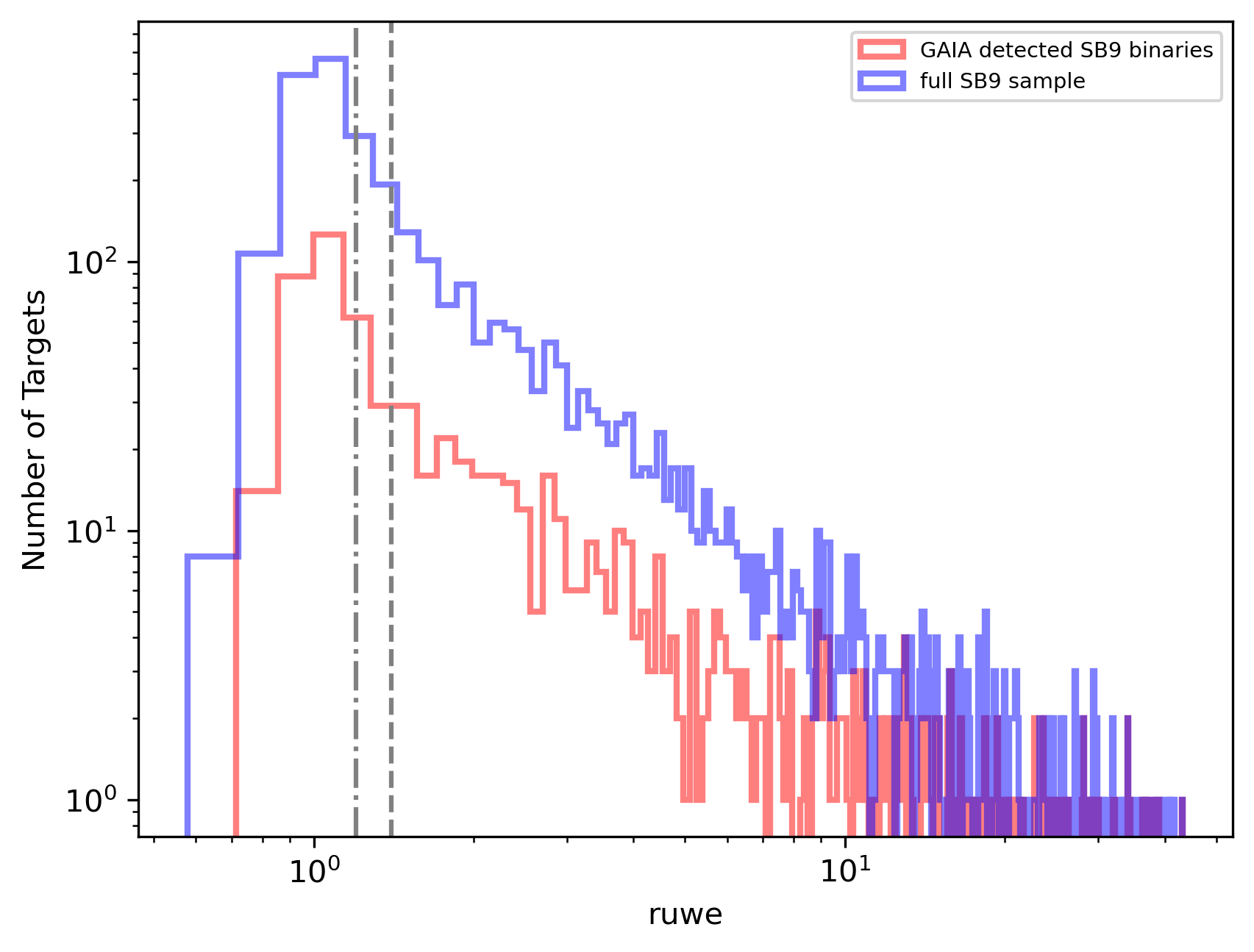}
    \caption{\Gaia astrometric detected SB9 binaries, based on   the \textit{ruwe} values for all \Gaia sources of the SB9 sample (blue) and the binary candidates in the TBO catalog  (red). The grey dash-dotted and dashed vertical lines mark the $ruwe$ suggested threshold values of 1.2 and 1.4, respectively.}
    \label{fig:ruweHist}
%\end{figure}
%\begin{figure}[t!]    
\includegraphics[width=\columnwidth]{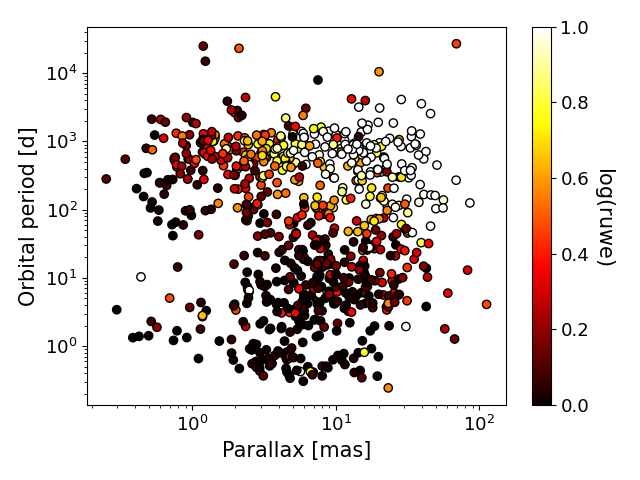}
    \caption{Distribution of \textit{ruwe} value of the SB9 systems in \Gaia in logarithmic scale as a function of the \Gaia parallax and orbital period from SB9 catalog.}
    \label{fig:ruweDist}
\end{figure}

The ruwe parameter presents the renormalized unit weight error (ruwe) of a star’s astrometry \citep[for details see][]{Arenou2022}. The analysis of the SB9 sample taught us that $\sim$40\% of all systems in the period-magnitude limited sample have $ruwe$ values below the threshold value of $\sim$1.4 (Fig.\,\ref{fig:ruweHist}), which is often discussed as a good general indicator for a binary detection. As shown in Fig.\,\ref{fig:ruweDist}, these systems are typically far away and their components’ displacement is therefore small, with respect to the astrometric precision of the \Gaia satellite. 

\subsection{Solar-like oscillators in \GaiaDR binary candidates}

\begin{figure}[t!]
    \centering
    \includegraphics[width=\columnwidth]{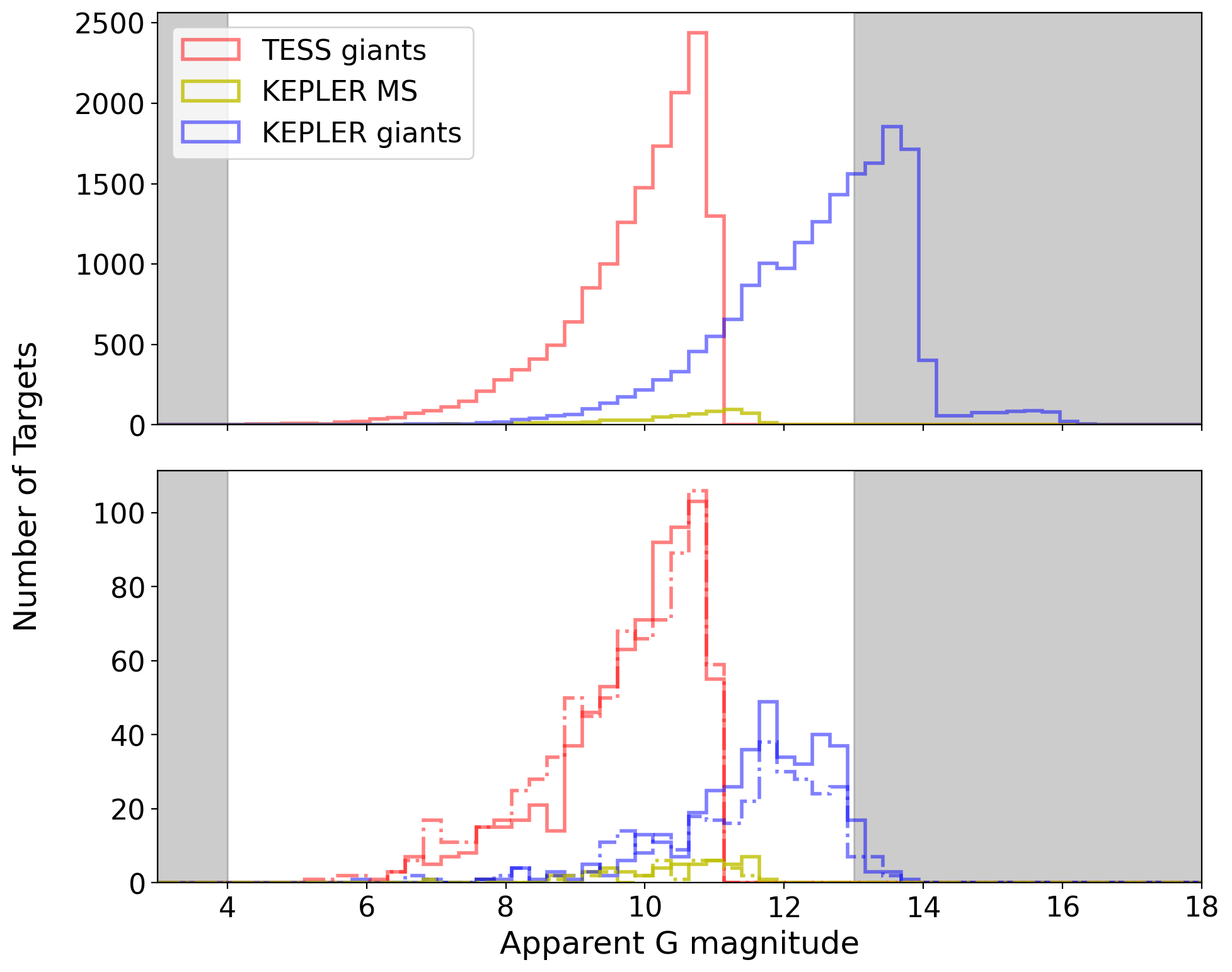}
    \caption{Distribution of the apparent G magnitude in the input samples for the search (top panel) and found binary candidate systems (bottom panel). The bin size is one-fourth of a magnitude. The grey-shade regions mark magnitudes outside the magnitude-limited sample. The dash-dotted lines indicate the reported candidates with acceleration and non-linear solutions for the respective samples.}
    \label{fig:magHistograms} 
\end{figure}

\begin{table*}[t!]
\centering
\caption{Input samples for oscillating main-sequence and red-giant stars and the corresponding binary yields.}
\tabcolsep=7pt
\begin{tabular}{lrrr|rrrr|r|r}
\hline \hline                                                                                                                   \multicolumn{2}{c}{Input samples}&       
\multicolumn{1}{c}{Oscillators} &       
\multicolumn{1}{c}{Magnitude} & 
\multicolumn{2}{c}{NSS TBO}     &       %Doublet        &       
\multicolumn{2}{c}{Binary fraction}     &       
%\multicolumn{1}{c}{Dyn.}       &       
\multicolumn{1}{c}{Astrom.}     &       
\multicolumn{1}{c}{NSS TBO}     \\

\multicolumn{1}{c}{Name} &
\multicolumn{1}{c}{Size} &      
\multicolumn{1}{c}{\GaiaDR}     &       
\multicolumn{1}{c}{limited}     &       
\multicolumn{1}{c}{Uni.}        &       
\multicolumn{1}{c}{Alt.}        &       
\multicolumn{1}{c}{Full} &
\multicolumn{1}{c}{Mag\,lim.}   &       
%\multicolumn{1}{c}{Mass}       &               
\multicolumn{1}{c}{Inclin.}     &
\multicolumn{1}{c}{Acc+NL}      \\
\hline

\hline  
\hline

Kepler giants   &       18\,824   &     17\,544   &     10\,374~~   &   376     &       3       &       2.2\%   &       3.6\%~~ &       31      &   {257} \\
TESS giants         &   15\,405   &     12\,016   &     12\,016~~   &   549     &       18      &       4.6\%   &       4.6\%~~ &               105     &       {633}   \\
MS + SG         &       624       &     595       &     595~~       &   45      &       1       &       8.1\%   &       8.1\%~~ &       14      &       {28}    \\ 
                
\hline
Lit. giants &   190     &       190     &       53~~    &       53      &       1       &       27.9\%  &       100\%
&       1       &       116     \\
SB9 sample      &       3\,413  &       $-$     &       2\,343$^\star$  &       668     &       0       &       19.6\%  &       28.5\%$^\star$  &       68      &       241     \\      \hline
\end{tabular}
\tablefoot{ The first four columns describe the input sample and construction of subsamples. \newline
1 and 2: The name of the sample and the input samples for the search and the number of all targets in the catalog;\newline
3: number of oscillating targets within this sample that have a \GaiaDR solution;\newline
4: number of targets in the magnitude limited sample within 4\,$\leq$\,G \,[mag]\,$\leq$\,13. \newline
The next 7 columns present the results of the search of the Non-Single-Star (NSS) catalog in \GaiaDR.\newline
5: gives the number of unique solutions returned by the ADQL query for the sample of oscillators with \gaia solutions (col.\,3);\newline
6: gives the number of alternative orbital solutions in case TBO contains multiple solutions;\newline
7: presents the binary detection fraction, calculated from the sample of all oscillating targets with a \GaiaDR solution (col.\,3);\newline
8: presents the binary detection, calculated from the magnitude limited sample (col.\,4) and binaries within the magnitude limit;\newline
9: presents the number of systems hosting an oscillating component with inclination reported in TBO; \newline
10:  number of sources with acceleration and non-linear solutions in TBO could point to longer-period binary systems.\newline
$\star$: Because the SB9 sample extends significantly further in period than the \Gaia\,TBO sample, the magnitude limited sample for SB9 has been further corrected for a cut-off period of P$\leq$1100\,days.} \label{tab:ArchiveResults}
\end{table*}

To guide our search for solar-like oscillators in binary star systems, we compiled a list of known solar-like oscillators whose global seismic parameters were reported in the literature (Table\,\ref{tab:ArchiveResults}). Due to their characteristic shape, the PSD of solar-like oscillations are typically described through the center frequency of the excess of oscillation power, \num. This quantity correlates with the stellar surface gravity, $\log g$, and the stellar luminosity, $L$.
The structure of oscillation modes in the power excess is very regular \citep{Tassoul1980,mosser2011b}. The average frequency separation between modes of the same spherical degree, $\ell$, but consecutive radial orders is described through the large-frequency separation, \dnu. The large-frequency separation is a proxy of the average speed of sound in the oscillating stars. We refer to the review by \cite{Aerts2021} for a recent overview of the two global-seismic parameters.  

The largest input sample comprises over eighteen thousand oscillating red giants, measured and identified from \Kepler photometry. The \textit{\Kepler}  giant sample was compiled from \cite{Yu2018}. 
Because of the long timebase of \kepler photometry, the frequency resolution allows for a clear identification of the mixed-mode pattern. Evolutionary states were extracted from seismology by \cite{Vrard2016}. 
In this sample, 17\,544 systems have valid astrometric solutions listed in the \GaiaDR catalog. 
The second sample is the \textit{TESS} giant sample provided by the giants in the TESS mission's Southern Continuous Viewing Zone (SCVZ). Of the fifteen thousand giants identified from photometric and spectroscopic calibration, \cite{Mackereth2021} detected oscillations in about two-thirds of the stars. 
It has been a matter of discussion in the literature \citep[e.g.,][]{SilvaAguirre2020} that while individual and well-proven asteroseismic pipelines agree on the position of the power excess within a few percent, there is an increased scatter on the determined large-frequency separation from TESS data. Therefore, we also accepted stars with missing or uncertain \dnu values into our catalog. We considered oscillations to be detected in a giant if at least two of the three pipelines used by \cite{Mackereth2021} report a power excess. For our analysis, we use the mean value of \num reported in their paper.
For 12\,016 oscillating giants, an astrometric solution exists in \GaiaDR. We follow the evolutionary states for this sample, determined by \cite{Vrard2021}.

Main sequence solar-like stars and subgiants with solar-like oscillations complement the giants samples (hereafter referred to as the MS+SG sample). Here, we have compiled a list of stars observed by \Kepler \citep[compiled by][]{Mathur2022}. Of the 624 oscillating dwarfs, an astrometric solution exists for 595 stars in \GaiaDR.

The distribution of the 30\,155 oscillating stars from the \Kepler, TESS, and MS+SG samples  (Table\,\ref{tab:ArchiveResults}), with an astrometric solution in \GaiaDR in the seismic Hertzsprung-Russell diagram (HRD) is depicted as the probability density map shown in the grey-scale of Fig.\,\ref{fig:seismicHRD}. 
The orbital and seismic parameters of the detected binary candidates are presented in Table\,\ref{tab:A1}. For reference and orientation, we show evolutionary tracks for a star of the solar metallicity of 1, 2, and 3\,M$_\odot$, calculated with the MESA evolutionary code \citep[][and references therein]{MESA2018, MESA2022}.

For our queries of the \Gaia data archive, we produced cross-identifier tables between the solution identifier of \GaiaDR, the target input catalogs for the \Kepler and TESS missions (KIC and TIC, resp.), and the Two-\textit{Micron All Sky Survey} (2MASS). For the entire sample of oscillating stars in \GaiaDR, we found 970 unique orbital solutions for binary system candidates in the \GaiaDR TBO catalog (Table\,\ref{tab:ArchiveResults}).
Between the input sample and the literature sample, we find a cross-section of 44 systems, of which 16 are detected. These are listed in the bottom panel of Table\,\ref{tab:A1}. Therefore, we report the detection of 954 solar-like oscillators in binary systems.

The position of these binary systems in the seismic HR diagram is shown in Fig.\,\ref{fig:seismicHRD}. 
For 22 candidate systems, the \Gaia catalog provides multiple solutions of the DR3 data productions that did not converge to a single solution. We have chosen not to keep them in our final sample.

\subsection{Potentially long-periodic systems}
Our search of the acceleration and non-linear solutions in TBO sub-catalogs delivered in total 918 results for the giants- and dwarf-star samples (see Table\,\ref{tab:ArchiveResults} and Fig.\,\ref{fig:magHistograms}). These solutions could indicate additional binary systems. However, no orbital parameters are given, as the orbital periods of these systems could substantially exceed the timebase covered by \Gaia DR3. Under the assumption that all these detections represent actual binaries, we would roughly double the binary detection rate. 

Confirming these suggested binary detections will require more extended timebases of \gaia observations with forthcoming data releases or ground-based RV monitoring. Both are beyond the scope of this paper. Therefore, we only list these systems in Table\,\ref{tab:B1} and we do not further explore their binary characteristics.

\subsection{Binary detection rates}
Differences between the input samples are found in the detection rate. 
Interestingly, the sample for giants seismically characterized with \Kepler (2.2\%) and TESS (4.6\%) differ by about a factor of 2. Several observational aspects can explain these differences.
The main difference between the samples is the distribution of the apparent magnitude of their targets. Figure\,\ref{fig:magHistograms} compares these data sets by depicting the distributions of the mean \Gaia\,G band magnitude for the three input catalogs (Table\,\ref{tab:ArchiveResults}). Indeed, the peak of the \Kepler giant sample is about 3.5 magnitudes fainter than that of the TESS giants. The \Kepler main-sequence sample peaks at about 11.5\,mag.

Therefore, we corrected the fractional count of the yields in Table\,\ref{tab:ArchiveResults}, calculated from the subsample of targets within the brightness limits (4\,$\leq$\,G\,[mag]\,$\leq$\,13) described in Appendix\,\ref{sec:SB9}.
As shown in Fig.\,\ref{fig:magHistograms}, correcting the sample sizes for the magnitude range mainly affects the sample size of the \Kepler giant sample, as about 40\% of all targets are fainter than 13$^{\rm th}$\,mag. Because the TESS telescope also shows significant saturation effects around 4$^{\rm th}$\,mag and the sample of \cite{Mackereth2021} was limited to TESS magnitudes brighter than 11$^{\rm th}$\,mag, all targets fall into the magnitude-limited range. 

To make the comparison of the binary rate more compatible between the space telescopes, we recalculated the fractional yields for a magnitude-limited sample, which leads to 3.6\% and 4.6\% for the \Kepler and TESS sample, respectively. Therefore, the corrected yields bring the results from the two giant samples into a closer agreement. The remaining difference could originate from the differences in the scanning law of the \Gaia satellite. As shown by \citet[in their Fig.\,7]{Arenou2022}, the \Kepler field of view has been substantially less intensively covered than the southern CVZ of TESS.

The \Kepler main-sequence sample achieves the highest detection rate with 8.1\%. At first glance, this is surprising, as their distribution mainly ranges between the bright TESS and the fainter \Kepler giant sample and suffers from the same lower number of scanning events as the \Kepler giants. 
This sample contains more stars on shorter orbital periods that are easily detected by the \Gaia satellite. 
These orbital periods are also located at periods untypical or physically impossible for giants in binary systems due to their extended radii. Therefore, the difference could already point to a different binary fractions as a function of the evolutionary state. 

\begin{figure}[t!] \centering
    \includegraphics[width=\columnwidth]{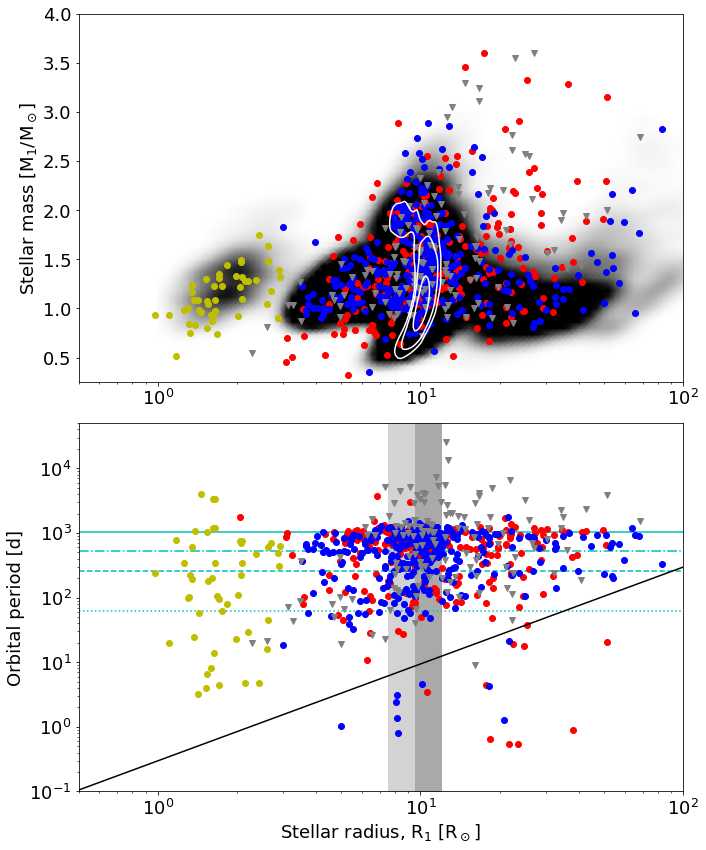}
    \caption{Characterisation of the set of binary candidates from \GaiaDR. 
    Top panel shows the distribution of the binary candidates in the radius-mass plane. The probability density map in the background shows the distribution of all oscillating stars in the \Kepler sample.
    Bottom panel provides a test of the feasibility of the orbital period as a function of the stellar radius. The black line resembles the estimated minimum period for a system to fill the Roche lobe of its giant. The dark and light shaded area mark the ranges in which RC and 2RC stars are expected, respectively.
    The meanings of the yellow, blue, and red symbols are similar to those of Fig.\,\ref{fig:seismicHRD} for the dwarfs and giants from the \Kepler sample and giants from the TESS sample, respectively. The literature sample is represented by grey triangles. The horizontal lines (from top to bottom) indicate the full, half, one fourth of the timebase, and the $\sim$63\,day precession period of \GaiaDR.
    }  
    \label{fig:periodSeismolgy}
\end{figure}

As discussed in the literature, based on the analysis of the large samples from the SB9 catalog and APOGEE data \citep[e.g.,][]{VanEylen2016, Badenes2018, Beck2022SB9}, most systems in the red giant phase are found at periods longer than several hundred to a few thousand days, which is beyond the baseline of \GaiaDR. Because we do not know the initial distribution of orbital periods, we cannot compile a magnitude-period limited sample, as done for the SB9 comparison in Appendix\,\ref{sec:completeness} and correct the yields.
However, we can estimate the approximate detection rates by including the 890 and 28 acceleration and non-linear solutions for giants and dwarfs (see Table\,\ref{tab:ArchiveResults}), respectively, in the calculation of the magnitude-corrected binary yields. Therefore, we gain tentative binary detection rates of $\sim$6.6\% for the \Kepler giants, $\sim$11.0\% for the TESS giants, and $\sim$14.1\% for the \Kepler dwarfs. 

These numbers are still low when compared with the expected binary fraction for the mass range of solar-like stars of about $\sim$50\% known from stellar population studies \cite[see the reviews by][and references therein]{MoeStefano2017, Offner2022}. In Appendix\,\ref{sec:completeness}, we showed that the binary yields are at least a factor of 3 too low. We argue that this high fraction of nondetections is due to insufficient data or the extensive orbital periods expected for these binaries.
Correcting these binary fractions by this factor for incompleteness puts us into the ballpark of about $\sim$30\% to $\sim$40\%, which approaches the expected value.

%=======================================

\section{General properties of the sample \label{sec:GeneralSampleProperties}}

From searching the catalogs in Table\,\ref{tab:ArchiveResults}, we find a total of 954 new binary-system candidates hosting a solar-like oscillating red giant (909) or either a main-sequence or a subgiant (45) star in the TBO. 
We note that \cite{Arenou2022} also presented a study of red giant binaries. The \Gaia sample was drawn from a cut of all TBO solutions, whereby the giants were identified from their 2MASS colors ($J-K >0$\,mag, and $M_K<0$\,mag). 
Because our sample was selected on the premise of detected oscillations, the sample's "purity", size, and distribution are different in these two works. 
The detection of oscillation in the target source has predetermined our sample. This additional selection criterion allows us to select giants with high accuracy and apply seismic techniques to exploit the properties of the sample.

\subsection{Masses and radii from asteroseismology \label{sec:scalingRelations}}

We calculate the mass and radius of an oscillating star in the binary candidates with the standard asteroseismic scaling relations \citep{Brown1991,Kjeldsen1995}. 
Using the solar values as a base reference, this homological formalism allows us to estimate the masses and radii from their measured global seismic parameters, the peak frequency of the global power excess, \num, and the large frequency separation, \dnu, and spectroscopically measured effective temperature:
\begin{eqnarray}
\frac{R_\star}{R_\odot} &=& 
\left(\frac{\nu_\mathrm{max}}{\nu_\mathrm{max,\odot}}\right)^3 
\cdot \left(\frac{\Delta\nu}{\Delta\nu_\odot}\right)^{-4} 
\cdot \left(\frac{T_\mathrm{eff,\star}}{T_\mathrm{eff,\odot}}\right)^{3/2}\,, \\
\frac{M_\star}{M_\odot} &=& 
\left(\frac{\nu_\mathrm{max}}{\nu_\mathrm{max,\odot}}\right) 
\cdot \left(\frac{\Delta\nu}{\Delta\nu_\odot}\right)^{-2} 
\cdot \left(\frac{T_{\rm eff,\star}}{T_{\rm eff,\odot}}\right)^{1/2}\,. 
\label{eq:mass}
\end{eqnarray}
We use the reference values of the A2Z pipeline \citep{Mathur2010} $\nu_\mathrm{max,\odot}$\,=\,3100\,$\mu$Hz,
$\Delta\nu_\odot$\,=\,135.2\,$\mu$Hz, and
$T_{\rm eff,\odot}$\,=\,5777\,K
to describe the oscillations in the Sun.

\begin{figure}[t!]
    \centering
    \includegraphics[width=\columnwidth]{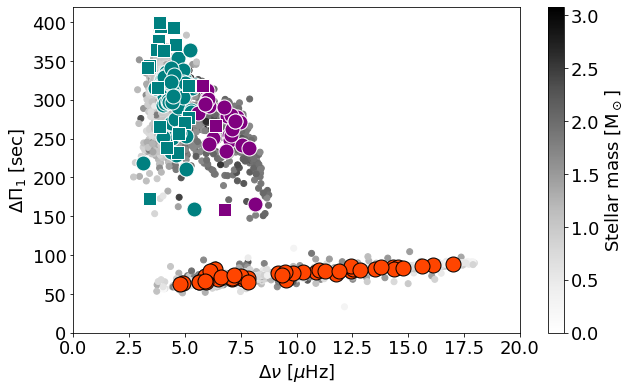}
    
    \caption{Identification of the evolutionary stage of the primary of the system. Teal, purple, and orange indicate RC, 2RC, and RGB stars, respectively. Dots indicate stars observed with the \Kepler mission. Squares mark stars observed by TESS. The points in grey mark the full input sample by \cite{Vrard2021}, whereby the greyscale value indicates the mass of the star, determined by seismology.}
    \label{fig:periodSpacingSample}
\end{figure}

\begin{figure*}[t!]
    \centering
    \includegraphics[width=0.90\textwidth]{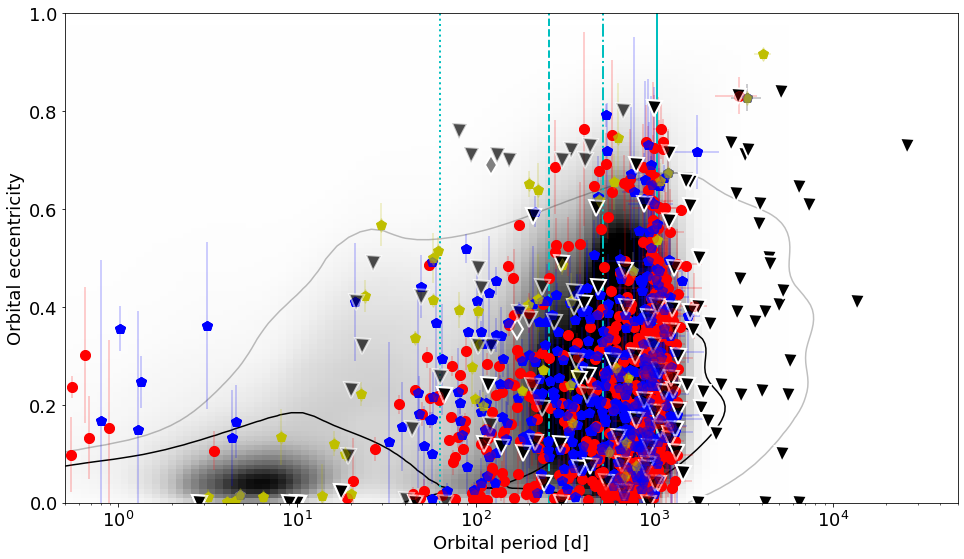}
    \caption{Distribution of the sample of binary-candidate systems from \GaiaDR, hosting evolved stars as their primary stellar component. 
    Colored dots mark systems with detected solar-like oscillations in the components primaries (blue and red mark stars from the sample of giants in \Kepler and TESS; yellow indicates main sequence stars observed with \Kepler). 
    Systems from the published \Kepler sample with an oscillating red giant primary are depicted as black triangles. Systems hosting two oscillating red giant components (PB2) are shown as grey diamonds.
    The background density plot represents the distribution of all systems in \GaiaDR TBO with full orbital solutions. 
    The thin and thick black contour lines show the isocontour regions of densely populated regions and the envelope of the distribution of SB9 binary systems, respectively. The vertical solid, dash-dotted, dashed, and dotted lines represent the 1034\,d baseline of \GaiaDR as well as one half and one-fourth of it and the $\sim$63\,day precession period of the satellite.}
    \label{fig:ePplane}
\end{figure*}

The seismic scaling relations assume the star's oscillations to be in the asymptotic regime of high radial-order modes, where oscillation modes are equally spaced in frequency by \dnu \citep{Tassoul1980}. It has been shown that with the decreasing frequency of the excess oscillation power, the stars oscillate in lower radial orders. Additionally, non-adiabatic effects in the stellar atmosphere have a stronger impact on the oscillation modes with increasing luminosity and decreasing density as the star approaches the luminous regime of the red giant branch (RGB). These effects lead to a departure of the measured global seismic properties with respect to the asymptotic assumptions built into the scaling relations. To correct for these effects, numerous techniques have been developed \citep[e.g.,][]{Sharma2017}. For the giants, we use the formalism of \cite{Mosser2013}, with the suggested value of $\zeta$=0.038 to correct for the observed a large frequency separation ($\Delta \nu_\mathrm{obs}$) for its asymptotic value of:
\begin{equation}
\Delta \nu_\mathrm{asy} = (1+\zeta)\Delta \nu_\mathrm{obs}\,.
\end{equation}
For stars on the main sequence, we use the uncorrected value as their structural properties and, consequently, their global seismic parameters are close to the solar reference values.

The top panel of Fig.\,\ref{fig:periodSeismolgy} shows the binary candidates' distribution of the obtained masses and radii. From the comparison with the masses and radii for giants and dwarfs from the \Kepler mission (Table\,\ref{tab:ArchiveResults}), we see that stars reported from the TESS mission that have radii larger than the clump show an excess of massive stars (2\,$\lesssim$\,$M/M_\odot$\,$\lesssim$\,4). 
This trend is more pronounced if \num and \dnu values from a single pipeline are used. For TESS giants, we therefore use the mean values reported by \hbox{\cite{Mackereth2021}}. 
We understand this as a problem of accurately determining the large frequency separation from TESS data if \num is low. 
In such circumstances, it is challenging to resolve the comb-like pattern of the power excess due to the contaminating signal from the granulation background, systematic effects introduced through multiple sectors with multiple CCD and pixel combinations, along with a smaller number of modes excited in the power excess as well as a lower frequency resolution due to the shorter timebase to resolve them.

\begin{figure*}[t!]
    \centering
    \includegraphics[width=2\columnwidth,height=100mm]{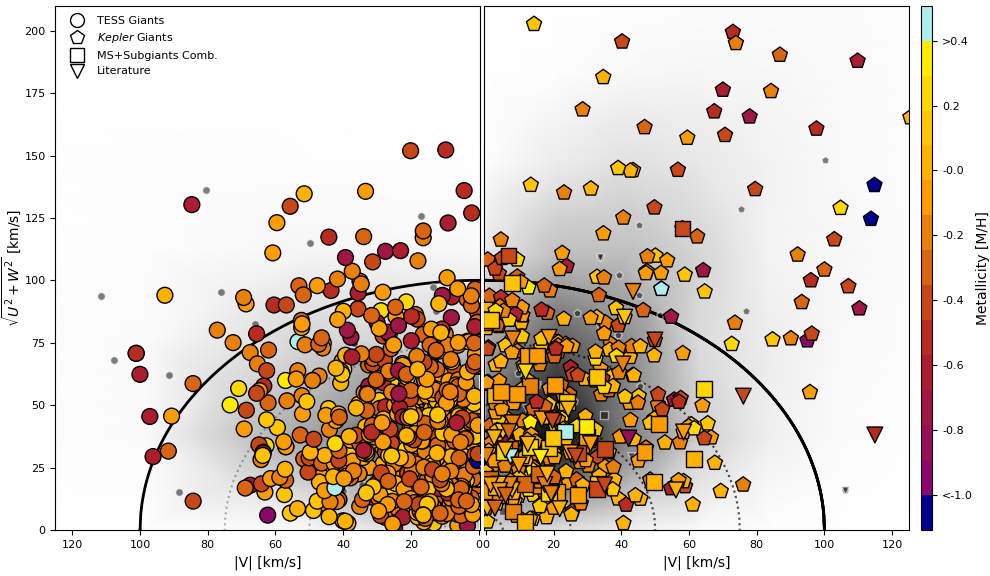}
    %{figures/Hist_metall.png}
    \caption{Distributions of the Galactic space velocities U, V, and W 
    for the red giants in the TESS, and red giants as well as main sequence stars in the \Kepler and literature samples are shown in the left and right Toomre diagram, respectively.
    The color of the markers indicates the metallicity.     
    We note that 9 candidate systems have velocity values outside the shown space velocity range in  the left panel and 11 in the right panel.
    The small grey symbols represent the stars without a solution for the metallicity in the \Gaia DR3. 
    The thick black line marks the approximative separation between the galactic thin  (X\,$\lesssim$\,100 km/s) and thick disk (X\,$\gtrsim$\,100 km/s).
    The form of the markers corresponds to the different samples as presented in Fig\,\ref{fig:seismicHRD}. The grey scale in the background shows the \hbox{distribution of all stars with seismic values in the \Kepler and \Tess samples.} 
    \label{fig:MetDist_Toomre}}
\end{figure*}

\subsection{Evolutionary states from asteroseismology \label{sec:evolStates}}
The evolution of a solar-like star includes several structurally distinct phases. It is straightforward to separate main sequence stars and subgiants from the red giant stars based on the peak frequency of their excess oscillation power. 

The red giant phase, however, is an apposition of several structurally distinct phases \citep[e.g.,][and references therein]{kippenhahn2013, Pinsonneault2023}. The first phase is the RGB. Once a star has consumed its core hydrogen content, the core will contract, and the envelope will expand. In this phase, the only energy source of the star is the fusion of hydrogen in a shell around the He core.
Consequently, its luminosity will rise until the inert He-core ignites. Depending on the star's mass, this will happen under degenerate conditions (for stars with $M_\star$\,$\lesssim$\,2\,$M_\odot$) and settle for quiescent core-He burning in the red clump. For more massive stars, the ignition temperature of He will be reached before forming a degenerate core. Core helium ignition then proceeds under non-degenerate conditions, with the star settling on the less luminous secondary clump (2RC).  
These phases are followed by the asymptotic giant branch (AGB) once the helium in the core has been exhausted.

These structural readjustments force substantial changes in the stellar radius and luminosity. We discuss the impact of the radius on the tidal forces in detail in Sect.\,\ref{sec:tides}. 
Figures\,\ref{fig:seismicHRD} and \ref{fig:periodSeismolgy} indicate the location of the MS, RGB, RC, and 2RC in the respective parameter space. Because the frequency pattern of mixed modes is sensitive to the density contrast between the surface and the stellar core, these modes can use the spacing in the dipole forest to unambiguously determine the evolutionary state of a giant
\citep{bedding2011}. The approach of \cite{Mosser2015}  reconstructs the value of the asymptotic-period spacing of dipole-gravity modes, $\Delta\Pi_{\ell=1}$ \citep{Tassoul1980}.

To identify binary candidates whose primaries have a determined evolutionary state, we correlated the published values of $\Delta\Pi_{1}$ for \Kepler and TESS giants \citep{Vrard2016, Vrard2021} with the binary candidate lists. The identification of RGB stars from TESS data was quite uncertain due to similar problems as discussed for the \dnu for the same mission. Therefore, we excluded any H-shell burning stars identified in the TESS sample.
In total, we have four samples composed of 45 stars on the main sequence and 41 on the RGB (H-shell burning). For the He-core burning phase, we get 80 stars in the RC and 25 in the 2RC. 
The colored dots in Fig.\,\ref{fig:periodSpacingSample} illustrate the identified stars in binary candidates from \GaiaDR. Because the RGB evolution is a continuous process, we believe the gap of binary candidates between 8$\lesssim$\,$\Delta\nu$\,[$\mu$Hz]\,$\lesssim$9 is purely a statistical artifact as it is not found in the larger sample of single stars.

\subsection{Validation and distribution of orbital parameters}

From the seismically inferred radius, we can test if the orbital period reported for the binary candidate in the \GaiaDR TBO is physical, as larger stellar radii require wider binary systems, and a very short period ($P_\mathrm{orb}$\,$\lesssim$\,10\,d) for a giant indicates a possible problematic solution for the period. If it is too short, the giant star would eventually fill up its Roche lobe, the boundary within which the material is gravitationally bound to the star. Mass transfer and a Roche-lobe overflow (RLOF) would lead to alternated orbital evolution on very short time scales \citep{Soberman1997}. As a red giant branch star further expands its envelope, RLOF would soon be followed by a common envelope phase, which ends in the ejection of the common envelope on even shorter time scales \citep{Han2002}.

\begin{figure*}[t!]
    \centering
    \includegraphics[width=\columnwidth]{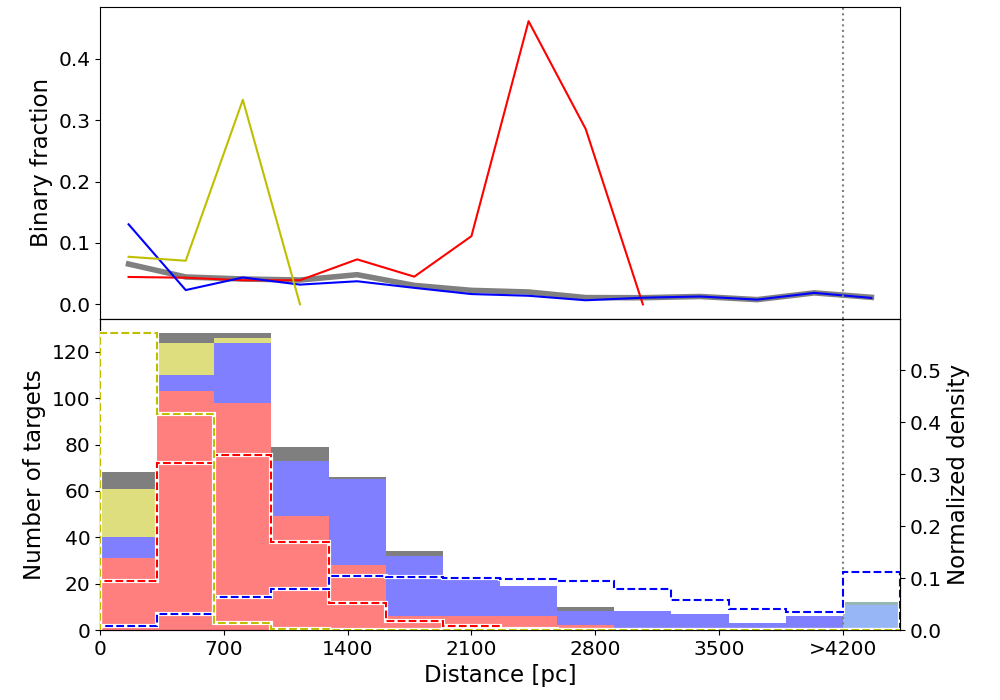}
    \includegraphics[width=\columnwidth]{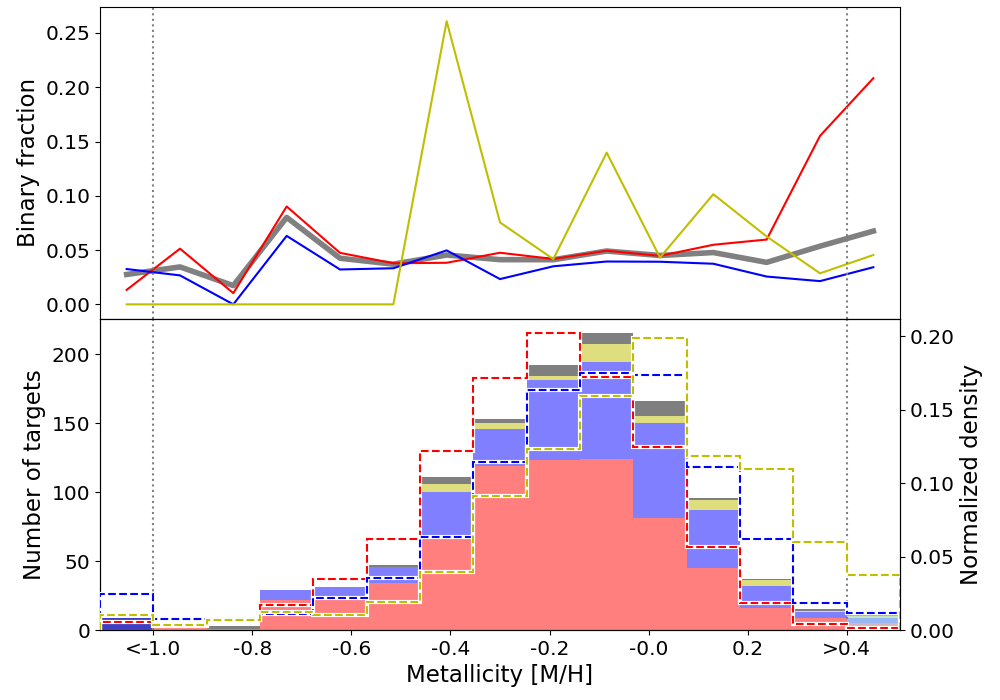}
    \caption{Binary fractions and distribution of the samples as a function of the distance (left panel) and metallicity (right panel). The top panels visualize the ratio of the number of binary candidates over the whole sample for each bin of the histogram in the bottom panel. The colors represent the samples of giants detected with TESS and the giants, and main sequence stars observed with \kepler as red, blue, and yellow, respectively. The thick grey line indicates the total binary fraction. 
    The bottom panels depict the distribution of the distances and metallicities in a histogram. The filled bars correspond to the number of binary solutions of the samples. The samples have the same color as in the top panel, and the black bars represent the literature sample. For comparison, the dashed lines show the normalized density of the corresponding input samples in the same color.
    The first and last bar of the histogram in the bottom left panel represents all stars with a metallicity lower than -1 and bigger than 0.4\,dex, respectively. For the histogram of the distance, the last bar indicates all systems with a distance higher than 4200\,pc
    \label{fig:MetDist_ratios}}
\end{figure*}

To identify such non-physically short-periodic binary systems, in Fig.\,\ref{fig:periodSeismolgy}, we show the Roche-lobe limit in the radius-period plane in the formulation of \cite{Arenou2022},
    \begin{eqnarray}
\log \left( \frac{P_d^\mathrm{Roche}}{365.25}\right) = 
\frac{3}{2} \frac{\log R_1}{216} 
- \frac{1}{2} \log (M_1+M_2) \nonumber -\\
- \frac{3}{2} \log \left(0.38+0.2 \log \frac{M_1}{M_2} \right) \,,
\end{eqnarray}
whereby $R_1$ is the radius of a primary and $M_1$ and $M_2$ are the primary and secondary masses, respectively.
In Fig.\,\ref{fig:periodSeismolgy} (bottom panel), this criterion is depicted for a hypothetical system of $M_1$=1.3\,M$_\odot$ and $M_2$=1\,M$_\odot$. For the primary's potential radius we assumed a range from less than 1 to 100\,R$_\odot$.

While \cite{Arenou2022} finds systems close to the Roche limit, the periods of most binary-candidate systems in our sample are substantially longer than the indicative Roche lobe limit shown in Fig.\,\ref{fig:periodSeismolgy}. This selection effect is explained by the selection criterion, as each primary component of systems depicted in the diagram oscillates. As shown in numerous papers \citep{Gaulme2014, Beck2018Tides, Tayar2022}, strong tidal interaction and the induced stellar activity suppress oscillation modes.
Tidal forces get enhanced as a star increasingly fills its Roche-lobe, leading to the shown selection effect.

A few systems fall below the Roche-lobe limit. While systems with orbits of less than 10 days are close to the limit of being realistic (but short-lived), they would have an orbital trajectory inside or very close to the giant. While these systems, on average, have a lower \textit{ruwe} value, this parameter is insufficient to identify these problematic systems. The documentation of the TBO reports an excess of orbital periods around four days, which could be connected to aliases that induce a spurious period, while the binary detection is valid. 
Therefore, we list these systems in the paper but flag them as unreliable and exclude them from our additional analysis.

Figure\,\ref{fig:ePplane} presents the orbital period and eccentricity of binary candidates that host a main-sequence or a red giant star. 
The same figure also compares the distribution of the full TBO catalog. 
From the \gaia systems, we find the same distribution as in the SB9 sample with two clear overdensities, whereby the one at shorter periods is populated with hot main sequence stars \citep{Torres2010AccurateMasses}. It was shown by \cite{Beck2022SB9} that oscillating giants mainly fall into the second overdensity between 500 and 1000\,days.

\subsection{Metallicity, distance, and space velocities}

The numerous detections provided by \Gaia allow us to view the rich dataset of the binary candidates in the much broader context of galactic archaeology. 
A standard tool to study the membership of stars to a particular galactic structure is the Toomre diagram.  It projects the distribution of the Galactic space velocities $U, V,$ and $W$, where V describes the absolute value of the velocity, $|V|,$ in the direction of the rotation of the Milky Way. The velocity, $U,$ represents the component of the motion in the direction from the sun toward the Galactic center, and W is the component perpendicular to the Galactic plane. To illustrate $U , V , W$ in two dimensions, we calculate $\sqrt{U^2+W^2}$, corresponding to the velocity perpendicular to $V$, which is the vector pointing away from the Galactic center. To collapse the diagram into one quadrant, we show $|V|$. Here, $U, V, W$ were corrected for the local standard of rest (LSR) ({$U_{\odot}=-8.63$, $V_{\odot}=4.76$, $W_{\odot}=7.26$} in $km\,s^{-1}$) taken from \cite{Ding2019}. The total velocity, which we refer to as $X$\,$=\sqrt{V^2+U^2+W^2}$, describes the radial distance to the origin of the plot and indicates if a star tends to belong to the younger thin disk, the older thick disk, or the halo of the galaxy.

The left and right panels of Figure\,\ref{fig:MetDist_Toomre}  depict the position of the binary candidates and the distribution of the TESS and \Kepler input sample as the background density map, respectively, in the parameter space of the Toomre diagram.
As input parameters, we used the astrometry provided by \GaiaDR and the radial velocity determined from spectra taken with the \Gaia \textit{Radial Velocity Spectrometer} \citep[RVS, ][respectively]{GaiaDR3Vallenari2022,Katz2022}. We note that the published RVS velocities are the average from all visits and are not the true systemic velocity. For simplicity, we used the inverse of the parallax as the proxy for the distance. 
Most binary candidates are located in the thin disk (X\,$\lesssim$\,100\,km/s). This result is in agreement with the location of the input sample and other ensemble studies \citep[e.g.,][]{Hon2021}. 
We note that a particular bias is set through the selection criterion of oscillations. Because detecting modes requires good signal-to-noise ratios (S/N) in the frequency analysis, the sample is biased toward closer and bright objects. The larger mirror size of \Kepler allows for the detection of oscillations in fainter stars, resulting in a richer population of stars located in the halo. In total, 159 binary candidates are located in the range of 100\,$\lesssim$\,X [km/s]\,$\lesssim$\,200, which indicates membership to the thick disk. We also find 22 candidate systems that are likely halo stars (X\,$\gtrsim$\,200\,kms). Giants in the halo are typical of spectral type K to produce the amplitudes needed to be detectable over such large distance \citep{Mathur2016, Hon2021}.

The binary yields as a function are depicted in Fig.\,\ref{fig:MetDist_ratios}. As expected, we find a clear trend in the binary detection rate (top panel) as a function of the distance (left panels in Fig.\,\ref{fig:MetDist_ratios}). As described before, this is connected to the decreasing integrated brightness of the source. This trend is best seen in the combined and \Kepler sample. The large increase in the binary yields from TESS and \Kepler dwarfs is due to small number statistics. The histograms of the binary detection (bottom panel) also shows that, as expected from the apparent magnitudes (Fig.\,\ref{fig:magHistograms}), the TESS sample contains stars generally closer to earth (within $\sim$1\,kpc) while the \Kepler giant sample is rich in stars in the kilo-parsec range. Naturally, dwarfs have a low luminosity. Therefore, all dwarfs that are bright enough to allow for the detection of solar-like oscillations are known to be close ($\lesssim$\,700pc).

Previous works have reported a strong trend in the binary occurrence rate as a function of the stellar metallicity,  whereby metal poor F, G, and K stars are more likely to be found in binary systems  \citep[e.g.,][and references therein]{MoeStefano2017, Badenes2018, Offner2022}.
For further interpretation of the sample, we used the metallicity ([M/H]) derived from the \Gaia RVS spectra \citep[\texttt{mh\_gspspec}][]{ RecioBlanco2022, Creevey2022, Fouesneau2022}. 
The right panels of Fig.\,\ref{fig:MetDist_ratios} show the binary yields and the histograms of the detection (top and bottom, respectively). For the \Kepler giants, a weak trend towards higher binary rates with lower metallcities is found, with the peak at [M/H]\,$\simeq$\,-0.7\,dex. 
For TESS, we find the same low-metallicity peak as for the \Kepler giants. However for the TESS giants, we find a flat distribution. The increase for [M/H]\,$\gtrsim$\,0.2\,dex is likely due to small number statistics. The binary yields of the \Kepler dwarf sample appear paticularly noisy, again due to small number statistics. Therefore, we do not find the expected trend as clearly as it was described in the literature.

\section{Orbital evolution through tidal interaction and stellar activity \label{sec:tides}}

\begin{figure}
    \centering
    \includegraphics[width=\columnwidth]{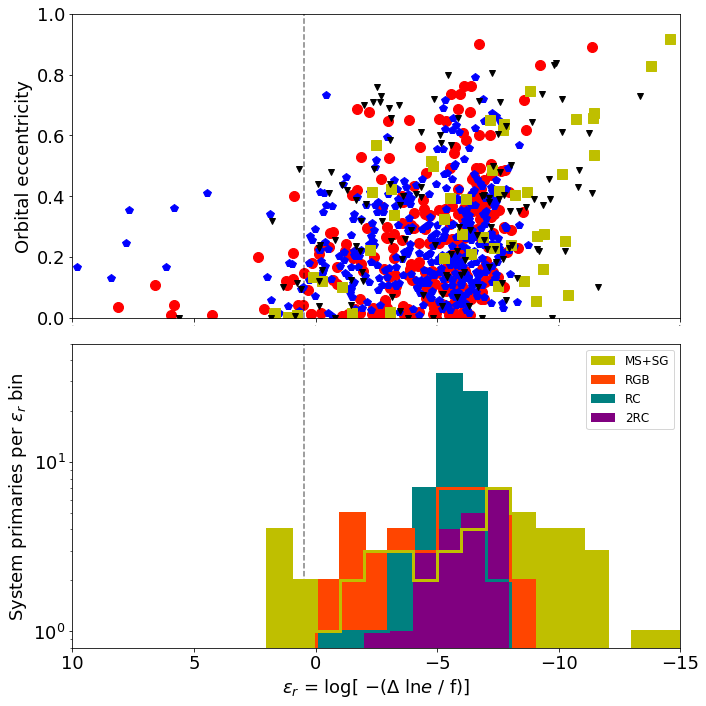}
    \caption{Orbital eccentricity of the parameterized time scale of the tidally driven orbital circularization, $\varepsilon_{\rm r}$. The top panel shows the position of all binary candidates in the parameter plane. 
    The histogram in the bottom panel depicts the distribution of $\varepsilon_{\rm r}$ for the systems with an oscillating primary, for which the evolutionary state could be determind from seismology.
    The color distinguishes between the evolutionary states of secondary clump (2RC), red clump (RC), red giant branch (RGB) in purple, teal, and orange, respectively. 
    Because of their small number and nearby evolutionary state, main sequence and subgiant primaries are shown as one group in yellow (MS+SG).
    The grey dashed line marked the proposed value of $\varepsilon_{\rm crit}$.
    }
    \label{fig:VerbuntPhinney}
\end{figure}

Stars in binary systems provide many constraints that simplify the ill-constrained parameter space for stellar modeling. However, for a detailed analysis, the effects of the interaction between the two stars need to be considered. The dissipation of tidal energy in the stellar structure influences the parameters of the system and the interior of its stellar components \citep[for comprehensive reviews of tidal theory, see ][and references therein]{Zahn2013, Ogilvie2014, Mathis2019}. 
The tidal forces lead to the circularization of the orbit and the alignment and synchronization of the orbital and rotational spins.  
In the case of a strong tidal interaction, the additional heat induced by dissipating the kinetic energy of the tides into the stellar structure might lead to an inflation of the stellar radius \citep{Mathis2013LNP}. From the first principles of physics, such extra heat will force the star to adjust its radius to stay in the thermal equilibrium and lead to the expansion of the stellar radius. Because the seismic scaling relation is based on the unperturbed solar case, such departure could lead to overestimating the seismically inferred stellar mass and radius. Therefore, it is essential to test the systems used for calibrating the seismic scaling relations for having negligible levels of tidal interaction.

\subsection{Strength of the equilibrium tide}

\begin{figure*}[t!]
    %\centering
    \includegraphics[width=0.98\textwidth]{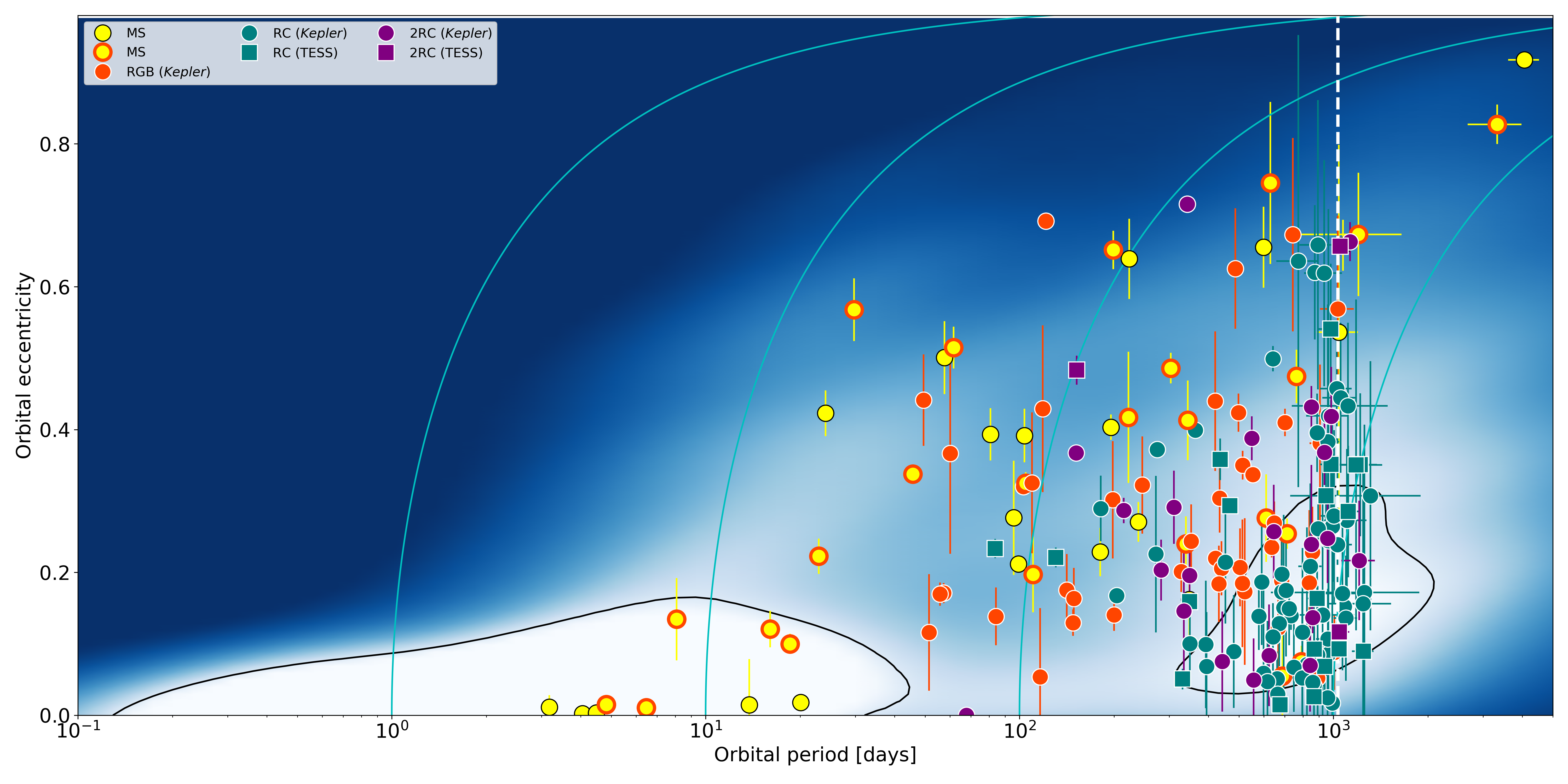}
    \caption{Orbital periods and eccentricities of the binary systems hosting a primary with identified evolutionary stages. The particular color and the shape of the data points indicate the seismically inferred evolutionary stage and the space mission this star has been observed with, respectively.  The light-blue lines indicate the arcs of constant angular momentum in the e–P plane for circular orbital periods for 1, 10,
100, and 1\,000 days. 
    The background color map represents the normalized probability-density distribution of the full SB9 sample. The black lines envelop the regions with a density of at least seven times the median probability density. The white vertical dashed line represents the 1034\,d timebase of \GaiaDR.
    \label{fig:periodSpacingEPPlane}}
%\end{figure*}
\vspace{3mm}
%\begin{figure*}[t!]
    \centering
    \includegraphics[width=\textwidth]{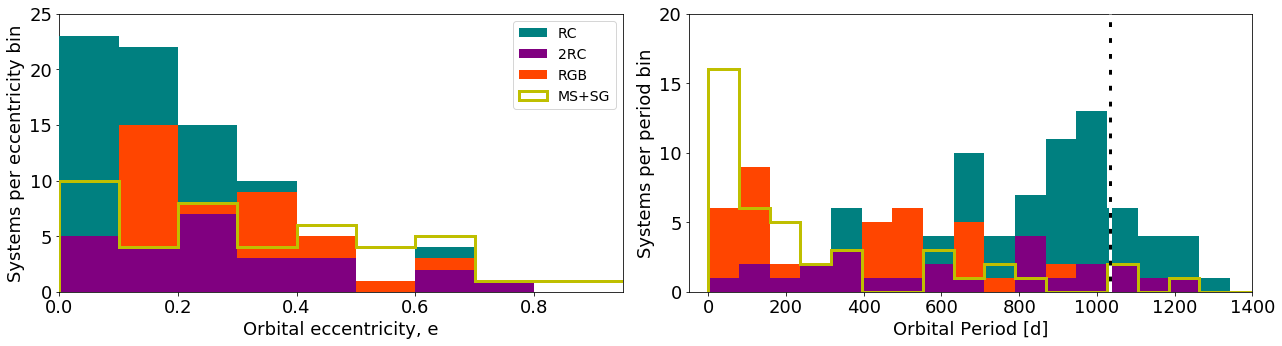}

    \caption{Distribution of orbital eccentricities, separated by evolutionary stage and channel. The left and right panel give the number of systems per eccentricity and period bin, respectively.  
    The color distinguishes between the evolutionary states of secondary clump (2RC), red clump (RC), red giant branch (RGB) in purple, teal, and orange respectively. 
    Because of their small number and nearby evolutionary state, main sequence and subgiant primaries are shown as one group in yellow (MS+SG).
    The vertical dashed line represents the 1034\,days timebase of \GaiaDR.}    
    \label{fig:eccentricityPeriodDistributions}
\end{figure*}

Because giant stars have deep convective envelopes and reach large stellar radii, the dominating mechanism for dissipation of tidal energy is expected to be the equilibrium tide \citep[e.g.,][]{mathis2015,remus2012,Gallet17}. It was confirmed by \cite{Beck2018Tides} that the dynamical tide could have a small contribution in the subgiant phase and on the very low-luminosity RGB, but is overall negligible for the orbital evolution of the binary system.

Using the formalism of \cite{Verbunt1995} \citep[based on][]{Zahn1977, hut1981} to quantify the efficiency of the dissipation of the equilibrium tide, we calculated the rate of the eccentricity reduction in a binary system, which we refer to as $\varepsilon_r $, %For the benefit of the reader, we refer to this term as $\varepsilon_r $,
\begin{equation}
    \varepsilon_\mathrm{r} = \log \left[- \frac{\Delta \ln e}{f}\right]\,. \label{eq:vareps}
\end{equation}
In this notation, the parameter $f$ is an unknown normalization factor on the order of unity. The change in eccentricity, 
\begin{eqnarray}
\label{eq:deltaLnE}
\frac{\Delta \ln e}{f}
 &=& \frac{-1.7}{ 10^{5}}\cdot\left(\frac{M_1}{M_\odot}\right)^{-11/3}\cdot 
 \frac{q}{(1+q)^{5/3}} \cdot I(t) \cdot \left(\frac{P_{\rm orb}}{{\rm day}}\right)^{-16/3}\,.
\end{eqnarray} 
with $q$=$M_2$/$M_1$ as the mass ratio between the system's secondary and primary components, 
incorporates the third Keplerian law and the circularization function,
\begin{equation}{I\left(t\right) = \int_0^t\left(\frac{T_{\rm eff}}{4500\,K}\right)^{4/3} 
\cdot \left(\frac{M_{\rm env}(t')}{M_\odot}\right)^{2/3}
\cdot \left(\frac{R_1(t')}{R_\odot}\right)^{8}
dt'\,[yr]}\,.
%\nonumber
\label{eq:IofT}
\end{equation} 
This function is dependent on the effective temperature ($T_\mathrm{eff}$), the mass in its convective envelope ($M_\mathrm{env}$), and most importantly, the stellar radius ($R_1$) of the primary. 
Because the circularization function depends on the eighth power of the primary radius, the equilibrium tide will become increasingly important as a star advances the red giant branch. Following the approach of \cite{Beck2018Tides}, we used the asteroseismic masses and radii, calculated as described in Section\,\ref{sec:scalingRelations}, for this analysis.

The top panel of Fig.\,\ref{fig:VerbuntPhinney} depicts the distribution of the binary systems as a function of the change in eccentricity. By construction, large values of $\varepsilon_\mathrm{r}$ indicate strong tidal interaction in a system. \cite{Verbunt1995} suggested that an $\varepsilon_\mathrm{crit}$\,=\,0.478 approximates the segregation between the systems with strong and less efficient tides. As $\varepsilon_\mathrm{crit}$ is not a sharp limit, eccentric binaries with values slightly above this value could be short-lived systems that are about to circularize. \cite{Kroupa1995} coined the term "forbidden binaries" for this group. Previous studies \citep[e.g.,][]{Verbunt1995,Beck2018Tides,Benbakoura2021} did not show eccentric systems at values of $\varepsilon_\mathrm{r}$\,$\gtrsim$\,3. These outliers are very likely unphysical and originate from significantly underestimated orbital periods in the TBO. Therefore, most of the system candidates depicted in Fig.\,\ref{fig:VerbuntPhinney} have reported periods in the TBO that are physically meaningful. 

The lack of systems with values $\varepsilon_r$\,$>$\,$\varepsilon_{crit}$ in Fig.\,\ref{fig:VerbuntPhinney} is a result of the selection of only oscillating objects. Several papers have demonstrated \citep{Gaulme2016, Beck2018Tides, Mathur2019, Benbakoura2021} that the tidally driven spin of the outer stellar layers increases the dynamo action. The increased magnetic activity suppresses solar-like oscillations. 
\cite{Beck2018Tides} indeed showed that this limit also separates between system hosting giants with oscillating ($\varepsilon_\mathrm{r}$\,$\lesssim$\,$\varepsilon_\mathrm{crit}$) and non-oscillating giant primaries ($\varepsilon_\mathrm{r}$\,$\gtrsim$\,$\varepsilon_\mathrm{crit}$).
Furthermore, a dependency on the orbital eccentricity can be assumed due to increased tidal strength and additional effects if the system encounters Roche-lobe overflow at the periastron. The resulting distribution of $\varepsilon_\mathrm{crit}$ supports the test using the approximation of the Roche-lobe radius, shown in Fig.\,\ref{fig:periodSeismolgy}.

The main sequence stars are typically located at intermediate to shorter periods (Fig.\,\ref{fig:ePplane}). For this class of stars, the formalism presented in Eq.\,\ref{eq:vareps}\,to\,\ref{eq:IofT}  separates well the circularized from the eccentric systems. It is interesting to note that in Fig.\,\ref{fig:VerbuntPhinney} the main-sequence dwarfs are the group of stars that extends to the lowest values of $\varepsilon_\mathrm{r}$. This is a consequence of the much smaller radii and convective envelopes of solar-like dwarfs compared to red giants. In such stars, radiative structures become significant, necessitating the inclusion of the dynamical tide to accurately describe the tidal budget of the system \citep{Ahuir2021,Barker2022}. For stellar objects cooler than the Kraft-break limit and orbital periods less than $\sim$10\,days, the dynamical tide is so efficient that systems are quickly circularized and synchronized \citep[][and references therein]{Offner2022}.

%\newpage
\subsection{Distribution of orbital eccentricities and periods}

Binary systems hosting solar-like stars are expected to be born with a flat distribution of eccentricities over a wide range of eccentricities \citep{MoeStefano2017, Mirouh2023}. The subsequent tidal evolution is a function of dwell time and the advances in stellar evolution of both binary components. Therefore, we can study the impact of tidal forces from the distributions of the parameters in distinct evolutionary stages.

Hereby, the efficiency of the dissipation of tidal energy has been debated in the literature. 
In the current understanding, eccentric binaries with an evolved component should have circularised during phases of their evolution when they have expanded to large radii. 
\cite{Verbunt1995} argued that because of the high radius dependence, they expect all systems hosting a red-clump star or an AGB to be circularized. Similar behavior is expected long before the onset of RLOF \citep[e.g., see][]{Vos2015}. Yet, contrary to this prediction, \cite{Beck2018Tides} pointed out the existence of red clump stars in eccentric binary systems. This finding agrees with the eccentricity distributions found in the large sample study of APOGEE time series spectroscopy by \cite{Badenes2018}. Furthermore, wide hot-subdwarf binaries are almost all eccentric, even though the primary is a post-RGB star that underwent a mass-loss episode near the tip of the RGB \citep[e.g.,][and references therein]{Vos2013}. Further examples are Ba stars, and Tc-poor S stars \citep[e.g.][]{Vanderswaelmen2017}, symbiotic stars \citep[e.g.,][]{Merc2019}, RV Tauri binaries \citep[e.g.,][]{Escorza2020} or dusty post-AGB stars \citep[e.g.,][]{Gorlova2014}, all of which can occur in binary systems with significantly non-zero eccentricities, even though they should have circularised if their periods are shorter than $\sim$3000 days \citep{Pols2003}. To explain the eccentricities found in red giant binary systems, \cite{Nie2017} suggested from modelling the binary evolution that the efficiency of the current formalism could be overestimated by a factor of 100. 

Solar-like oscillators cover a wide range of evolutionary phases and channels.
From such a sample, we can probe the distribution of the orbital eccentricities and periods as a function of the evolutionary states to learn more about the efficiency of the equilibrium tide. 
Figure\,\ref{fig:periodSpacingEPPlane} shows the position of systems with a solar-like oscillator with a known evolutionary state in the $e$-$P$ plane. The sample sizes of the various evolutionary stages, presented in Section\,\ref{sec:evolStates}, are sufficient to derive general conclusions. To better quantify and discuss the distributions of these four groups, we present their distribution in histograms (shown in  Fig.\,\ref{fig:eccentricityPeriodDistributions}).  

Stars on the main sequence are significantly less evolutionarily advanced than any star in the giant phase and consequently closer to the system's initial conditions. As discussed above, these stars have smaller radii which in wider binary systems ($P_\mathrm{orb}$\,$\gtrsim$10\,d) results in lower tidal interaction compared to the equilibrium tide. This is the reason why systems with primaries in this evolutionary phase, which lasts two orders of magnitude longer than the red giant phase, retain a relatively flat eccentricity distribution between 0\,$\lesssim$\,$e$\,$\lesssim$\,0.9, originating from the birth distribution of $e$.

The situation changes for systems hosting a post-main sequence star. Even considering that the eccentricities have rather large error bars compared to typical values from ground-based RV monitoring, a general trend is found to show that red giant stars have lower eccentricities than main sequence stars. 
Overall, RGB stars are found between 0.1\,$\lesssim$\,$e$\,$\lesssim$0.7. The lack of circularized systems ($e$\,$\lesssim$\,0.1) with RGB primaries is in agreement with previous studies \citep{Beck2014a,Beck2022SB9, Gaulme2014, Benbakoura2021}.
Clear differences are found among the two stages of the quiescent helium-core burning, which occupy different regions in the HRD. While the RC stars have their highest occurrence rate below $e$\,$\lesssim$0.2, the more massive 2RC stars show a flat distribution between 0\,$\lesssim$\,$e$\,$\lesssim$0.8, similarly to the main sequence stars.

These differences in the distributions are likely to be the product of the accumulated tidal history along the stellar evolution. If the mass of a star is $M_\star$\,$\lesssim$2\,$M_\odot$, the inert core will degenerate before it reaches the ignition temperature of He. To obtain the energy to again lift the core degeneracy, the star needs to reach a high luminosity, which forces the star to keep expanding \cite[see][and references therein]{Hekker2020} until the core ignites and the star's envelope readjusts. Due to the degeneration, all cores of RGB stars of a given luminosity are similar independent of their mass and metallicity \citep[$\log(L/L_\odot)$\,$\simeq$\,3.4,][]{Serenelli2017TRGB}. At the tip of the RGB, a star has $\sim$175\,$R_\odot$. 
In stars with masses $M_\star$\,$\gtrsim$2\,$M_\odot$, the core reaches the ignition temperature before the central regions degenerate. Consequently, such a star will ignite He in its core much earlier and at smaller radii ($\sim$30\,$R_\odot$), thus settling in the less luminous secondary clump.

\begin{table*}[t!]
\centering
\caption{Rotation and photospheric activity for primaries of detected binary systems. \label{tab:rot}}
\tabcolsep=4pt
\vspace{-2mm}
\begin{tabular}{r|rrrrr|rrrrrrrr}
\hline\hline
\multicolumn{1}{c}{KIC} &
\multicolumn{1}{c}{P$_\mathrm{orb}$} &
\multicolumn{1}{c}{$e$} &
\multicolumn{1}{c}{P$_\mathrm{rot}$} &
\multicolumn{1}{c}{P$_\mathrm{orb}$/P$_\mathrm{rot}$} &
\multicolumn{1}{c}{Type} &
\multicolumn{1}{c}{\num} &
\multicolumn{1}{c}{\dnu} &
\multicolumn{1}{c}{S$_\mathrm{ph}$} &
Ref \\
\multicolumn{1}{c}{~} &
\multicolumn{1}{c}{[d]} &
\multicolumn{1}{c}{~} &
\multicolumn{1}{c}{[d]} &
\multicolumn{1}{c}{~} &
\multicolumn{1}{c}{~} &
\multicolumn{1}{c}{[$\mu$Hz]} &
\multicolumn{1}{c}{[$\mu$Hz]} &
\multicolumn{1}{c}{[ppm]} &
\multicolumn{1}{c}{~} \\[0.5em]
\hline

\,4914923   & 99.24$\pm$0.07     & 0.21$\pm$0.01 & 24.5$\pm$3.8  & 4.04     & ASB1 & 1825$\pm$107 & 88.57$\pm$2.46  & 125.3$\pm$3.7    & {[}1{]}        \\
\,5516982   & 24.12$\pm$0.02     & 0.42$\pm$0.03 & 24.5$\pm$2.0  & 0.98     & SB1 & 1699$\pm$109 & 85.0$\pm$1.91   & 134.0$\pm$4.6    & {[}1{]}        \\
\,5696625   & 611.35$\pm$7.2     & 0.28$\pm$0.06 & 10.6$\pm$1.3  & 57.89    & ABS1 & 696$\pm$4    & 39.24$\pm$3.22  & 105.1$\pm$5.9    & {[}1{]}        \\
\,5814512   & 222.54$\pm$1.34    & 0.42$\pm$0.09 & 5.6$\pm$0.7   & 39.39    & ORB & 995$\pm$21   & 53.35$\pm$1.66  & 82.5$\pm$6.5     & {[}1{]}        \\
(h)~~~7206837  & 4.05$\pm$0.0       & 0.0$\pm$0.01  & 4.0$\pm$0.3   & 1.0     & SB1  & 1645$\pm$120 & 79.17$\pm$1.76  & 234.2$\pm$16.1   & {[}1{]}        \\
(b)~~~7668623  & 4.82$\pm$0.0       & 0.01$\pm$0.01 & 5.5$\pm$0.4   & 0.88    & $-$  & 822$\pm$45   & 46.82$\pm$1.51  & 843.1$\pm$46.9   & {[}1{]}        \\
\,8016496   & 711.7$\pm$11.52    & 0.25$\pm$0.02 & 10.9$\pm$0.8  & 65.23   & ASB1  & 1045$\pm$29  & 53.99$\pm$1.5   & 93.6$\pm$5.3     & {[}1{]}        \\
\,8414062   & 791.36$\pm$7.68    & 0.07$\pm$0.02 & 11.0$\pm$0.8  & 71.94   & ASB1  & 1100$\pm$13  & 72.74$\pm$2.27  & 104.2$\pm$5.3    & {[}1{]}        \\
\,8677016   & 347.4$\pm$4.32     & 0.16$\pm$0.06 & 29.6$\pm$2.5  & 11.74   & SB1  & 1882$\pm$53  & 92.64$\pm$2.37  & 78.5$\pm$2.7     & {[}1{]}        \\
\,9025370   & 239.12$\pm$0.45    & 0.27$\pm$0.03 & 23.2$\pm$3.5  & 10.3    & ORB  & 3045$\pm$75  & 131.43$\pm$3.42 & 250.9$\pm$6.5    & {[}1{]}        \\
\,9098294   & 20.1$\pm$0.0       & 0.02$\pm$0.01 & 19.9$\pm$1.3  & 1.01    & SB1  & 2368$\pm$102 & 110.0$\pm$3.16  & 223.8$\pm$7.0    & {[}1{]}        \\
\,9209245   & 22.92$\pm$0.0      & 0.22$\pm$0.02 & 22.6$\pm$1.4  & 1.02    & SB1  & 1017$\pm$19  & 53.86$\pm$1.5   & 155.2$\pm$5.3    & {[}1{]}        \\
\,9225600   & 343.75$\pm$2.01    & 0.41$\pm$0.06 & 9.0$\pm$0.7   & 38.24   & ASB1  & 1192$\pm$29  & 64.01$\pm$1.77  & 32.7$\pm$2.7     & {[}1{]}        \\
(f)~~~9328372  & 95.9$\pm$0.58      & 0.28$\pm$0.08 & 3.9$\pm$0.7   & 24.47   & SB1  & 1379$\pm$48  & 68.66$\pm$1.82  & 307.2$\pm$22.2   & {[}1{]}        \\
(c)~~~9390670  & 1041.87$\pm$156.35 & 0.54$\pm$0.26 & 3.6$\pm$0.2   & 293.48  & SB1  & 1351$\pm$26  & 71.07$\pm$1.97  & 705.4$\pm$49.1   & {[}1{]}        \\
\,9702369   & 195.76$\pm$0.3     & 0.4$\pm$0.02  & 11.6$\pm$0.9  & 16.88   & ASB1  & 2039$\pm$118 & 94.05$\pm$2.36  & 50.5$\pm$2.5     & {[}1{]}        \\
(e)~~~9898385  & 13.78$\pm$0.01     & 0.01$\pm$0.06 & 3.4$\pm$0.3   & 4.1    & SB1   & 1352$\pm$33  & 69.29$\pm$1.92  & 368.9$\pm$27.5   & {[}1{]}        \\
(g)~10355856 & 4.49$\pm$0.0       & 0.0$\pm$0.01  & 4.5$\pm$0.3   & 1.0     & SB1  & 1320$\pm$79  & 67.41$\pm$1.31  & 299.8$\pm$19.6   & {[}1{]}        \\
(a)~10775748 & 6.48$\pm$0.0       & 0.01$\pm$0.01 & 6.4$\pm$0.5   & 1.01    & SB1  & 997$\pm$10   & 60.88$\pm$1.82  & 843.2$\pm$43.8   & {[}1{]}        \\
\,11862497  & 598.79$\pm$7.41    & 0.66$\pm$0.06 & 4.6$\pm$0.4   & 129.89  & ORB  & 1888$\pm$47  & 90.7$\pm$2.09   & 88.6$\pm$7.0     & {[}1{]}        \\
\,12317678  & 80.84$\pm$0.06     & 0.39$\pm$0.04 & 3.7$\pm$0.7   & 21.97   & ORB  & 1244$\pm$79  & 63.49$\pm$1.18  & 51.4$\pm$4.2     & {[}1{]}        \\
\hline
\,3942719   & 339.26$\pm$1.41    & 0.24$\pm$0.04 & 38.0$\pm$5.0  & 8.94    & ASB1  & 788$\pm$27   & 45.2$\pm$1.66   & 532.8$\pm$6.45   & {[}2{]}        \\
\,3952580   & 45.74$\pm$0.02     & 0.34$\pm$0.01 & 89.8$\pm$7.6  & 0.51    & SB1  & 636$\pm$22   & 37.09$\pm$1.11  & 140.6$\pm$1.16   & {[}2{]}        \\
\,6587236   & 688.01$\pm$17.57   & 0.05$\pm$0.09 & 17.9$\pm$2.2  & 38.46  & ASB1   & 499$\pm$2    & 32.09$\pm$1.57  & 89.8$\pm$2.28    & {[}2{]}        \\
\,8408931   & 18.57$\pm$0.0      & 0.1$\pm$0.01  & 18.4$\pm$1.3  & 1.01    & SB1  & 609$\pm$4    & 33.98$\pm$1.54  & 192.34$\pm$3.4   & {[}2{]}        \\
(d)\,9163769   & 3.17$\pm$0.0       & 0.01$\pm$0.02 & 3.2$\pm$0.2   & 1.0    & SB1   & 1573$\pm$11  & 80.62$\pm$1.88  & 451.47$\pm$17.89 & {[}2{]}        \\
\,10732098  & 199.15$\pm$0.46    & 0.65$\pm$0.03 & 25.6$\pm$6.0  & 7.79   & ASB1   & 1055$\pm$42  & 60.11$\pm$1.85  & 85.65$\pm$1.76   & {[}2{]}        \\
\,11709205  & 57.9$\pm$0.02      & 0.41$\pm$0.01 & 58.0$\pm$3.3  & 1.0    & SB1   & 271$\pm$12   & 18.95$\pm$0.87  & 224.65$\pm$2.28  & {[}2{]}        \\
\hline
\,3437031   & 623.74$\pm$10.24   & 0.08$\pm$0.07 & 138.8$\pm$7.2 & 4.49    & ASB1  & 63$\pm$3     & 5.6$\pm$0.18    & $-$              & {[}3{]}          \\
\,4358067   & 139.5$\pm$0.58     & 0.34$\pm$0.06 & 67.5$\pm$4.4  & 2.07   & ASB1   & 4$\pm$0      & 0.71$\pm$0.04   & $-$              & {[}3{]}          \\
\,4758020   & 290.39$\pm$6.41    & 0.18$\pm$0.18 & 124.2$\pm$4.7 & 2.34    & SB1  & 92$\pm$6     & 7.5$\pm$0.14    & $-$              & {[}3{]}          \\
\,5087190   & 959.44$\pm$106.84  & 0.33$\pm$0.15 & 66.1$\pm$9.5  & 14.51   & SB1  & 70$\pm$3     & 6.08$\pm$0.13   & $-$              & {[}3{]}          \\
\,5382824   & 653.29$\pm$4.31    & 0.11$\pm$0.02 & 92.2$\pm$7.9  & 7.08    & SB1  & 100$\pm$4    & 8.0$\pm$0.23    & $-$              & {[}3{]}{[}4{]} \\
%\,5382824   & 653.29$\pm$4.31    & 0.11$\pm$0.02 & 92.2$\pm$7.9  & 7.08      & 100$\pm$4    & 8.0$\pm$0.23    & $-$              & {[}3{]}          \\
\,5439339   & 98.04$\pm$0.07     & 0.03$\pm$0.02 & 60.4$\pm$5.3  & 1.62    & SB1  & 99$\pm$5     & 7.89$\pm$0.28   & $-$              & {[}3{]}{[}4{]} \\
%\,5439339   & 98.04$\pm$0.07     & 0.03$\pm$0.02 & 60.4$\pm$5.3  & 1.62      & 99$\pm$5     & 7.89$\pm$0.28   & $-$              & {[}3{]}          \\
\,5534910   & 851.96$\pm$38.35   & 0.24$\pm$0.11 & 94.8$\pm$6.7  & 8.99    & SB1  & 112$\pm$5    & 8.26$\pm$0.24   & $-$              & {[}3{]}          \\
\,5707338   & 885.17$\pm$14.19   & 0.27$\pm$0.07 & 60.4$\pm$4.3  & 14.65   & SB1  & 81$\pm$4     & 6.49$\pm$0.18   & $-$              & {[}3{]}          \\
\,6032639   & 979.76$\pm$45.02   & 0.29$\pm$0.07 & 123.4$\pm$4.1 & 7.94    & SB1  & 46$\pm$2     & 4.73$\pm$0.15   & $-$              & {[}3{]}{[}4{]} \\
%\,6032639   & 979.76$\pm$45.02   & 0.29$\pm$0.07 & 123.4$\pm$4.1 & 7.94      & 46$\pm$2     & 4.73$\pm$0.15   & $-$              & {[}3{]}          \\
\,6933666   & 48.34$\pm$0.11     & 0.18$\pm$0.08 & 99.2$\pm$2.6  & 0.49    & SB1  & 34$\pm$2     & 3.83$\pm$0.12   & $-$              & {[}3{]}{[}4{]} \\
%\,6933666   & 48.34$\pm$0.11     & 0.18$\pm$0.08 & 99.2$\pm$2.6  & 0.49      & 34$\pm$2     & 3.83$\pm$0.12   & $-$              & {[}3{]}          \\
\,7661609   & 106.58$\pm$0.08    & 0.19$\pm$0.01 & 104.6$\pm$3.1 & 1.02    & SB1  & 23$\pm$1     & 2.7$\pm$0.4     & $-$              & {[}3{]}          \\
\,8365782   & 702.84$\pm$5.19    & 0.07$\pm$0.02 & 116.0$\pm$4.1 & 6.06    & SB1  & 81$\pm$3     & 6.3$\pm$0.13    & $-$              & {[}3{]}          \\
\,8825444   & 128.52$\pm$0.21    & 0.04$\pm$0.02 & 83.7$\pm$7.6  & 1.54    & SB1  & 83$\pm$4     & 6.57$\pm$0.15   & $-$              & {[}3{]}          \\
\,8936339   & 796.03$\pm$54.97   & 0.05$\pm$0.12 & 103.2$\pm$5.0 & 7.71   & SB1  & 44$\pm$2     & 4.62$\pm$0.62   & $-$              & {[}3{]}          \\
\,9086060   & 453.2$\pm$11.43    & 0.21$\pm$0.09 & 81.0$\pm$4.0  & 5.59   & SB1   & 48$\pm$2     & 4.77$\pm$0.16   & $-$              & {[}3{]}          \\
\,9240941   & 254.08$\pm$1.04    & 0.13$\pm$0.04 & 125.2$\pm$6.3 & 2.03    & SB1  & 110$\pm$5    & 8.34$\pm$0.18   & $-$              & {[}3{]}          \\
\,9469212   & 992.55$\pm$141.7   & 0.26$\pm$0.14 & 68.4$\pm$6.7  & 14.5    & SB1  & 45$\pm$2     & 4.31$\pm$0.12   & $-$              & {[}3{]}          \\
\,9898373   & 617.06$\pm$16.97   & 0.05$\pm$0.08 & 128.6$\pm$6.8 & 4.8     & SB1  & 45$\pm$3     & 4.73$\pm$0.63   & $-$              & {[}3{]}          \\
\,10148118  & 388.24$\pm$2.65    & 0.06$\pm$0.03 & 89.7$\pm$9.1  & 4.33    & SB1  & 62$\pm$3     & 5.46$\pm$0.19   & $-$              & {[}3{]}          \\
\,10935853  & 882.61$\pm$17.57   & 0.3$\pm$0.07  & 62.6$\pm$5.6  & 14.11   & SB1  & 65$\pm$3     & 5.76$\pm$0.13   & $-$              & {[}3{]}          \\
\,11650041  & 107.17$\pm$0.29    & 0.04$\pm$0.05 & 53.0$\pm$7.9  & 2.02    & SB1  & 8$\pm$0      & 1.41$\pm$3.07   & $-$              & {[}3{]}          \\
\,12314910  & 69.53$\pm$0.06     & 0.02$\pm$0.02 & 104.5$\pm$3.7 & 0.67    & SB1  & 24$\pm$1     & 2.91$\pm$0.08   & $-$              & {[}3{]}{[}4{]}\\
\hline 
\end{tabular}
%\vspace{-3mm}
\tablefoot{The star's identifier in the \Kepler input catalog is given in the first column. The next columns report the period P$_\mathrm{orb}$ and eccentricity $e$ of the orbit as reported in the \Gaia DR3 TBO catalog. The next column corresponds to the period of the surface rotation P$_\mathrm{rot}$ of the primary, determined from the light curve analysis. P$_\mathrm{orb}$/P$_\mathrm{rot}$ reports the ratio between the orbital and rotation period, whereby P$_\mathrm{orb}$/P$_\mathrm{rot}$=1 suggests that orbit and surface rotation are synchronized. The next three columns give the global seismic parameters of the power excess and large frequency separation as well as the photospheric activity indicator S$_\mathrm{ph}$. The last column reports the main references for the target, where [1] refers to \cite{Santos2021}, [2] to \cite{Garcia2014}, [3] to \cite{Ceillier2017}, and [4] to \cite{Beck2022SB9}.
}
\end{table*}

The important aspect between RC and 2RC in the context of the tidal analysis is the difference in their maximum radius on the RGB. Because the equilibrium tide depends strongly on the stellar radius, systems hosting 2RC stars are expected to be far less circularised than those hosting RC primaries. Another effect that could lead to lower eccentricities is mass transfer if the system had episodes of RLOF. 

The range of higher eccentricities in the RGB sample results from an observational bias. By selecting oscillating giants with resolved seismic parameters and evolutionary states, we limit ourselves to giants on the lower part of the RGB with radii 4\,$\lesssim$\,$R/R_\odot$\,$\lesssim$12 (see Fig.\,\ref{fig:seismicHRD} and \ref{fig:periodSeismolgy}). Therefore, the stars in the RGB sample are preferentially smaller and observed prior to maximal tidal interactions.

\begin{figure}[t!]
    \centering %\vspace{-0.75mm}
    \hfill\includegraphics[width=0.977\columnwidth]{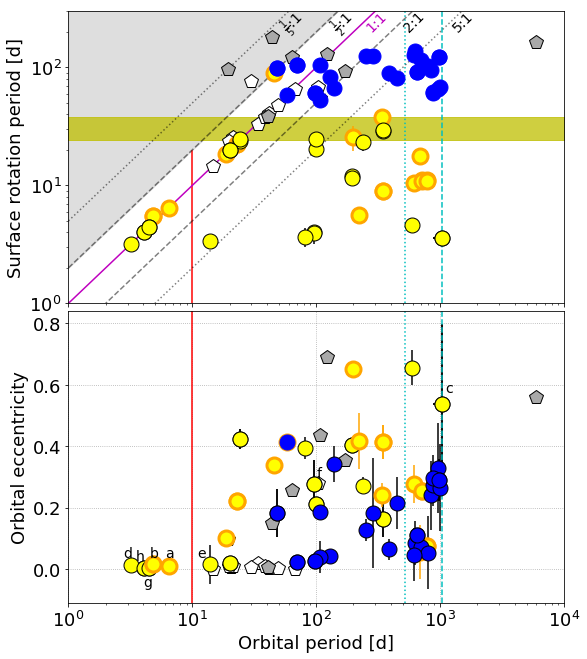}
    \includegraphics[width=1.02\columnwidth]{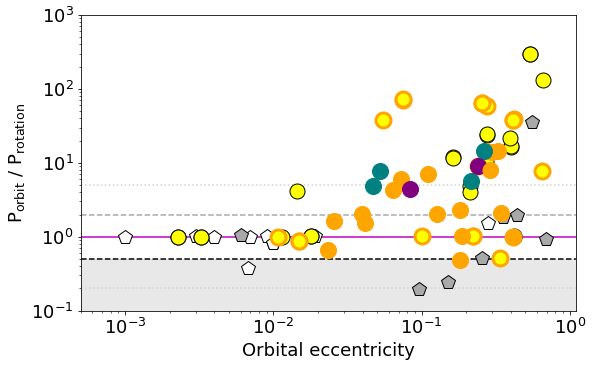}\hfill
    
    \caption{Synchronization and circularization of binary systems, hosting solar-like oscillators. 
    The top and middle panel show the surface rotation periods of the primary and orbital eccentricities for binary systems as a function of their orbital period, respectively. The bottom panel presents the ratio $P_\mathrm{orbit}/P_\mathrm{rotation}$ as a function of the orbital eccentricity. 
    Yellow, orange, and blue dots indicate dwarfs, subgiants and red giants, observed with \Kepler, respectively.
    Grey pentagons indicate systems reported previously in the literature. Filled and open symbols denote oscillating and non-oscillating primaries, respectively. 
    The inclined lines represent the resonance ratios P$_\mathrm{orb}$\,:\,P$_\mathrm{rot}$ as indicated to the right of the line. The solid magenta line indicates the synchronisation of the surface rotation with the orbit (P$_\mathrm{orb}$\,=\,P$_\mathrm{rot}$). The grey shaded area depicts the region in which the dynamical tide cannot be excited (2$\cdot$P$_\mathrm{orb}$\,$<$\,P$_\mathrm{rot}$). The yellow shaded area indicates the range of the solar siderial surface rotation.  
    The solid vertical red line indicates the limiting period for synchronization and circularization on the main sequence (P$_\mathrm{circ}$\,$\simeq$10\,d). The vertical dashed and dotted lines indicate the length and half the length of the \Gaia DR3 mission timebase. 
    }
    \label{fig:surfaceRotations}
\end{figure}

\begin{figure}
    \centering
    \includegraphics[width=\columnwidth]{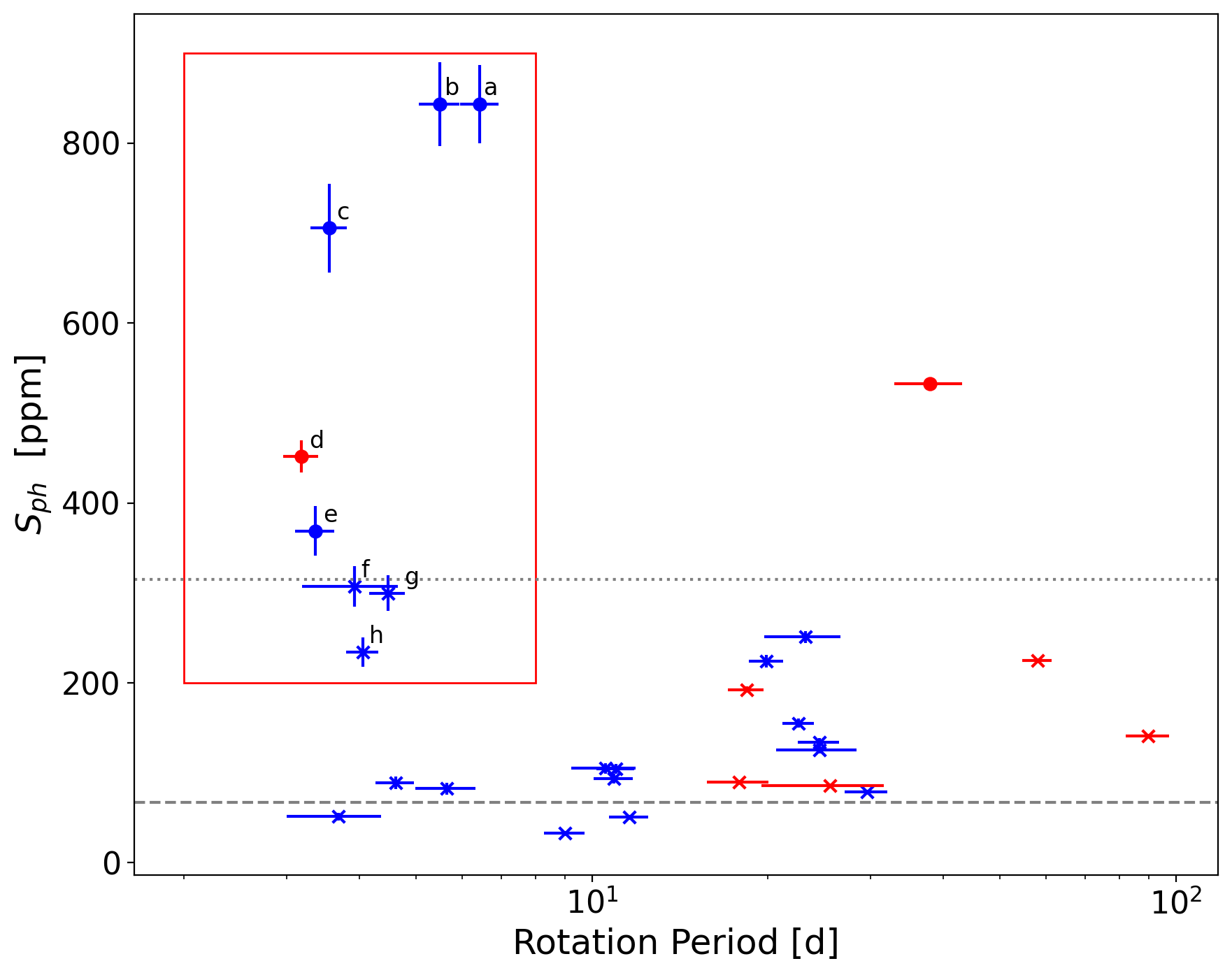}
    \caption{Photospheric activity of main sequence stars in binary candidates of \GaiaDR as a function of the rotational period. Red and blue markers indicate rotation periods reported by \cite{Garcia2014} and \cite{Santos2021}, respectively. If measurements in both references are given, the \cite{Santos2021} value is shown. For comparison, the minimum and maximum  S$_\mathrm{ph}$ value of solar cycle 23 are shown as dashed horizontal lines.
    The red box frames the distinct group stars of short rotation periods and mostly super-solar levels of stellar activity. The labels a-h provide cross identification with the stars in Fig.\,\ref{fig:surfaceRotations} and Table\,\ref{tab:rot}.
    }
    \label{fig:sph}
\end{figure}

While many systems are expected to be found betweeen 1\,000 and 2\,000\,d, the pronounced peak of periods of 1000\,days is likely to be an artifact of the \GaiaDR solutions because periods longer than that are often underestimated (Fig.\,\ref{fig:SB9residuals}, and \ref{fig:periodSeismolgy}). 
However, \Fig{fig:periodSpacingEPPlane} strongly indicates that the excess of giants in system periods around 1\,000\,days results from He-core burning stars. We interpret the reduced eccentricity scatter for RC primaries (0.1\,$\lesssim$\,$e$\,$\lesssim$\,0.5) compared to RGB primaries to be the result of tidal interactions.

The effect of the radius dependence on the tidal strength is also seen in the period distributions as a function of the four evolutionary stages. 
Because RC stars have already reached their maximum radius on the RGB, many systems with periods below $\sim$500 days could have undergone a common-envelope phase, potentially leading to the destruction of the giant primary. Therefore, we only see RC systems at longer periods, while RGB, 2RC, MS, and SG are also found at short periods. This is also seen in the typical tidal strength $\varepsilon_{\rm r}$ for both evolutionary stages. Figure\,\ref{fig:VerbuntPhinney} depicts that the remaining RC systems, because of their smaller radii and wider orbits have indeed much lower tidal interaction than RGBs.

\subsection{Surface rotation and stellar activity}

The rotational behavior of stellar components of a binary system is strongly connected to tides \citep{Zahn2013,Ogilvie2014}. The subsequent phases of pseudo-synchronization and full synchronization of the stellar rotation with the orbital motion of the binary are steps of the evolutionary path to the equilibrium state. Particularly for red giant stars, with their slow rotation period, the tidal spin-up of the envelope will give rise to strong magnetic fields through the triggered dynamo action. Lately, from photometric and chromospheric-emission measurements, \citet{Gaulme2020} and \citet{Gehan2022} showed that red giants belonging to binary systems in a configuration of spin-orbit resonance display an enhanced magnetic activity compared to single stars with the same rotation rate. Therefore, stellar rotation and activity are key observables of stars that provide information about tidal interaction. 

The level of activity and the rotational period can be estimated from the rotationally modulated flux signal introduced by dark stellar spots being rotated in and out of the line of sight of the observer \citep[][and references therein]{Mathur2019}. For the solar-like main-sequence and red giant stars in our binary candidate sample, rotation periods were derived from the dominant period of the brightness modulation, determined through auto-correlation and wavelet analysis of the \Kepler photometry by \cite{Garcia2014b}, \cite{Ceillier2017} and \cite{Santos2021}. We assume that the spots originate from the more luminous, oscillating component. The values are reported in Table\,\ref{tab:rot}.

The full new picture of the extended sample is shown in Fig.\,\ref{fig:surfaceRotations}. As a reference, the range of the equatorial and polar rotation of the Sun is shown in this picture (27 and 35\,days, respectively). 
Most of the main-sequence primaries with determined rotation periods spin significantly faster than the Sun or at least with solar-like rotation rates. 
Depending on the orbital period of the systems, we find three distinct forms of appearance related to the interplay between rotation and orbital eccentricity. As expected from theoretical predictions, the surface rotation of stars in systems with orbital periods below ten days is synchronized and the orbit is circularized \citep{OgilvieLin2007,Barker2022}. 
Dwarf-hosting systems with periods up to $\sim$30\,days are still pseudo-synchronized but have a wider range of eccentricities. 
For these groups, the rotational period is influenced by the tidal interaction. Therefore, the age of the primary cannot be determined from the relation between the surface rotation period and stellar age, known as gyrochronology \citep{Barnes2007}.
We do not see any relation between the rotational and orbital periods for systems with orbits longer than the solar rotation rate. Given that these stars have rotation periods shorter than the solar rate, we can estimate that these systems are typically younger than the Sun. 

The analysis depicted in Fig.\,\ref{fig:surfaceRotations} also reveals a fundamental difference between dwarfs and giants. As mentioned above, no oscillations are found in giants in circularized and synchronized systems \cite[typicall with $P_\mathrm{orb}$\,$\lesssim$\,20\,days][]{Gaulme2014, Beck2018Tides}. The eight dwarfs and subgiants in circularized and synchronized systems on much shorter periods (stars a-h), however, do oscillate.  A detailed analysis of this finding is beyond the scope of this paper.

%\newpage~\newpage
For 28 of the main sequence stars, the catalogs by  \cite{Garcia2014b} and \cite{Santos2021} provide an estimate of the average photospheric activity. The $S_\mathrm{PH}$ value is the mean, standard deviation of the photometric variation in a sliding boxcar, with a timescale of five times the rotation period \citep{Mathur2014, Garcia2014}. The top panel of Fig.\,\ref{fig:sph} compares these values with the minimum and maximum value of $S_\mathrm{PH}$ from solar cycle 23, determined by \cite{Salabert2017}. As can be seen from this depiction, about 80\% of the systems exhibit a solar-like activity level. These are typically longer periodic systems. One separate group of eight highly active stars at short periods stands out. To better reference individual members of this group, we labeled them with letters from $a$ to $h$ in Fig.\,\ref{fig:sph} (see red box).

To separate if this activity is caused by tides or is simply a young, rapidly spinning star, we present these primaries in the context of their orbital parameters in  the middle panel of Fig.\,\ref{fig:surfaceRotations}. If tides are efficient and lead to a tidal spin up this leads to more efficient excitation (and dissipation) of tidal inertial modes and thus a more efficient synchronisation and circularisation. Six of the eight active stars are indeed found in short periodic and circularized systems ($e$\,$\simeq$\,0, $P_\mathrm{orb}$\,$\lesssim$11\,days). For these stars, the reason for the enhanced activity is very likely rooted in the tidal interaction. 
Two stars from this active group are found in wide orbits.   
If the orbital period listed in the TBO catalog of $\sim$1\,000\,days for star $c$ is correct, even at such high eccentricity ($e$\,$\simeq$\,0.55) for such wide orbits, tides will not produce a lasting effect on the rotation. Similar can be assumed for star $f$ with a period of $\sim$100\,days, and an eccentricity of 0.3.
Because no significant tidal interaction is present in these systems, their high activity is likely an effect of young and rapidly rotating stars \citep{Skumanich1972}. All other stars showing solar-like activity values are at orbital periods longer than 11 days.

For the red giants, the picture is a different one. For the new systems, the orbital periods range from 30 to 1000 days, while their primaries rotate with periods between 50 to about 200 days. These rotation periods are typical for giants on the less-luminous part of the RGB. Typically, giants have low spot-filling factors. Only a few systems show the signature of spots in their light curves \citep{Ceillier2017}. The rapid rotators among them could be the product of stellar mergers \citep[e.g.,][]{Tayar2015, Patton2023}.
In the bottom panel of Fig.\,\ref{fig:surfaceRotations}, we show the ratio of the orbital period to the surface rotation period as a function of the orbital eccentricity. This form shows that hardly any of the systems are synchronized. 
Only a few systems with shorter periods and stronger equilibrium tides are nearly pseudo-synchronized. Also, their measured spot signature could originate from internal processes that trigger the dynamo \citep{Charbonnel2017}.

\section{Orbital inclinations and eclipsing binaries \label{sec:EclipsingInclinations}}

The mass of a star is the single most fundamental parameter for understanding its structure and evolution. Depending on the type of star, different techniques can be applied to determine the stellar mass \citep[see][for a review]{Serenelli2021}. 
While many mass-determination techniques report high precision, the question of how accurate is the mass can only be answered from the cross-calibration of independent techniques.

The most basic technique is the derivation of the dynamical masses of the stellar components in a binary system \citep[for details see][and references therein]{Prsa2018}. The precision of this technique relies on well-determined orbital parameters, radial-velocity amplitudes for both components (SB2), and the orbital inclination. The method's bottleneck is the knowledge of the inclination, which is traditionally determined through modeling eclipsing binary systems.
Only for 17 eclipsing systems hosting an oscillating component dynamical masses are available from the literature \citep{Gaulme2016, Benbakoura2021}. From such an analysis, \cite{Gaulme2016} suggested that the seismic scaling relations overestimate mass and radius by 15\% and 5\%, respectively. New systems with known orbital inclinations are needed to solve this dichotomy between seismic and dynamical masses. 
To increase the sample size, we tested whether \GaiaDR contains oscillators in yet unknown eclipsing systems or systems with determined orbital inclinations.

\begin{figure}[t!]
    \centering
    
    \includegraphics[width=\columnwidth]{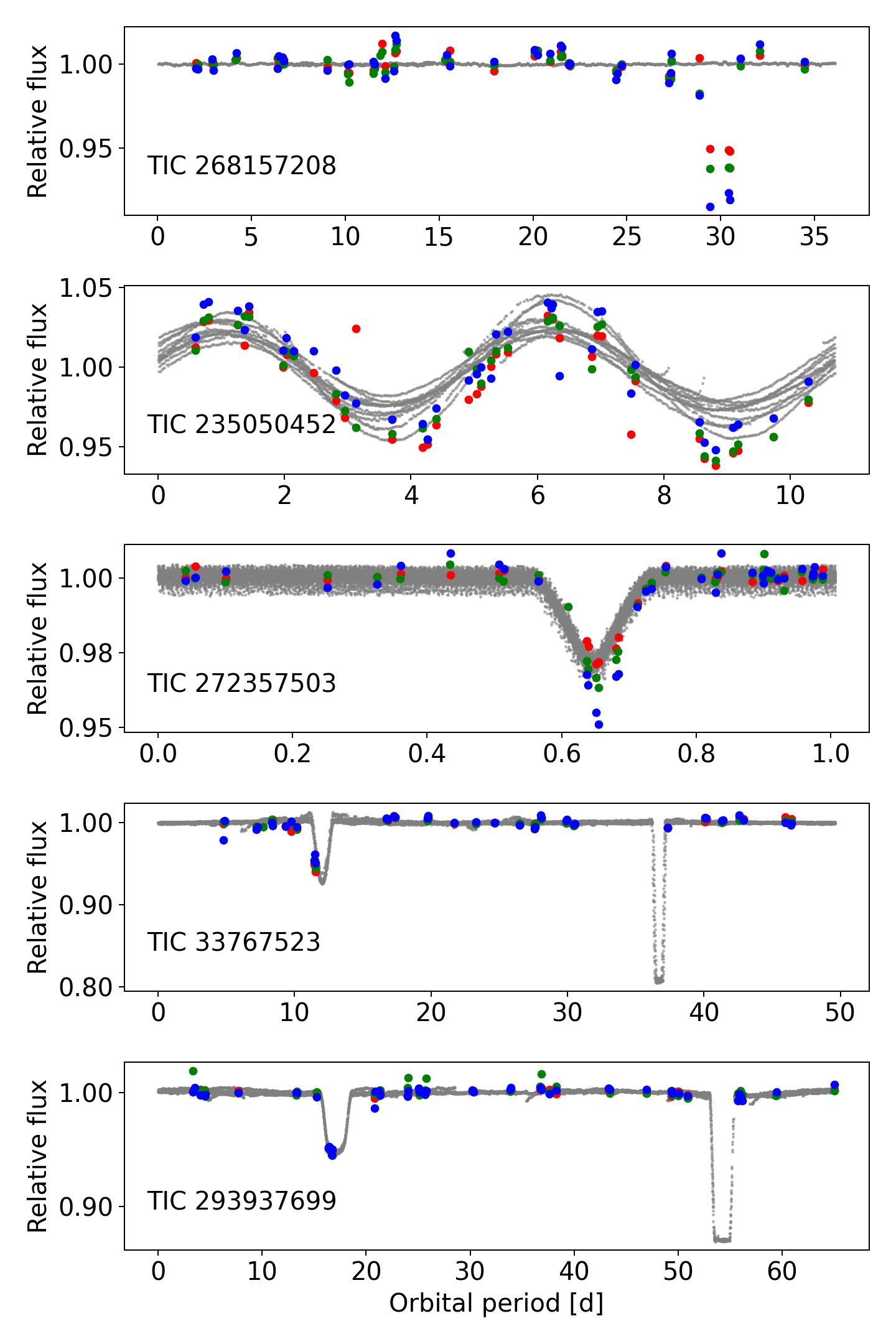}
    
    \caption{Phase-folded TESS light curves of the \GaiaDR eclipsing binary system candidates, TIC\,268157208 (top panel), TIC\,235050452 (second panel), TIC\,272357503 (third), TIC\,33767523 (fourth), and TIC\,293937699 (bottom). Relative \Gaia epoch photometry in the 
    \gmag, \gbp, and \grp 
    passbands are shown in green, blue, and red, respectively.
    For TIC\,268157208, TIC\,235050452, TIC\,272357503 the period from the \GaiaDR eclipsing binary catalog was used to fold the light curve. For TIC\,3376752 the orbital period from the SB1 solution, provided in the TBO catalog is used.
    For TIC\,293937699 we used our orbital period, determined from the TESS data. 
    \label{fig:eclipsingBinaries}
    }
\end{figure}

\begin{table*}[t!]
\centering
\caption{Candidates from the eclipsing binary \GaiaDR catalog with at least one oscillating component.}
\begin{tabular}{lccclll}
\hline\hline
TIC & Frequency (geom.) & $\nu_{max} $  & $\Delta\nu$  & Reference for  & Data &Comment \\ 
& [$d^{-1}$] & [$\mu Hz$] & [$\mu Hz$] &  seismic values  \\
\hline
235050452   & 0.09336 $\pm$ 0.00002 & 98\,$\pm$\,10   & 9.2\,$\pm$\,0.1   & \cite{Mackereth2021} & 6S&  Surface or synchronized rotation?\\
272357503     & 0.99314 $\pm$ 0.00001 & 51\,$\pm$\,13 & 5.7\,$\pm$\,0.2   & \cite{Mackereth2021} &23S& \\
293937699       & 0.26058 $\pm$ 0.00006 & 47\,$\pm$\,4    & 6\,$\pm$\,20  & \cite{Mackereth2021}  &16S& Eclips.Bin., actual period:  65.112\,d\\
33767523 & 0.04607\,$\pm$\,0.00003 & $-$ & $-$ & $-$ & 18S & $P_\mathrm{orb}$\,=\,49.63\,$\pm$0.02, $e$\,=\,0.33\,$\pm$0.01\\

268157208      & 0.02770 $\pm$ 0.00008 & 47.49     & 4.804   & \cite{Yu2018} & 1Q, 3S &  =\,KIC\,8646982\\
\hline
\end{tabular}
\tablefoot{The first column reports the identifier of the eclipsing binary candidate in the \textit{TESS Input Catalogue} (TIC). The column 'Frequency' specifies the frequency of geometric model of the eclipsing binary light curve. The third and fourth column report the mean seismic parameters $\nu_{max}$ and $\Delta\nu$ from \cite{Mackereth2021}, respectively. Those seismic parameters were extracted from the references shown on the next columns. the column 'Data' reports how many Sectors (S) of TESS, or Quarters (Q) of data was available for these systems at the time of the analysis. Comments on the system are provided in the last column: the priode reported for TIC\,293937699 was redetermined from TESS data; the orbital parameters for TIC\,33767523 are reported in the TBO catalog  by \cite{Arenou2022}. }
\label{tab:EclipsingBinaries}
\end{table*}

\subsection{\Gaia epoch and TESS time series photometry \label{sec:Eclipsing}}

The probability, $\theta_\mathrm{Ecl}$, of a randomly orientated binary system to show eclipses is a function of the sum of the radii of the components ($R_1$, and $R_2$) and the average distance $a$ (assuming $e$=0) between \citep{Deeg2018}. Using Kepler's third law, we can express this as a function of the orbital period and the sum of the mass of the components ($M_1$, and $M_2$):\begin{eqnarray}    
    \theta_\mathrm{Ecl} = \frac{R_1+R_2}{a} = \frac{R_1+R_2}{\sqrt[3]{\left(M_1+M_2\right)P^2_\mathrm{orb}}}\,.
\end{eqnarray}
The maximum radius of a solar-like star is less than 1AU. Due to the high dependence on the mass difference between both components, it is unlikely, that both targets reach the maximum expansion simultaneously. Typical binary systems are found with values of $a$ up to several thousand AU (corresponding to orbital periods of   tens of thousands of years), which provides very low probabilites for detecting eclipsing binary systems.

Finding eclipsing binaries becomes increasingly challenging with longer orbital periods, as the projected surface that is eclipsed is becoming smaller and requires a nearly perfect alignment at 90\,degrees. 
On the contrary, observing a binary system through RV monitoring, a few well-spread spectroscopic measurements are sufficient to confirm the object's binary nature and even fit the orbital parameters. However, an eclipsing binary can only be detected during the eclipsing phases through direct observations of the eclipses in the photometry or the Rossiter-McClaughlin effect on the radial velocities. The successful photometric search for yet unknown eclipsing binary systems requires continuous monitoring. Consequently, almost all systems hosting an oscillating red giant detected in the \Kepler data have periods shorter than the mission duration of four years. 

As discussed by \cite{Beck2022SB9}, such photometric detection strategy introduces a bias where eclipsing systems identified by satellites tend to have short periods. Such relatively short-period binary systems are not wide enough for luminous red giants at the tip of the red giant branch to remain in a detached configuration. Therefore, the previous literature sample is strongly biased toward the assumption of young RGB stars as the primary component, thus hardly containing RC stars.

The quasi-random single-epoch observing strategy of \Gaia, originating from the scanning law, is not optimal for detecting long-periodic eclipsing binaries and explains why hardly any such eclipsing binaries have been found by the mission \citep[see Fig.\,4 in][]{Arenou2022}. See Appendix~A of \cite{2017arXiv170203295E} for more details on the time sampling of \Gaia. During a field-of-view transit of a star three photometric measurements are obtained within less than a minute: a broadband visual \gmag, and two narrower blue \gbp and red \grp pass bands \citep{2018A&A...616A...3R,2021A&A...649A...3R}. The latter two are derived from integrating low-resolution photo-spectroscopic measurements of the BP and RP instrument, respectively \citep[for more details see Sect.~3.3.6 of][]{GaiaMissionMainReference2016}. This data is quasi-randomly sampled over the mission duration. 
\cite{Mowlavi2022} provided a catalog (\texttt{vari\_eclipsing\_binary}) of 
2.2 million eclipsing binary candidates in \GaiaDR, of which a subset of 86\,918 stars were fitted for astrophysical parameters and published in \texttt{gaiadr3.nss\_two\_body\_orbit} with \texttt{nss\_solution\_type} set to `EclipsingBinary' (EB). The search of the catalog of eclipsing binaries for the full seismic sample returned five candidates, which are presented in Table\,\ref{tab:EclipsingBinaries} and Fig.\,\ref{fig:eclipsingBinaries}. 

To validate these candidates, we extracted light curves from space photometry. All four targets were observed in multiple sectors by the TESS Mission. At the time of the analysis, data up to Sector 53 was available. The data were extracted from the full-frame images (FFI) using mostly the point-spread function fitting module of the \textsc{Eleanor} package \citep{Feinstein2019}. These observations provide a cadence of 30 and 10 minutes, depending on which sector they were taken from. The \Gaia multi-color epoch photometry \citep{2021A&A...649A...3R} and monochromatic space photometry from and TESS mission for these four candidates is shown in Fig.\,\ref{fig:eclipsingBinaries}. 
For the target TIC\,268157208 (=\,KIC\,8646982), a sub-quarter of \Kepler data exists, which is $\sim$1.5 times the length of the proposed orbit. Due to the crowded field in which the target is located, we prefer the \Kepler data due to its smaller pixel plate scale. The \Kepler light curve was taken from the \texttt{KEPSEISMIC} database on the MAST archive\footnote{\href{https://archive.stsci.edu/prepds/kepseismic/}{https://archive.stsci.edu/prepds/kepseismic/}} \cite[for details see][]{garcia2011,Garcia2014}.

TIC\,268157208 (=\,Gaia\,DR3\,2079109044266147328) is reported in \GaiaDR as an eclipsing binary system with a period of 36.1\,d. The clear eclipse in the \Gaia epoch photometry of about $\sim$10\% is not found in \Kepler and TESS data and is clearly an artifact.  
For TIC\,235050452 (=\,Gaia\,DR3\,4797117284359411712), the \Gaia and TESS light curves show a good agreement in phase and amplitude of the long periodic variations with a period of 10.7 days. The sinusoidal flux modulation with variable amplitude in TESS data indicates rotational-modulated spots.
For TIC\,272357503 (=\,Gaia\,DR3\,5214824569250240128), the \Gaia epoch photometry suggests a binary with a period of nearly 1 day and eclipses of about $\sim$2 to $\sim$4\% in the red and the blue passband, respectively. The shape of the feature indicates a partial eclipse. The analysis of the TESS photometry does not exclude that the primary and secondary eclipses are very similar, and the system's period is $\sim$2\,days. Such a system is too small to host a red giant. To determine more information about the components, radial velocities for this system are required.

From \Gaia epoch photometry, the system TIC\,293937699 (=\,Gaia\,DR3\,5490280956749756928) shows a clear drop of the flux of about 5\%. From TESS data, we can confirm the presence of the 5\% eclipse, which is actually the secondary eclipse. In contrast, the well-pronounced primary eclipse with an eclipse depth of $\sim$15\% was missed by chance by the sparse sampling of the epoche photometry. 
However, the period of 65.112\,$\pm$\,0.05\,days, determined from a period analysis of the TESS data, does not agree with the geometrical period of $\sim$3.8\,d, reported in the \GaiaDR eclipsing star catalog. This system is similar to the cases with large residuals comparing the literature values for period and eccentricity from the SB9 catalog. TIC\,293937699 is one of the few systems hosting an oscillating red giant in an eclipsing binary system. Because radial velocities have yet to be obtained for this system, we have postponed a deeper analysis of this system for a later paper.

Similarly, TIC\,33767523 (=\,Gaia\,DR3\,4627969652492312320) is an eclipsing binary for which also the TBO catalog presents an orbital solution of $P_\mathrm{orb}$\,=\,49.63\,$\pm$0.02, and $e$\,=\,0.33\,$\pm$0.01. Typically, such systems are wide enough to allow for an RGB star to oscillate. However, no oscillations are detected. This is probably due to the faintness of the (V\,$\simeq$\,11\,mag) target. 

\begin{figure}[t!]
    \centering
    \includegraphics[width=\columnwidth]{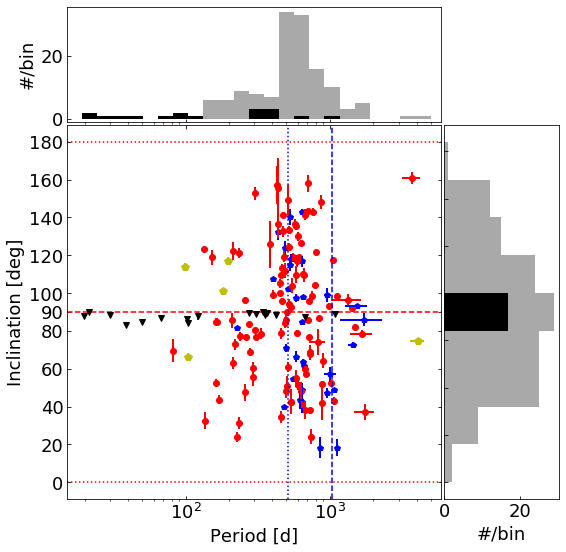}
    \caption{Distributions of the inclinations and orbital period of astrometric binaries hosting a solar-like oscillating primary. The red dots indicate oscillating giants, observed by TESS; blue and yellow symbols mark oscillating giants and main-sequenc stars observed by \Kepler. Black triangles indicate giants in eclipsing binary systems from the \Kepler literature sample The vertical dashed and dotted blue lines mark the full and half length of the \GaiaDR timebase, respectively. The horizontal dashed line marks the inclination for the edge-on orientation, while the dotted lines indicate the plane-on orientation of an orbit. The grey and black histogram indicate the distribution of the sample of astrometric binaries from \GaiaDR and the literature sample of eclipsing binaries from \Kepler. }
    \label{fig:inclinations}
\end{figure}

\subsection{Inclinations from astrometric solutions \label{sec:inclinations}}
For systems that do not have edge-on orientations ($i$\,$\simeq$90$^\circ$), the orbital inclination cannot be determined from the light curve. The precise and time-resolved astrometry of the \gaia mission now allows for the determination of the orbital inclination and provides photometric or spectroscopic constraints on the primary and secondary mass.

\begin{figure*}[t!]
\centering
    \includegraphics[width=\textwidth]{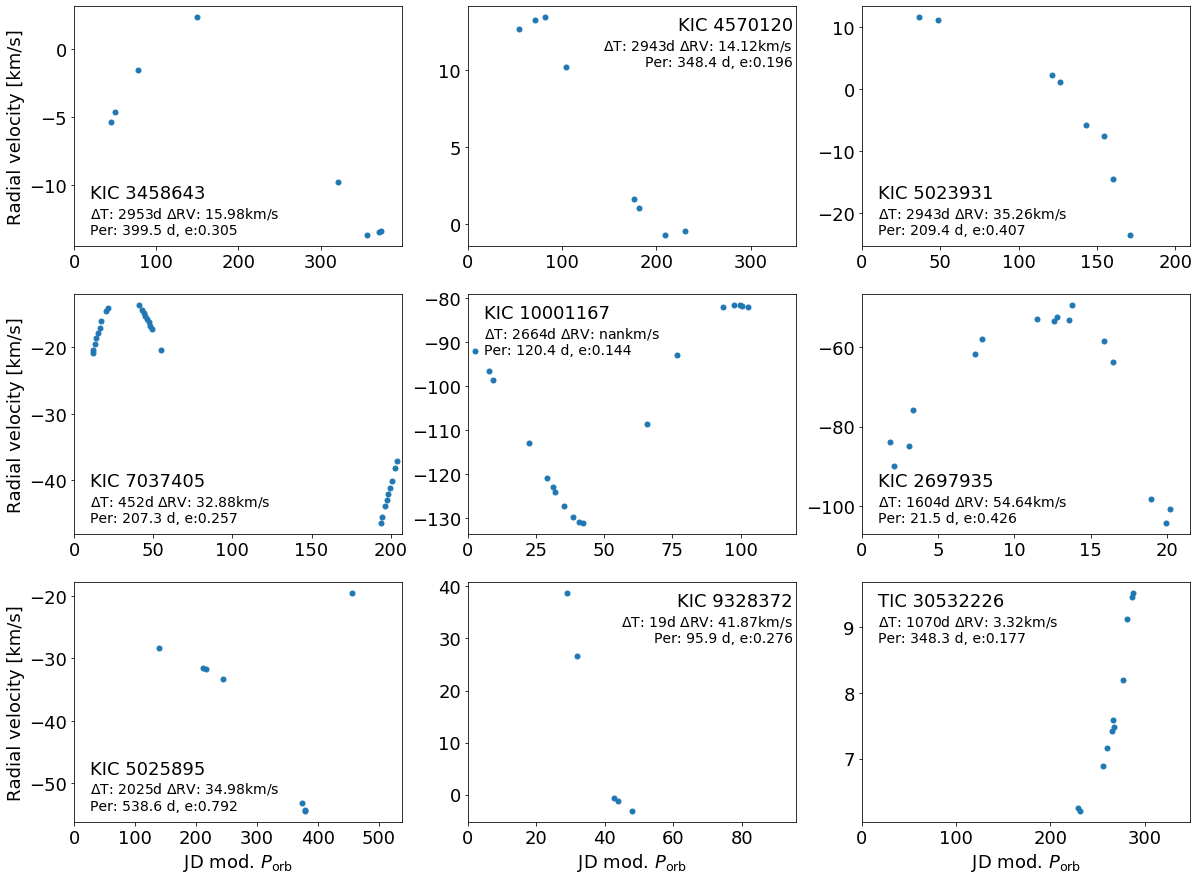}
    
\caption{Example radial-velocity curves from APOGEE spectroscopy of \GaiaDR binary candidates. The panels show the RV values phasefolded with the period from \GaiaDR, which is also given in the annotated text.    \label{fig:ApogeeRvCurves}}
%\vspace{5mm}
\end{figure*}

{For 146 systems hosting an oscillating component, astrometric solutions were found to provide an orbital inclination.
The solutions for astrometric orbits of \cite{Halbwachs2022,Holl2022} 
and listed in \GaiaDR provide the Thiele-Innes coefficients describing the orbital solutions and implicitly contain the inclination.} The conversion from these orbital elements to the elements in the Campbell formalism, which explicitly contain the inclination was performed with a python tool\footnote{We used the standard conversion formalism \citep[e.g.,][]{Halbwachs2022}. The NSS software tools have been developed by N.\,Leclerc and C.\,Babusiaux and are available at \href{https://www.cosmos.esa.int/web/gaia/dr3-nss-tools}{https://www.cosmos.esa.int/web/gaia/dr3-nss-tools} \label{fn:inclinationTool}} provided by \cite{Arenou2022}. 

As shown in Fig.\,\ref{fig:inclinations}, inclinations are found for 9 main-sequence targets, as well as 31, and 106 giants from the \Kepler and TESS samples, respectively. We note that  inclinations range from 0 to 180 (both plane-on orientations). This notation allows us to distinguish among prograde and retrograde movements with respect to the line of sight. Similarly to the dynamical masses, these systems have periods starting at about 100\,days and range beyond 1\,000\,days. While the analysis of the full sample by \cite{Arenou2022} shows a maximum in distribution of the inclination at 0 and 180\,degrees, we find the maximum for stars in our sample around 90\,degrees (Fig.\,\ref{fig:inclinations}) compared to literature.
If an inclination for a system is given in \GaiaDR, we list it in Table\,\ref{tab:A1}. 
Of the systems with an inclination from \GaiaDR, only KIC\,7103951 has been previously reported in the literature \citep{Gaulme2020}, but without measured radial velocities. 

Once these systems have accurate radial velocities from ground-based follow-up, these inclinations will be valuable information to extend the sample of calibrators for the scaling relations. For comparison, the \Kepler literature sample of red giant eclipsing binary systems, is shown in Fig.\,\ref{fig:inclinations}. 
The wide range of orbital periods and the sheer number of targets suggest that the astrometric sample contains a sufficient number of giants in the more advanced RC or 2RC status, which will help to break the evolutionary bias for the calibration of the seismic scaling relations.

\subsection{Searching for eclipses in complimentary data}

From the inclinations reported in \GaiaDR, we identified for binaries with a quasi edge-on orientation to search for additional eclipsing systems that originally have been missed from \Gaia epoch photometry. From the systems with known inclinations that are hosting a solar-like oscillator, we found 6 that fall into a range of 90$\pm$3 degrees, which are
KIC\,10732098, TIC\,379953111, TIC\,38843858, TIC\,308539721, TIC\,142053145, and TIC\,237973654. These systems have periods between $\sim$200 and $\sim$1420 days.
To search for eclipses in these targets, we extracted light curves from TESS FFIs. Given that in most cases the orbital period exceeds the timebase of TESS data, it is not surprising that no eclipses are found. 

Next, we searched the
\textit{All-Sky Automated Survey for Supernovae} \citep[ASAS-SN,][and references therein]{ASASSN2017}\footnote{\href{https://www.astronomy.ohio-state.edu/asassn/}{https://www.astronomy.ohio-state.edu/asassn/}}. ASAS-SN began surveying the entire sky in V-band in 2014 with 2-3\,day cadence and swapped to nightly monitoring in the g-band in 2018. The existing timebase therefore exceeds multiple orbits for all of the candidate systems. However, most of the six targets are brighter than the ASAS-SN saturation limit at Johnson V around 10\,$-$\,11\,mag.
We started by cross-matching with the ASAS-SN V- and g-band variables catalogs \citep{Jayasinghe2021, Christy2022, Rowan2022}. While identifying several matches, the variability is consistent with rotational variability instead of eclipses. 
We then extracted the light curves of the sample mentioned above. Neither the light curve nor the phase curve, produced from the orbital period reported by \GaiaDR, revealed any signature of the eclipses. The range of three degrees around the edge-on configuration might be too wide, given the decreasing angular size of the binary components to allow for eclipses.

We conclude that these targets are most likely non-eclipsing. Given the wide orbits, the range of three degrees around the edge-on configuration might be too wide (given the decreasing angular size of the binary components) to allow for eclipses.

\section{Confirmation through radial-velocity monitoring
\label{sec:RadialVelocities}}

\begin{table*}[t!]
\centering \small
\caption{Reported orbital parameters for confirmed and candidate symbiotic binary stars in \GaiaDR. }
\tabcolsep=4pt

\begin{tabular}{lr|rrrr|rrr}
\hline\hline
\multicolumn{1}{c}{Star} &
\multicolumn{1}{c}{2MASS} &
\multicolumn{1}{c}{\GaiaDR} &
\multicolumn{1}{c}{$e$} &
\multicolumn{1}{c}{$P$} &
\multicolumn{1}{c}{Type} &
\multicolumn{1}{c}{$e_{\rm Lit}$} &
\multicolumn{1}{c}{$P_{\rm Lit}$} &
\multicolumn{1}{c}{Ref}\\
\multicolumn{1}{c}{identifier} &
\multicolumn{1}{c}{~} &
\multicolumn{1}{c}{~} &
\multicolumn{1}{c}{~} &
\multicolumn{1}{c}{[d]} &
\multicolumn{1}{c}{~} &
\multicolumn{1}{c}{~} &
\multicolumn{1}{c}{[d]} \\[0.5em]
\hline
StHA 32                      &       J04374563-0119118 &  3229441606998725888 &  0.12\,$\pm$0.03 & 618.47\,$\pm$10.60  & SB1 & $-$ & 612 & [1]\\
IV Vir                       &       J14163429-2145500 &  6276714894852124032 &  0.03\,$\pm$0.02 & 279.98\,$\pm$1.13 & SB1 & 0 & 281.6\,$\pm$1.2 & [2]\\

AG Dra                       &       J16014101+6648101 &  1642955252784454144 &  0.28\,$\pm$0.03 &   502.77\,$\pm$6.01 & SB1 & 0 & 548.65\,$\pm$0.97 & [3]\\
Hen 3-1213                   &       J16351508-5142274 &  5934206543151802752 &  0.17\,$\pm$0.06 &   530.11\,$\pm$3.76 & SB1 & 0.183\,$\pm$0.034 & 533\,$\pm$2 & [4]\\
YY Her                       &       J18143419+2059213 &  4528063078197198848 &  0.15\,$\pm$0.03 &   607.73\,$\pm$8.37 & SB1 & $-$ & 589.5\,$\pm$0.3$^a$ & [5]\\
StHA 176                     &       J20224225-2107546 &  6859282948915521664 &  0.07\,$\pm$0.07 &   246.64\,$\pm$1.43 & SB1 & $-$&$-$ \\
LT Del                       &       J20355722+2011275 &  1817300516637652352 &  0.40\,$\pm$0.10 &   462.81\,$\pm$6.88 & SB1 & $-$ & 465.6 & [6]\\[0.5em]\hline
%\medskip
TYC 1371-69-1                &       J07573112+2017347 &   670455944074475008 &  0.01\,$\pm$0.01 &   119.13\,$\pm$0.06 & SB1 & 0.024\,$\pm$0.015 & 119.18\,$\pm$0.07&[7]\\
GaSS 1-4                     &       J11121548-3207193 &  5403474822973970816 &  0.06\,$\pm$0.04 &   458.92\,$\pm$4.25 & SB1 & $-$ & 235$^a$&[8]\\
SkySyC 1-3                   &       J15265734-7003104 &  5796098502440628864 &  0.08\,$\pm$0.02 &   701.13\,$\pm$3.81 & SB1 & $-$ & 482.78$^{a,b}$ & [9]\\
IGR J15293-5609              &       J15292939-5612133 &  5883707000513657216 &  0.09\,$\pm$0.04 &    31.50\,$\pm$0.03 & SB1 & $-$&$-$\\
GaSS 1-20                    &       J16005485-1628325 &  6250366095129668992 &  0.09\,$\pm$0.14 &    10.51\,$\pm$0.00 & SB1 & $-$&$-$\\
SS 295                       &       J17073816-0744485 &  4360702354583742080 &  0.17\,$\pm$0.10 &  471.11\,$\pm$13.94 & SB1 & $-$&471.00$^{a,b}$ & [10]\\
Gaia DR3 217... &        J21180196+5721343 &  2178988199495779456 &  0.28\,$\pm$0.10 &  753.92\,$\pm$26.14 & SB1 & $-$&$-$\\
Gaia DR3 533... &        J11240425-6013342 &  5339026227414066432 &  0.07\,$\pm$0.12 &    17.15\,$\pm$0.02 & SB1 & $-$&$-$\\[0.5em] \hline
%LHA 115-S 50 & J01063239-7217195 & 4687489481109769216 & / & / & / & / & / \\
CGCS 5926 & J23454464+6252511 & 2016034975622911360 & $-$ & $-$ & $-$ & $-$ & $-$ & $-$ \\
Gaia DR3 553... & J08070625-4308520 & 5533253788183484672 & $-$ & $-$ & $-$ &$-$ & $-$ & $-$ \\
\hline
\end{tabular}
\label{tab:symbiotics}
\tablefoot{Confirmed symbiotic systems are presented in the top panel of the table. Candidate systems for symbiotic stars with orbital solution and non-linear or acceleration solutions are reported in the middle and bottom panel, respectively. The first three columns report the commonly used identifier as well as the identifiers in the 2MASS and \GaiaDR catalog. The next three columns report the orbital parameter and type of solution, presented in the TBO catalog \citep{Arenou2022}. The last three columns report the most recent values in the literature and their reference. 
$^{a}$ Period obtained from photometric variability. $^{b}$ Orbital origin of the photometric variability with this period was suggested by \cite{Merc2019}. References in the Table: 1: \cite{Murset1999}, 2: \cite{Smith1997}, 3: \cite{Fekel2000}, 4: \cite{Fekel2015}, 5: \cite{Mikolajewska2002}, 6: \cite{Munari2002}, 7: \cite{Tang2012}, 8: \cite{Munari2021}, 9: \cite{Pojmanski2002}, 10: \cite{Jayasinghe2020}.}
\end{table*}

APOGEE \citep{Majewski2017} is an all-sky survey, consisting of two nearly identical multi-object fiber-fed spectrographs mounted on the northern 2.5\,m Sloan Foundation Telescope at Apache Point Observatory  and the southern 2.5\,m Irénée du Pont Telescope of Las Campanas Observatory to perform near-infrared spectroscopy in the H-Band with a resolution of R$\sim$22\,500. It typically visits a source multiple times in the course of the project. Several papers have utilized the millions of single, homogeneous spectra to successfully search for large quantities of binaries in the red giant phase from RV variations \citep[e.g.,][]{Badenes2018, GaulmeGuzik2019, Daher2022}

To test the binary-candidate detection from \GaiaDR, we searched for significant radial-velocity variations in the spectra contained in APOGEE DR17 \citep{ApogeeDR16}. 
Hereafter, we adopt the simple significance criterion of \cite{Patton2023}, which flags a source as potential binary if the scatter around the average radial velocity (\texttt{VSCATTER}) is greater than three times the average uncertainty of its RV measurement (\texttt{VERR\_MED}) for a target with at least two visits.

From the 382 giant oscillators in the \Kepler sample with orbit solutions in the TBO, 181 were visited multiple times, leading to 149  binary detections. DR17 is the first APOGEE data release to include a substantial set of observations of TESS targets in the Southern Continuous Viewing Zone. However, most targets were only visited once yet. Therefore, we only could test 7 binary candidates for RV variations, of which 5 exceed the significance limit for RV; for the 45 dwarfs and subgiants observed by the \Kepler mission and reported orbital parameters from \GaiaDR , 27 sources had at least two spectra, of which 25 showed significant RV variations. 

All systems with at least two spectroscopic observations by the APOGEE project are listed in Table\,\ref{tab:ApogeeBinaryDetection}. In the last column, we indicate if a candidate system exceeded our significance threshold for a binary candidate. Because of the limited number of spectra typically only the binary nature can be confirmed. For systems that are particularly rich in RVs, we folded the APOGEE RVs with the period reported in \GaiaDR. 
As illustrated by the selected systems in Fig.\,\ref{fig:ApogeeRvCurves} for which a good agreement was found.

A non-detection of significant RV variation from APOGEE data does not prove a proposed binary candidate from \GaiaDR wrong. In many cases the multiple visits of a source in APOGEE occur in close temporal proximity, within a few nights. Therefore, the timebase for the spectroscopic observations can be short, compared to orbital periods of several hundred to thousands of days, leading to insignificant results from the RV variations. In particular, the RV variation over small ranges in phase can hardly show any variation for eccentric systems.

An additional source of RVs is provided by 
\cite{Beck2017Lithium}, who reported six binary systems with an oscillating solar-analog primary from RV monitoring with the \textsc{Hermes} spectrograph \citep{Raskin2011}, mounted on the 1.2\,m \textit{Mercator} telescope on La Palma.  
The two systems KIC\,4914923, and KIC\,9098294 were reported to have a peak to peak RV amplitude of 2.11, and 41.35\,km/s. KIC\,4914923 is also confirmed as binary from APOGEE spectroscopy. 
The star KIC\,3241581 is found in the non-linear solution. Indeed, the RV time series of \cite{Beck2016} points to an orbital period of $\sim$1500\,d.

\section{Extended science cases for this dataset \label{sec:extendedScienceCase}}

This data set can also be used to test  science cases, related to red giant stars in binary systems. 

\subsection{Symbiotic binaries}

Symbiotic stars are interacting binary systems composed of a cool red giant star and a hot white dwarf or, in some cases, even a neutron star. Such systems move on orbits typically between hundreds and a few thousands of days embedded in an environment of circumstellar gas. The spectra of these systems are often showing strong emission lines due to the photoionization of the nebula by the radiation of the hot component \citep[and references therein]{Munari2019}. 
Because these systems are highly photometrically variable \citep{Merc2023}, it is interesting to test how known or suspected symbiotic binaries perform in \GaiaDR. We conducted a crossmatch, as described in Sect.\,\ref{sec:Search}, between the \GaiaDR database and the catalog of symbiotic binaries, published by \cite{Merc2019,Merc2019b}. 

\begin{figure}[t!]
    \centering
    \vspace{-1mm} 
    \includegraphics[width=\columnwidth]{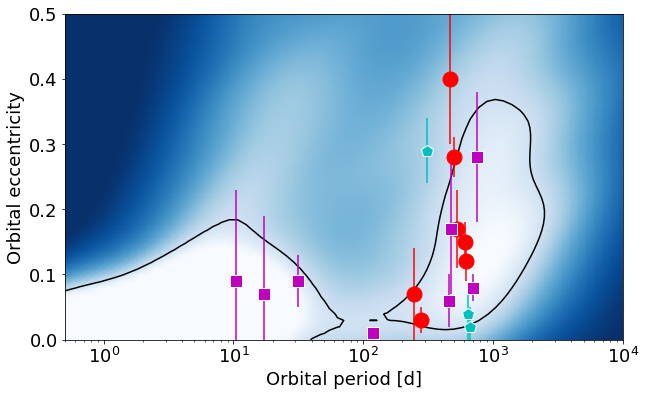}
    \caption{Orbital parameters of systems hosting symbiotic stars and red giants that exhibit anomalous peaks in the PSD (cyan pentagrams). For the symbiotic systems, red dots and magenta squares depict the confirmed and candidate symbiotic stars, respectively. The background density map depicts the distribution and mark the most common combinations of orbital solutions in the SB9. }
    \label{fig:symbiotics}
    %\vspace{5mm}
\end{figure}

Out of 141 confirmed galactic symbiotic systems within the range of the magnitude-limited sample (4\,$\leq$\,G\,[mag]\,$\leq$\,13), seven targets ($\sim$5\%) were found in the \GaiaDR with orbital solutions (Table\,\ref{tab:symbiotics} and Fig.\,\ref{fig:symbiotics}). 
For the six systems for which literature values for the orbital parameters exist, a good agreement is found. 
Only for AG\,Dra, a K-type giant and a hot white dwarf on a well-established circular orbit and an orbital period of $\sim$550 days \citep{Fekel2000}, \GaiaDR underestimates the orbital period by $\sim$9\%. 
During the time covered by \GaiaDR, AG\,Dra was in its active stage, showing at least two outbursts at the time when the data for \GaiaDR were collected \citep[see][]{Merc2019c,Galis2019}. 

Although the majority of symbiotic systems are found on orbits of 200 to 500 days \citep{Merc2019,Merc2019b}, which are suited for \Gaia, the detection rate for this class of interacting-binary systems is nearly an order of magnitude lower than for non-interacting, giant-hosting  systems (Sect.\,\ref{sec:Search}). Because of the large variations that influence the photocenter as well as the stellar spectrum, these targets are difficult to \hbox{determine in an automated way}. 

In addition to the confirmed symbiotic stars, the crossmatch revealed that another eight out of 744 galactic symbiotic candidates have orbital solutions listed in \GaiaDR. Half of the eight candidates have periods reported in the literature. For those systems for which orbital periods are known, we again find good consistency between the literature values and the periods reported by \GaiaDR. For one system the period of the spectroscopic solution in the TBO catalog is twice the photometric period, reported in the literature. Such a difference between the photometric and spectroscopic orbital period is often seen in the case when the giant fills or nearly fills its Roche radius and is ellipsoidally distorted. As a consequence, two minima per orbital period are observable in the light curve and the period search might return half the value of the true period. 
Three sources in the sample of candidates have rather short
Gaia orbital periods (10 to 32 days) that are substantially shorter
than the minimum periods of $\sim$200 d found for symbiotic stars
\citep{Merc2019,Merc2019b}. If they are true orbital periods of these
systems, this would rule out the symbiotic classification of these
targets.

Among the candidates from the New Online Database of Symbiotic Variables, there are also 337 targets newly identified as possible symbiotic stars from a supervised machine-learning classification by \cite{Rimoldini2022} of the color and variability in \GaiaDR of 12.4 million sources. Only three of the \Gaia symbiotic candidates are reported in \GaiaDR as binary candidates, whereby one of them has a period of $\sim$17\,d. Also these numbers show a very low detection rate and that it is very challenging to identify symbiotic binaries from the existing observational data in \GaiaDR. 

Comparing these 15 systems with the distribution of the orbital period shows that they mostly follow the distribution of other red giant stars. As for the other red giant binaries (see Fig.\,\ref{fig:ePplane}), the systems are found with periods less than 1\,000\,days. 
Searching the non-linear and acceleration solutions as well, we find two additional systems, listed in the bottom panel of Table\,\ref{tab:symbiotics}.

\subsection{Testing giants with anomalous peaks in the PSD}

\cite{Colman2017} published a collection of 168 oscillating red giant stars, in which the power-spectral density (PSD) reveals anomalous peaks. These peaks occur with frequencies very different and outside of the classical power excess. Furthermore, the shape of these peaks in the PSD do not resemble a Lorentz profile but seem to resemble a delta function. This suggests that these peculiar peaks are not stochastically excited but correspond to a periodic variation. 

For about half of the cases, contamination with background stars was found as the most likely explanation. However, in 81 cases the source of the peculiar frequencies appears to coincide with the giant star. The authors suggested that such frequencies could be produced by the presence of close stellar components within the convective envelope of the red giant or due to a close binary in a hierarchical triple system.   

\begin{table}[t!]
\centering
\caption{Binaries from \Gaia DR3 in the sample of stars with anomalous peaks, reported by \cite{Colman2017}. 
\label{tab:Colman}}
\tabcolsep=10pt

\begin{tabular}{r|rrr}
\hline\hline
\multicolumn{1}{c}{KIC} & 
\multicolumn{1}{c}{$P_{\rm peak}$} &
\multicolumn{1}{c}{$P_{\rm orb}$} &
\multicolumn{1}{c}{$e$} \\
\multicolumn{1}{c}{~} & 
\multicolumn{1}{c}{[d]} &
\multicolumn{1}{c}{[d]} &
\multicolumn{1}{c}{} \\[0.5em]
\hline
\multicolumn{4}{c}{Possible physical associations}\\[0.5em]
\hline
2449020     & 0.83  & 310\,$\pm$2      & 0.29\,$\pm$0.05 \\
10936814    & 4.45  & 665\,$\pm$6      & 0.02\,$\pm$0.03 \\
7596350     & 0.26  & 647\,$\pm$11     & 0.04\,$\pm$0.04 \\
\hline
\multicolumn{4}{c}{presumed chance alignments}\\[0.5em]
\hline
5556726     & 0.48  & 172\,$\pm$1 & 0.24\,$\pm$0.02 \\
12117138    & 4.40  & 685\,$\pm$18 & 0.39\,$\pm$0.03 \\
\hline
\multicolumn{4}{c}{confirmed chance alignments}\\[0.5em]
\hline
2167774     & 0.35  & 137.0\,$\pm$0.2  & 0.21\,$\pm$0.06 \\
1872210     & 0.67  & 540\,$\pm$6 & 0.20\,$\pm$0.07 \\
\hline

\hline
\end{tabular}
\tablefoot{The period and eccentricity of the orbit is taken from \Gaia DR3 TBO. P$_\mathrm{peak}$ is the period of the dominant anomalous peak, reported by \cite{Colman2017}.
The categorisation of the anomalous peak, given by the authors is reported.}
\end{table}

We searched the \Gaia DR3 TBO catalog  for these systems to test if these are actually binary systems, where the found period indeed coincides with the anomalous peak. In total we found seven objects with peculiar peaks to be binary candidates in \Gaia, which are listed in Table\,\ref{tab:Colman} and shown in Fig.\,\ref{fig:symbiotics}. 
In two of these systems the peaks were identified as contamination and in two additional ones contamination could not be excluded. Three system were found to be possible physical associations. For all seven objects, the period of the anomalous peak is below 4.5\,days, while all orbital periods are reported to be between 137 and 685 days on moderately eccentric orbits (0.03\,$\lesssim$\,$e$\,$\lesssim$0.39). These periods are also too long to excite tidal forces. We therefore suggest that these anomalous frequencies are unlikely to be excited by binary interaction.

\section{Discussion and conclusions \label{sec:Discussion}}

In this work, we presented the successful search for solar-like oscillating stars in binary systems, revealed through photometric, spectroscopic, and astrometric solutions in the \GaiaDR catalog of Two-Body-Orbit solutions, and tested it for completeness and purity. 

To test the TBO, we used the SB9 catalog of orbital solutions. We introduced a magnitude-limited sample to account for observational biases due to partial saturation (4\,$\leq$\,G\,[mag]\,$\leq$\,13). Because the sample contains systems with periods of several ten-thousand days, which are too long to be resolved by the \GaiaDR baseline of 1034 days, we limited our sample further in periods (P$_{\rm orb}$\,$\leq$\,1100\,days). We found an overall completeness factor of 28.3\% for the complete SB9 catalog. 

The ruwe measures the astrometric likelihood for a source to be a single star. About 40\% of the detected binaries from the SB9 sample had ruwe values below 1.4, a conservative limit for the astrometric binary detection, and were detected by other means. 
 
Performing the same searches of the TBO catalog for the lists of identified solar-like oscillators from the NASA\,\Kepler, and in the Southern Continuous Viewing Zone of the NASA\,TESS mission, we identified 970 binary system candidates that host solar-like oscillating stars, among which 954 systems are newly detected systems. The sample presented in this work increases the binary stars with oscillating components by an order of magnitude. The full wealth of asteroseismic information allows for a comprehensive study of the system and its oscillating component.

From the search results, we obtained a magnitude-limited completeness factor of about 4\% for the full red giant star sample.
Taking into account the unresolved binaries and the completeness factor, determined from our comparison of the SB9, we arrive at a binary yield similar to the expected value in the literature of 30-40\% \citep{MoeStefano2017, Offner2022}.

We assessed the mass and stellar radii ranges using the asteroseismic scaling relations. Our analysis shows that TESS suffers from noise and the limited length of the photometric time series, leading to an underestimated large-frequency separation. As a result, the masses for stars with radii larger than the typical radius of the red clump (\num\,$\lesssim$\,30\,$\mu$Hz, see Fig.\,\ref{fig:periodSeismolgy}, top panel) can be overestimated, leading to an excess of stars with $M_\star$\,$\gtrsim$2\,$M_\odot$. 

To test whether the orbits reported in the TBO catalog  are physically plausible, we compared the orbital periods and the seismically inferred radii with the radius limit for the Roche-lobe overflow (Fig.\,\ref{fig:periodSeismolgy}, bottom panel). Except for a few reported systems with periods $P_{\rm orb}$\,$\lesssim$\,10\,days where the orbit would be smaller than the radius of the primary, all reported values were found to be physically possible. However, these systems could be actual binaries with a significantly underestimated period. An additional argument for the realistic orbital periods comes from the location of the datapoints in the parameter plane, which are all well separated from the Roche-lobe limit. This gap is expected because systems were selected based on the criterion that the primary is oscillating. As a system approaches the limit for the RLOF, increased strength of tidal interactions starts to suppress the oscillations. 

Because of the robust residuals centered at zero found in the comparison of the TBO catalog  with the SB9, the large fraction of physically reasonable orbital periods, and the approximate agreement of the expected binary fractions for stars of about 1\,$M_\odot$, we consider most of the binary candidates reported in \GaiaDR TBO catalog as reliable new binary systems. 

The large amount of binary systems opens the door to studying binary star interaction and related activity. Using the seismically determined evolutionary stages, we could view the distribution of the orbital eccentricity and period as a function of stellar evolution. We showed that red clump stars have lower eccentricities and are biased towards longer periods than systems hosting the less evolved RGB stars. We attribute the lower eccentricities  as a result of the increased strength of the tidal interaction due to the larger radii at the tip of the RGB. The lack of periods below 500\,d originates from phases of intense star-star interaction, such as the RLOF or common envelope phase. 

For the oscillating dwarfs, we showed the correlation between high photospheric activity and tidal circularization and synchronization. We used the asteroseismic inferences for the oscillating giants to analyze the distribution of the orbital period and eccentricity as a function of the evolutionary state. Indeed, we could show differences that agree with the predictions for the tidally driven evolution of binary as they converge to the equilibrium state of circularized systems.

For 146 systems, the inclinations angles are reported in \GaiaDR. We converted the notation of the value and associated it with oscillating primaries. If the RVs for both components are reported from ground-based observations, these systems will provide additional valuable benchmark systems for calibrating the scaling relations. 
If in those systems rotational splitting of non-radial modes is measured, the inclination of the rotation axis can be measured and test the spin-orbit alignment. A first work in this direction, based on \GaiaDR was presented by \cite{Ball2023}

With an increasing orbital period, the probability of detecting transits in a binary system decreases for geometrical reasons, which explains the small number of eclipsing binary systems found in the vast datasets of space photometry. 
From our search of the \Gaia variability catalog, we found one previously unknown eclipsing binary system hosting an oscillating red giant primary.

Analyzing the radial velocities derived from APOGEE and \textsc{Hermes} spectroscopy, we could independently confirm 149 binaries out of 181 systems, proposed by \GaiaDR and with multiple APOGEE spectra. This low number of the sample viewed in APOGEE originates from the limited sampling in the APOGEE observations and is expected to be improved with forthcoming DRs. 
For most of the systems, which had RV measurements, binarity could be confirmed. Therefore, we see the majority of the binary candidates reported in \GaiaDR as bona fide candidates.

Given the numbers, this work is a first encouraging step into binary ensemble seismology. The work presented was only based on a subsample of well-characterized stars. Forthcoming data from the TESS mission will soon provide new detections of solar-like oscillations systems. 
With the launch of the ESA PLATO \citep[and Rauer et al. 2023 subm.]{Rauer2014}, scheduled for 2026, 
we will increase the sample of binary systems with a component characterized by seismology. 
Given the mission goals of observing bright stars \citep{Nascimbeni2022}, the expected yield of the seismic detection for the \Gaia binaries should be similar to those of the K2 mission described in this work, namely, offering an increased number of potential binary systems with seismic characterization.

The forthcoming data releases of the \Gaia mission\footnote{\href{https://www.cosmos.esa.int/web/gaia/release}{https://www.cosmos.esa.int/web/gaia/release}} will allow for a more complete census of the binary population and therefore a closer estimate of the actual binary occurrence rate. 
The 66 months or $\sim$2000\,days of DR4, which is projected to be made public before 2026, will double the current timebase, allowing for the detection of wider systems and substantially improving the residuals of the orbital parameters beyond 500\,days. Furthermore, DR5 is planned to cover all data collected during the entire mission duration.  
Such extended baseline will increase the number and reliability of orbital solution around 1\,000\,days. 
Because most systems with a He-core burning giant (RC) have orbital periods in this regime, this binary census extension will help reduce the current selection bias that produces an abundance of H-shell burning (RGB) primaries. 
The data set of the ESA\,\Gaia mission are truly a Rosetta stone for studying the evolution of binary systems and tidal interactions with evolved stellar components.

\vspace{5mm}

\begin{acknowledgements}

We thank the referee for useful comments that allowed
us to improve the article.

The authors thank the people behind the ESA \Gaia, NASA \Kepler, and NASA TESS missions.

PGB acknowledges support by the Spanish Ministry of Science and Innovation with the \textit{Ram{\'o}n\,y\,Cajal} fellowship number RYC-2021-033137-I and the number MRR4032204. Substantial research work for this paper was performed during a summer sabbatical research stay of PGB at the \textit{Ohio State University} (OSU), Columbus, Ohio, generously supported by the \textit{NAWI Graz Mobility Grant 2022}. PGB thanks OSU for the hospitality and scientific exchange during his stay.

PGB, DG, LS, LSS, and NM acknowledge the financial support by \textit{NAWI\,Graz}. JM acknowledges support from the Instituto de Astrofísica de Canarias (IAC) received through the IAC early-career visitor program. 
SM acknowledges support by the Spanish Ministry of Science and Innovation with the \textit{Ram{\'o}n\,y\,Cajal} fellowship number RYC-2015-17697. SM and DGR acknowledge support from the Spanish Ministry of Science and Innovation with the grant no. PID2019-107187GB-I00.
RAG and StM acknowledge support from the PLATO CNES grant. PG was supported by the German space agency (Deutsches Zentrum für Luft- und Raumfahrt) under PLATO data grant 50OO1501. PG and JJ acknowledge NASA grant NNX17AF74G for partial support. 
KH acknowledges support through NASA ADAP grants (80NSSC19K0594).
LS acknowledges financial support through the \textit{Marshall Plan Foundation Austria} with contract number 2056139429282022).
MV acknowledges support from NASA grant 80NSSC18K1582
 \\

This work has made use of data from the European Space Agency (ESA) mission
\Gaia (\url{https://www.cosmos.esa.int/gaia}), processed by the \Gaia
Data Processing and Analysis Consortium (DPAC,
\url{https://www.cosmos.esa.int/web/gaia/dpac/consortium}). Funding for the DPAC
has been provided by national institutions, in particular the institutions
participating in the \Gaia Multilateral Agreement. %\\
This paper includes data collected with the \textit{Kepler}\,\&\,\TESS missions, obtained from the MAST data archive at the Space Telescope Science Institute (STScI). 
Funding for these missions is provided by the NASA Science Mission Directorate and by the NASA Explorer Program respectively. STScI is operated by the Association of Universities for Research in Astronomy, Inc., under NASA contract NAS 5–26555. 
\\
\textit{Software:} \texttt{Python} \citep{10.5555/1593511}, 
\texttt{numpy} \citep{numpy,Harris_2020},  
\texttt{matplotlib} \citep{4160265},  
\texttt{scipy} \citep{2020SciPy-NMeth},
%\texttt{pandas} \citep{reback2020pandas, mckinney-proc-scipy-2010},
\texttt{Astroquery} \citep{Grinsburg2019_astroquery}.
This research made use of \texttt{Astropy} \citep{astropy:2013, astropy:2018}, a community-developed core Python package for Astronomy.

\end{acknowledgements}

%\newpage 
\bibliographystyle{aa}
\bibliography{AA202346810}

\begin{appendix}

\section{Characteristics of the TBO solutions for SB9 \label{sec:SB9}}

\begin{figure}[t!]
\includegraphics[width=\columnwidth,height=100 mm]{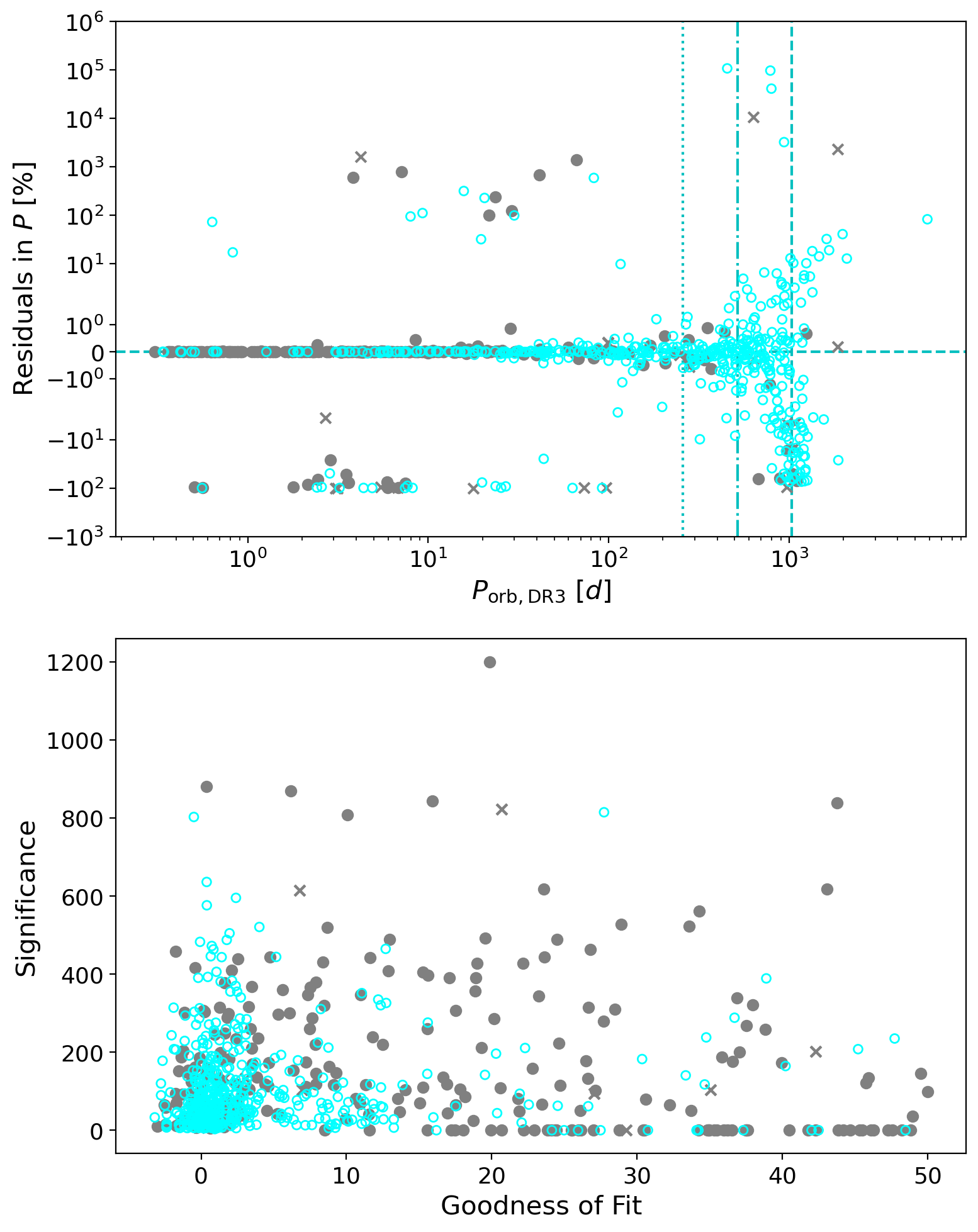}
     \caption{Fractional residuals of the period from the comparison of \Gaia DR3 and SB9 in percent is shown in the top panel. The bottom panel shows the goodness-of-fit and significance parameters originating from the binary solutions in \Gaia DR3 plotted against each other. The vertical lines from left to right indicate the precession period of the \Gaia satellite of 69\,days, as well as the one-fourth, half, and full duration of the \GaiaDR, $\sim$250, $\sim$500, and 1035\,days, respectively. The meaning of the used symbols and colors is similar to Fig.\,\ref{fig:SB9residuals}.}
    \label{fig:SB9comparison}
\end{figure}

\subsection{Definition of a magnitude-limited sample}

The current version of the SB9 \citep[version 2022/03]{Pourbaix2004} lists 4021 systems (2.85\,$\leq$\,G\,[mag]\,$\leq$\,18.77) and provides 5042 orbital solutions from ground-based radial-velocity monitoring. 
We limited the selection to one orbit solution per system to obtain a sample of unique stellar identifiers. For multiple entries for the same orbit of an object, we took the one with the highest SB9 grade. If two solutions had the same grade, we adopted the one with the more recent bibliographic code. The remaining sample was then screened for triples, where we adopted the orbit closer to the \Gaia value for the residual calculation. 
By excluding alternative solutions for binaries and the solutions for triple and quadruple systems, we arrive at 3\,413 unique stellar identifiers in the "full sample."

The \textit{Gaia catalog of Nearby Stars (GCNS)} is a clean catalog of 331\,312 sources within 100\,pc \citep{Smart2021GCNS}. Among the stars in the GCNS, the most common solutions in our sample are the orbital, the astrometric-spectroscopic SB1, and SB solution \citep[see Fig.\,10 in][]{Arenou2022}. The latter two, of which most of the solutions for our binaries consist, are abundant for magnitudes brighter than G\,[mag]\,$\lesssim$\,13. The fainter limit is set by the brightness limit of the spectrograph and radial-velocity (RV) measurements. On the bright end of the GCNS distribution (4\,$\lesssim$\,G\,[mag]), \Gaia observations are limited by saturation effects. 

To obtain the realistic corrected binary yield for the SB9 sample, we calculated the sample size within the magnitude limitations (4\,$\leq$\,G\,[mag]\,$\leq$\,13). The SB9 magnitude limited sample contains 2964 systems. Because the SB9 is compiled from decades of ground-based RV monitoring, it contains systems with periods up to several ten thousand days to which \GaiaDR is not sensitive. Therefore, we corrected the SB9 magnitude-limited sample with a maximum value for the orbital period of $P_{\rm orb, SB9}$\,$\leq$\,1\,100\,d. 
This magnitude-period-limited sample contains 2\,343 unique DR3 source identifiers.

\begin{figure}[t!]\centering
\includegraphics[width=\columnwidth,height=100 mm]{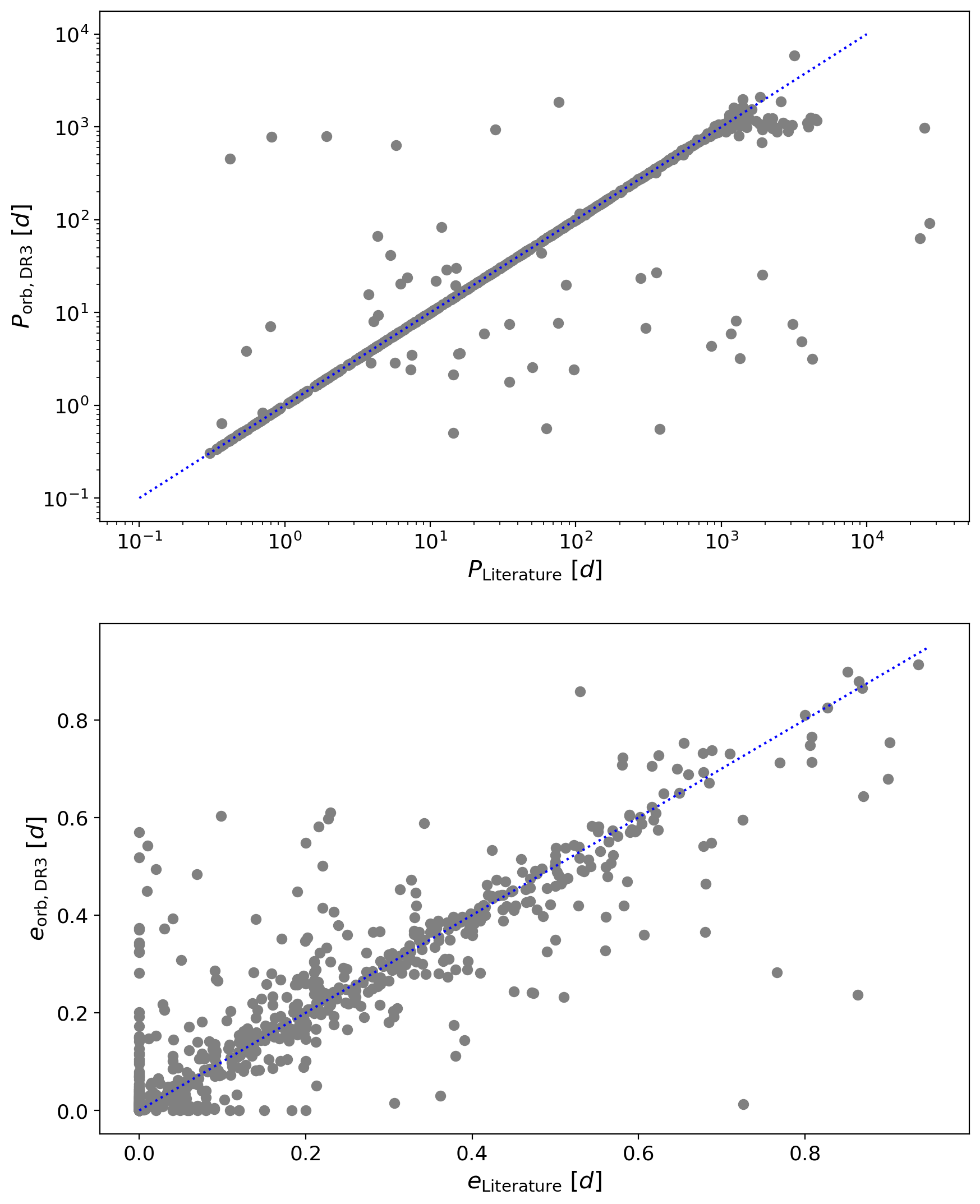}
     \caption{Direct comparison of the orbital period (top) and eccentricity (bottom) of the orbital elements from \Gaia DR3 and SB9. The dotted line represents the 1:1 ratio.}
    \label{fig:ePcomparison}
\end{figure}

\subsection{\GaiaDR TBO completeness factor and residuals \label{sec:completeness}}

The SB9 catalog literally provides the ground(-based) truth. We can assess the completeness and reliability of the solutions provided in the \GaiaDR TBO. The crossmatch of the TBO catalog  with the  SB9 catalog returned 743 matches for the full sample and 668 for the magnitude-period limited sample, corresponding to a corrected binary yield of 21.7\% and 28.5\%, respectively. Because all the searched sources are confirmed binary systems, we can estimate the overall completeness factor of \GaiaDR TBO catalog  to be $\sim$30\%. 

Because the binary fraction is a strong function of the primary's mass \citep[e.g.,][]{Offner2022}, we cannot compare these numbers to those from binary population studies. This is therefore a purely systematic yield of the mission.

This sample is also sufficiently large to test the reliability of the provided values for the orbital parameters. From the full sample, we find the period and eccentricity for 743 and 715 systems (magnitude corrected: 668 and 640), respectively. These numbers show that for about 95\% a complete set of period and eccentricity are reported.
The middle panel of Fig.\,\ref{fig:SB9residuals} shows the $e$-$P$ plane of all 715 systems for which both parameters, eccentricity and period are provided in the SB9 and the \GaiaDR catalog. Consequently, such a system is represented by two data points. This comparison presents an overall good agreement between the SB9 and \GaiaDR solutions.  From the eccentricities' residuals of the full sample (see bottom panel of Fig.\,\ref{fig:SB9residuals}), we find a mean value of 0.011\,$\pm$0.104. Accounting only for systems where the period residual between TBO and SB9 is less than 10\%, the mean residual in eccentricity reduces to 0.004 \,$\pm$0.060. 

A more complex picture is found for the orbital periods (Fig.\,\ref{fig:SB9comparison}, top panel). The periods reported in the \GaiaDR are quite consistent with the periods in the SB9 up to  P$_\mathrm{orb}$\,$\lesssim$250\,d, which is one-fourth of the timebase of \Gaia DR3 of 1035\,d (dashed and dash-dotted cyan vertical lines in Fig.\,\ref{fig:SB9comparison}, respectively). 
%The overall standard deviation is \red{XXX\,$\pm$XXX\,d}. 
In Fig.\,\ref{fig:SB9comparison}, it can be seen that the general scatter in the residuals increases with the increasing orbital period. As these large residuals in the long-periodic regime of the data set would dominate any overall metric, we split the period range into three different regions. 

To test the reliability of the \Gaia periods we show in Fig.\,\ref{fig:SB9comparison} the fractional residuals as the function of the \Gaia period, $P_\mathrm{orb,DR3}$. First we inspected the region of $P_\mathrm{orb,DR3}$\,$\leq$\,250\,days. Following a sigma-clipping approach, we excluded 54 solutions from the magnitude-unconstrained sample with period deviations between TBO and SB9 larger than 10\%. From this comparison, we conclude that $\sim$90\% of the periods in DR3 below 250 days are reliable. 
For the remaining systems, the typical errors (mean and standard deviation) in period and eccentricity are 0.006\,$\pm$0.610\,d, and 0.006\,$\pm$0.057, respectively.

From the magnitude-unconstrained sample in the range of 250\,$\leq$\,P$_\mathrm{orb,DR3}$ [d]\,$\leq$500,  76 out of the 77 systems have period residuals, better than 10\%. This leads us to the conclusion that 99\% of the solutions between 250\,$\leq$\,P$_\mathrm{orb,DR3}$ [d]\,$\leq$500 are reliable. To avoid skewing the distribution by rejecting too many solutions, we report the mean residual values in the unclipped sample. In this period range 
we find a mean residual of 5\,$\pm$52\,days and -0.003\,$\pm$0.075 for the orbital period and eccentricities, respectively.

For all systems with periods in DR3 longer than 500  days $-$ half of the timebase of \Gaia DR3, a mean period scatter of -273\,$\pm$1818\,d is found.  Only 74\%, 147 out of 198 systems with $P_{\rm orb}$\,$\geq$\,500\,days have residuals, better than 10\%.

As it can be seen from Fig.\,\ref{fig:SB9residuals} many periods with SB9-periods longer than P$_\mathrm{orb,DR3}$\,$\gtrsim$\,1100 are largely underestimated in DR3, which leads to the escalating residuals around and above 1000\,days. We look forward to the forthcoming data releases, DR4 and DR5, by the \Gaia mission, with extended timebases.

In the range of $P_\mathrm{orb,SB9}$\,$\leq$\,250\,days the SB9 magnitude limited sample contains 1855 unique systems, of which \GaiaDR TBO catalog  reported 461 as candidates. This corresponds to a completeness factor of 24.9\%. In the period range between 250\,$\leq$\,P$_\mathrm{orb,SB9}$ [d] $\leq$500 of the magnitude-limited sample, contains 41.0 \%. Therefore, the completeness in the shorter periodic regime is lower than the completeness factor for the overall period-limited  sample.

\subsection{Potentially long-periodic systems}
The non-single star (NSS) acceleration catalog reports solutions for objects for which a solution for the proper motion with a higher-order term fits significantly better than a linear solution. As the trajectory of binary systems with long orbital periods is perceived as a parabolic arc, adding constant acceleration to the linear single-star model would enable the detection of such objects as binaries, as discussed in \gaia DR3 documentation. For binary systems with slightly shorter periods, the time derivation of the acceleration has to be considered an additional parameter, as described in the \GaiaDR documentation. This results in the solution types of constant acceleration (Acceleration7) and variable acceleration (Acceleration9), as seen in the \gaia DR3 documentation. 

Figure\,\ref{fig:SB9residuals} depicts the distribution of the 241 systems %(of which \red{XXX}are in the magnitude limited sample) 
found with acceleration and non-linear solutions. As it can be seen, they mostly populate the regime of orbital periods, which extends from  periods just a little longer than the timebase of \GaiaDR to more than 10\,000 day orbits. Interestingly, a few very short-periodic  and circularized systems ($P_\mathrm{orb}$\,$<$ $10$\,days) are also found among these solutions. 

Including these systems with both orbital parameters unresolved in the count of detected systems gives exactly 1\,000 systems out of the 3529 unique stellar identifiers comprised in SB9, which were reported in the \GaiaDR TBO. However, we do not include these systems in the comparison of the fractional binary yields, as the magnitude-period-limited sample of the SB9 was constructed to account for this bias.

%\begin{figure}[t!]
%    \centering
%    \includegraphics[width=\columnwidth]{figures/ruwehisto.png}
%    \caption{\Gaia astrometric detected SB9 binaries, based on   the \textit{ruwe} values for all \Gaia sources of the SB9 sample (blue) and the binary candidates in the TBO catalog  (red). The grey dash-dotted and dashed vertical lines mark the $ruwe$ suggested threshold values of 1.2 and 1.4, respectively.}    \label{fig:ruweHist}\end{figure}

\subsection{Quality metrics in \GaiaDR TBO}

The TBO catalog  provides numerous measures to describe the quality of a solution. The main two parameters are the goodness of fit (GoF) and significance (s). The GoF parameter describes how well a model's prediction matches the observed data \citep{Halbwachs2022}. Due to its definition in Eq.\,1 in \cite{Halbwachs2022}, combining the \chisq-metrics and the degrees of freedom, values range from (small) negative to large positive values, whereby larger GOF values (GOF\,$\gtrsim$\,25) are considered as unreliable.  The significance \textit{s} describes the S/N of the semi-amplitude for spectroscopic binaries and the semi-major axis for astrometric orbits, respectively, as described in \cite{Arenou2022}. 
We also used the SB9 sample to improve our understanding of the quality metrics in the \GaiaDR. 
The bottom panel of Fig.\,\ref{fig:SB9comparison} shows the quality metrics used to select the systems accepted for the catalog entries of confirmed binaries. We find most systems with single-digit GoF values and a wide range of significance.

Filtering data based on the GoF and $s$ parameter has already been done in the post-processing of the \Gaia DR3 catalog. Binary solutions possessing a GoF\,$< 50$ and a $s > 5$ were accepted into the catalog with the exception of binaries possessing the \textit{OrbitalTargetedSearch*} solution, which were accepted possessing a GoF\,$< 50$ and a $s > 2$, as is discussed in \cite{Arenou2022}. In our further search for red giant binaries, we adopted all solutions provided in \GaiaDR without further filtering on these parameters.

Another parameter that was reported to be a good indicator for identifying binary stars is the ruwe of the astrometry. By definition, a ruwe\,$\simeq$\,1 suggests that the astrometric solution satisfies a single star model, while a larger value suggests a binary star solution. We tested this metric based on the SB9 magnitude- and period-limited sample. For the TBO-SB9 comparison we find 59\% with ruwe\,$\geq$\,1.4 (dashed vertical line in Fig.\,\ref{fig:ruweHist}), a solid threshold for binary detection from the proper motions, according to \cite{Arenou2022}. A less conservative limit of  ruwe\,$\geq$\,1.2 still includes 65\% of all confirmed binaries. The distribution peak is located at a ruwe\,$\simeq$\,1.0, while the maximum value is found at ruwe\,$\simeq$\,40. 

While it shows that ruwe is a good indicator, we find that a significant number of the confirmed binaries are found with low ruwe values suggesting (likely) single-star objects.  As shown in Fig.\,\ref{fig:ruweDist}, these systems are likely to be distant systems in shorter orbits, where the projected motion of the stellar component due to their binarity is small compared to the proper motion of the system. A small ruwe also means that the astrometry is very accurate, providing a good parallax for this system which provides better constraints for the stellar model.

\subsection{Known oscillating giants in binaries}
In addition to the SB9 sample, the second comparison sample consists of binary systems with an oscillating red giant binary component. This catalog, referred to here as the "literature sample," was recently published by \cite{Beck2022SB9} and contains eclipsing binaries reported by 
\citet[]{Hekker2010}, 
\citet[]{Frandsen2013},
\citet[]{Gaulme2014, Gaulme2016, Gaulme2020},
\citet[]{Rawls2016},
\citet[]{Handberg2017},
\citet[]{Brogaard2018}, 
\citet[]{Themessl2018},
\cite{GaulmeGuzik2019},  and
\citet[]{Benbakoura2021},
as well as the catalogs of red giant heartbeat stars, compiled by \citet[]{Beck2014a,Beck2015Toulouse}. In addition to those 81 systems, \cite{Beck2022SB9} identified 99 systems, previously unknown to host an oscillating giant components from \Kepler, TESS and BRITE data, using the inventory of the SB9 catalog.
Both samples are presented in the bottom panel of Table\,\ref{tab:ArchiveResults}.
The position of the red giant binaries in the literature sample is also shown in Fig.\,\ref{fig:seismicHRD}.

Of these 190 systems  53 are located in the magnitude limited sample. All systems in this limited sample were identified as binaries by \GaiaDR. Additionally, 116 were listed as possible binaries in the non-linear and acceleration solutions. Since half of them are also listed in the SB9, it is challenging to destill these detection rates into an independent metric.

%\newpage
\section{Tables of orbital and seismic values \label{sec:AppendixTables}}
This section presents a limited version of the online tables for the full binary sample in \GaiaDR. %A more extended version is available on CDS (LINK will be given) 
In addition to the values provided in Tables\,\ref{tab:A1}\,and\,\ref{tab:B1}, the online version of this table, in a machine-readable format on CDS (if available) contains the effective temperature and metallicity from apogee. 

\subsection{Systems with orbital parameters in \GaiaDR}

%\red{give a short explanation for content and values in} Table\,\ref{tab:A1}.
Table\,\ref{tab:A1} presents the catalog of binary-system candidates that host a solar-like oscillator, for which the orbital period or eccentricity are reported in \GaiaDR.
The first three identifiers for the star in the relevant catalogs are provided.
The first column reports the source ID in the \GaiaDR catalog. The second column indicates the stars identifier in the \textit{Kepler} or TESS input catalog (KIC or TIC, respectively). The 2MASS identifier is given in the third column.  

The next four columns report values from the TBO  \GaiaDR catalog for the binary-system candidate. The ruwe indicates the astrometric error. The $P_{\rm orb}$ and $e$ values give the period and eccentricity of the orbit. The inclination of the orbit is given after the conversion of the Thiele-Innes coefficients to the Campbell formalism$^{\ref{fn:inclinationTool}}$.

The type of the TBO catalog  solution is listed in the next column, whereby
SB1 stands for Single Lined Spectroscopic binary model,
ASB1 
(\textit{=\,AstroSpectroSB1} in the official documentation\footnote{the more detailed documentation is found \href{https://gea.esac.esa.int/archive/documentation/GDR3/Gaia_archive/chap_datamodel/sec_dm_non--single_stars_tables/ssec_dm_nss_two_body_orbit.html}{https://gea.esac.esa.int/archive/documentation/GDR3/Gaia\_archive/ chap\_datamodel/sec\_dm\_non--single\_stars\_tables/ ssec\_dm\_nss\_two\_body\_orbit.html}}) for combined astrometric + single lined spectroscopic orbital model, 
ORB \textit{(=\,Orbital)} for an orbital model for an astrometric binary, and OTS \textit{(=\,OrbitalTargetedSearch)} for orbital model for a priori known systems, with a subset containing the suffix Validated.

The final three columns report the asteroseismic parameters. The global seismic parameters $\nu_{\rm max}$ and $\Delta\nu$ describe the mean frequency of the excess of oscillation power and the mean large frequency separation between consecutive modes of the same spherical degree, $\ell$. The asymptotic period spacing between pure gravity dipole ($\ell$=1) modes, $\Delta\Pi_1$ is given in the last column.

\subsection{Systems with non-linear and acceleration solutions in \GaiaDR}
Table\,\ref{tab:B1} presents the catalog of binary-system candidates that host a solar-like oscillator, for which no orbital elements are reported in \GaiaDR. Because the found solution for the star's proper motion does not agree with that of a single star, these are flagged as acceleration and non-linear solutions in the non-single star catalog of \GaiaDR. %\red{Acceleration7 = ACC7, Acceleration9 = ACC9, FirstDegreeTrendSB1 = FSB1, SecondDegreeTrendSB1 = SSB1}

The type of the TBO catalog  solution is listed in the next column, whereby
ACC7 (\textit{=\,Acceleration7} in the official documentation) and ACC9 (=\,\textit{Acceleration9}) stand for
an acceleration model\footnote{\href{https://gea.esac.esa.int/archive/documentation/GDR3/Gaia_archive/chap_datamodel/sec_dm_non--single_stars_tables/ssec_dm_nss_acceleration_astro.html}
{https://gea.esac.esa.int/archive/documentation/GDR3/Gaia\_archive/ chap\_datamodel/sec\_dm\_non--single\_stars\_tables/ ssec\_dm\_nss\_acceleration\_astro.html}} with 7 and 9 parameters, respectivley. 
FSB1 (=\,\textit{FirstDegreeTrendSB1}), and
SSB1 (=\,\textit{SecondDegreeTrendSB1}) stand for specialized solutions\footnote{\href{https://gea.esac.esa.int/archive/documentation/GDR3//Gaia_archive/chap_datamodel/sec_dm_non--single_stars_tables/ssec_dm_nss_non_linear_spectro.html}{https://gea.esac.esa.int/archive/documentation/GDR3//Gaia\_archive/ chap\_datamodel/sec\_dm\_non--single\_stars\_tables/ ssec\_dm\_nss\_non\_linear\_spectro.html}}.
Similarly to Table\,\ref{tab:A1}, we report the three relevant source identifiers, the ruwe of the source and the key asteroseismic parameters $\nu_{\rm max}$, $\Delta\nu$, and $\Delta\Pi_1$.

\subsection{Confirmed \GaiaDR binary systems from APOGEE radial velocities}
Table\,\ref{tab:ApogeeBinaryDetection} presents the catalog of binary-system candidates in the \textit{Kepler} field of view, that have been confirmed or supported through RV-variations, derived from single epoch spectroscopy of the APOGEE project, released in the DR16.
The first and second column report the source identifiers in the \textit{Kepler} and 2MASS input catalog. The orbital period for the system candidate, as reported in \GaiaDR is given. 
The next three columns indicate the number of spectra taken, the timebase between the first and the last observation, and the mean uncertainty of the RV measurements. The next column reports the mean difference between spectra in the time series. 
If the source is indeed a binary, according to our significance criterion it is marked with an "x."

\clearpage

\tabcolsep=1pt

\clearpage
\onecolumn
%\scriptsize % 
\tiny

\begin{longtable}{rrr|rrrrr|rrr}

\caption{Orbital parameters of binary systems reported in \GaiaDR, and the seismic parameter of solar-like oscillating primary.
\label{tab:A1}}\\

\hline\hline
\multicolumn{1}{c}{Gaia DR3} & 
\multicolumn{1}{c}{KIC/TIC}	& 
\multicolumn{1}{c}{2MASS}	&
\multicolumn{1}{c}{ruwe} &
\multicolumn{1}{c}{$P_{\rm orb}$} &
\multicolumn{1}{c}{$e$} &
\multicolumn{1}{c}{$i$} &
\multicolumn{1}{c}{Type} &
\multicolumn{1}{c}{$\nu_{\rm max}$} &
\multicolumn{1}{c}{$\Delta\nu$} &
\multicolumn{1}{c}{$\Delta\Pi_1$}\\
\multicolumn{1}{c}{~} & 
\multicolumn{1}{c}{~}	& 
\multicolumn{1}{c}{~}	&
\multicolumn{1}{c}{~} &
\multicolumn{1}{c}{[d]} &
\multicolumn{1}{c}{~}	&
\multicolumn{1}{c}{[deg]} &
\multicolumn{1}{c}{~} &
\multicolumn{1}{c}{[$\mu$Hz]} &
\multicolumn{1}{c}{[$\mu$Hz]} &
\multicolumn{1}{c}{[s]}\\
[1mm] \hline
\endfirsthead
\caption{Orbital and seismic parameters of \GaiaDR binary systems. (cont.)}\\

\hline
\multicolumn{1}{c}{Gaia DR3} & 
\multicolumn{1}{c}{KIC/TIC}	& 
\multicolumn{1}{c}{2MASS}	&
\multicolumn{1}{c}{ruwe} &
\multicolumn{1}{c}{$P_{\rm orb}$} &
\multicolumn{1}{c}{$e$} &
\multicolumn{1}{c}{$i$} &
\multicolumn{1}{c}{Type} &
\multicolumn{1}{c}{$\nu_{\rm max}$} &
\multicolumn{1}{c}{$\Delta\nu$} &
\multicolumn{1}{c}{$\Delta\Pi_1$}\\
\multicolumn{1}{c}{~} & 
\multicolumn{1}{c}{~}	& 
\multicolumn{1}{c}{~}	&
\multicolumn{1}{c}{~} &
\multicolumn{1}{c}{[d]} &
\multicolumn{1}{c}{~}	&
\multicolumn{1}{c}{[deg]} &
\multicolumn{1}{c}{~} &
\multicolumn{1}{c}{[$\mu$Hz]} &
\multicolumn{1}{c}{[$\mu$Hz]} &
\multicolumn{1}{c}{[s]}\\
[1mm] \hline
\endhead
\hline
\endfoot
2050236453034915840 & KIC 1026084 & J19241348+3642145 & 0.93 & 393$\; \pm$6 & 0.10$\; \pm$0.09 & $-$ & SB1 & 41$\; \pm$1 & 4.4$\; \pm$0.1 & 251\\
2051023943874834688 & KIC 1569823 & J19222790+3711159 & 2.25 & 283.5$\; \pm$1.5 & 0.42$\; \pm$0.10 & $-$ & SB1 & 44$\; \pm$5 & 5.4$\; \pm$0.5 & $-$\\
2051764945994896000 & KIC 1572049 & J19242968+3711000 & 1.19 & 361.3$\; \pm$4.1 & 0.01$\; \pm$0.05 & $-$ & SB1 & 52$\; \pm$3 & 5.19$\; \pm$0.05 & $-$\\
2051723679950384256 & KIC 1872210 & J19291111+3718547 & 1.71 & 549.7$\; \pm$6.5 & 0.39$\; \pm$0.04 & $-$ & SB1 & 72$\; \pm$2 & 6.24$\; \pm$0.04 & 251\\
2051786760127550208 & KIC 2016676 & J19274515+3727100 & 1.31 & 842.5$\; \pm$29.3 & 0.07$\; \pm$0.10 & $-$ & SB1 & 81$\; \pm$2 & 6.95$\; \pm$0.05 & 255\\
2051797450308126336 & KIC 2161731 & J19270486+3734137 & 7.93 & 1066.0$\; \pm$12.6 & 0.42$\; \pm$0.01 & 49$\; \pm$1 & ASB1 & 102$\; \pm$2 & 8.1$\; \pm$0.1 & $-$\\
2051656918979678592 & KIC 2167774 & J19321701+3731246 & 1.39 & 137.0$\; \pm$0.2 & 0.24$\; \pm$0.02 & $-$ & SB1 & 60$\; \pm$1 & 5.79$\; \pm$0.01 & $-$\\
2099089472639454208 & KIC 2283075 & J19040434+3737566 & 1.28 & 57.35$\; \pm$0.03 & 0.17$\; \pm$0.02 & $-$ & SB1 & 69$\; \pm$1 & 6.47$\; \pm$0.02 & 69\\
\end{longtable}
%\newpage ~ \newpage
%\end{landscape}
%}
\clearpage
\normalsize

%second table

\clearpage
\onecolumn
\tabcolsep=10pt
%\scriptsize % 
\tiny

\begin{longtable}{rrr|rr|rrr}
%\centering
\caption{Seismic parameters for potential binary systems in \GaiaDR, with non-linear and acceleration solutions.
\label{tab:B1}}\\

\hline\hline
\multicolumn{1}{c}{Gaia DR3} & 
\multicolumn{1}{c}{KIC/TIC}	& 
\multicolumn{1}{c}{2MASS}	&
\multicolumn{1}{c}{ruwe} &
\multicolumn{1}{c}{Type} &
\multicolumn{1}{c}{$\nu_{\rm max}$} &
\multicolumn{1}{c}{$\Delta\nu$} &
\multicolumn{1}{c}{$\Delta\Pi_1$}\\
\multicolumn{1}{c}{~} & 
\multicolumn{1}{c}{~} & 
\multicolumn{1}{c}{~} &
\multicolumn{1}{c}{~} &
\multicolumn{1}{c}{~} &
\multicolumn{1}{c}{[$\mu$Hz]} &
\multicolumn{1}{c}{[$\mu$Hz]} &
\multicolumn{1}{c}{[s]}\\
[1mm] \hline
\endfirsthead
\caption{Seismic parameters for potential binary systems in \GaiaDR, with non-linear and acceleration solutions.  (continued) }\\

\hline
\multicolumn{1}{c}{Gaia DR3} & 
\multicolumn{1}{c}{KIC/TIC}	& 
\multicolumn{1}{c}{2MASS}	&
\multicolumn{1}{c}{ruwe} &
\multicolumn{1}{c}{Type} &
\multicolumn{1}{c}{$\nu_{\rm max}$} &
\multicolumn{1}{c}{$\Delta\nu$} &
\multicolumn{1}{c}{$\Delta\Pi_1$}\\
\multicolumn{1}{c}{~} & 
\multicolumn{1}{c}{~} & 
\multicolumn{1}{c}{~} &
\multicolumn{1}{c}{~} &
\multicolumn{1}{c}{~} &
\multicolumn{1}{c}{[$\mu$Hz]} &
\multicolumn{1}{c}{[$\mu$Hz]} &
\multicolumn{1}{c}{[s]}\\
[1mm] \hline
\endhead
\hline
\endfoot
2050253151867756160 & KIC 1160867 & J19233983+3650426 & 2.34 & ACC7 & 4.7$\; \pm$0.1 & 0.89$\; \pm$0.02 & $-$  \\
2051742028048983296 & KIC 1163359 & J19255838+3650557 & 1.34 & SSB1 & 21.4$\; \pm$0.5 & 2.64$\; \pm$0.02 & $-$  \\
2051760410508998912 & KIC 1431599 & J19250117+3705112 & 1.16 & FSB1 & 2.30$\; \pm$0.05 & 0.5$\; \pm$- & $-$  \\
2051751030300733440 & KIC 1433593 & J19264298+3704199 & 0.88 & FSB1 & 69$\; \pm$1 & 6.24$\; \pm$0.02 & $-$  \\
2051714265379610624 & KIC 1435467 & J19281984+3703353 & 5.73 & FSB1 & 1369$\; \pm$57 & 70.8$\; \pm$1.5 & $-$  \\
2051816245082293632 & KIC 1725815 & J19293996+3716306 & 4.04 & ACC9/SSB1 & 1040$\; \pm$29 & 56$\; \pm$2 & $-$  \\
2051819062581067776 & KIC 2019396 & J19300833+3726004 & 1.01 & FSB1 & 37$\; \pm$1 & 4.2$\; \pm$0.1 & $-$  \\
2099171691193663872 & KIC 2284679 & J19060736+3740100 & 1.82 & SSB1/ACC7 & 198$\; \pm$1 & 16.15$\; \pm$0.03 & $-$  \\
\hline
\end{longtable}

\clearpage
\onecolumn
\tabcolsep=10pt

\small

\begin{longtable}{rrr|crr|rccc}

\caption{APOGEE radial velocity measurements for \GaiaDR binaries.\label{tab:ApogeeBinaryDetection}} \\

\hline\hline 
\multicolumn{1}{c}{KIC}	&
\multicolumn{1}{c}{2MASS}	&
\multicolumn{1}{c}{$P_{\rm orb, DR3}$} &
\multicolumn{1}{c}{N} &
\multicolumn{1}{c}{$\Delta$T} &
\multicolumn{1}{c}{$\sigma_{\rm RV}$} &
\multicolumn{1}{c}{mean $\Delta$RV} &
\multicolumn{1}{c}{Binary}\\
\multicolumn{1}{c}{~} & 
\multicolumn{1}{c}{~} &
\multicolumn{1}{c}{[d]} &
\multicolumn{1}{c}{~} &
\multicolumn{1}{c}{[d]} &
\multicolumn{1}{c}{[km/s]} &
\multicolumn{1}{c}{[km/s]} &
\multicolumn{1}{c}{[d]} &
\multicolumn{1}{c}{~}\\
[1mm] \hline
\endfirsthead
\caption{APOGEE radial velocity measurements for \GaiaDR binaries (cont.)}\\

\hline
\multicolumn{1}{c}{KIC}	&
\multicolumn{1}{c}{2MASS}	&
\multicolumn{1}{c}{$P_{\rm orb, DR3}$} &
\multicolumn{1}{c}{N} &
\multicolumn{1}{c}{$\Delta$T} &
\multicolumn{1}{c}{$\sigma_{\rm RV}$} &
\multicolumn{1}{c}{mean $\Delta$RV} &
\multicolumn{1}{c}{Binary}\\
\multicolumn{1}{c}{~} & 
\multicolumn{1}{c}{~} &
\multicolumn{1}{c}{[d]} &
\multicolumn{1}{c}{~} &
\multicolumn{1}{c}{[d]} &
\multicolumn{1}{c}{[km/s]} &
\multicolumn{1}{c}{[km/s]} &
\multicolumn{1}{c}{[d]} &
\multicolumn{1}{c}{~}\\
[1mm] \hline
\endhead
\hline
\endfoot
%KIC	&	2MASS_ID	&	period	&	NVISITS	&	Days between first/last obs	&	VSCATTER	&	VERR_MED	&	Days between first/last obs	&	Binary	\\		
2991448	&	J19271797+3806475	&	3328.4	&	2	&	16	&	0.064	&	0.013	&			x	\\		
2991448	&	J19271797+3806475	&	1201.2	&	2	&	16	&	0.064	&	0.013	&			x	\\		
3630240	&	J19060409+3843351	&	561.8	&	5	&	11	&	0.209	&	0.024	&			x	\\		
3730801	&	J19042876+3848423	&	57.6	&	3	&	31	&	6.804	&	0.019	&			x	\\		
3942719	&	J19115818+3902135	&	339.3	&	3	&	31	&	0.828	&	0.018	&			x	\\		
3952580	&	J19240489+3900592	&	45.7	&	2	&	16	&	19.004	&	0.014	&			x	\\		
4446300	&	J19012602+3933549	&	104.7	&	3	&	31	&	0.751	&	0.011	&			x	\\		
4914923	&	J19163489+4002501	&	99.2	&	2	&	74	&	4.068	&	0.009	&			x	\\
\end{longtable}

\clearpage
\normalsize

\end{appendix}

\end{document}